\documentclass{article}
\usepackage[utf8]{inputenc}
\usepackage{jheppub}

\usepackage{amsmath,amssymb,amsthm,amscd,mathrsfs,bm}
\usepackage{mathtools}

\usepackage{xcolor}
\usepackage{url}
\usepackage{listings}
\usepackage{tabularx}
\usepackage{diagbox}
\usepackage{tikz-cd}
\usepackage{nicematrix}
\usepackage{pifont}
\usepackage{subfig}
\usepackage{booktabs}
\usepackage{float}
\newcommand{\CS}{\mathcal{S}}

\usepackage{tikz,textcomp}
\usetikzlibrary{decorations.pathreplacing,calc,fadings,fit,shapes,arrows,positioning,chains,matrix}

\newtheorem{theorem}{Theorem}[section]

\newtheorem{definition}[theorem]{Definition}

\newcommand{\bq}{\boldsymbol{q}}

\DeclareMathOperator{\ord}{ord}
\DeclareMathOperator{\disc}{disc}

\newcommand{\IZ}{\mathbb{Z}}
\newcommand{\IC}{\mathbb{C}}
\newcommand{\IP}{\mathbb{P}}
\newcommand{\IN}{\mathbb{N}}
\newcommand{\IR}{\mathbb{R}}
\newcommand{\IQ}{\mathbb{Q}}

\newcommand{\cA}{{\cal A}}
\newcommand{\cO}{{\cal O}}
\newcommand{\cD}{{\cal D}}

\newcommand{\cM}{{\cal M}}
\newcommand{\cS}{{\cal S}}

\newcommand{\cF}{{\cal F}}
\newcommand{\cB}{{\cal B}}
\newcommand{\cL}{{\cal L}}
\newcommand{\cN}{{\cal N}}

\newcommand{\sF}{\mathscr{F}}

\newcommand{\sS}{\mathscr{S}}

\newcommand{\fm}{\mathfrak{m}}
\newcommand{\e}{\mathrm e}

\newcommand{\ii}{\mathrm{i}}
\newcommand{\dd}{\mathrm{d}}

\newcommand{\DT}{\mathrm{DT}}
\newcommand{\PT}{\mathrm{PT}}
\newcommand{\mum}{\mathrm{MUM}}

\newcommand{\X}{X}
\newcommand{\gammaMirror}{\gamma^\circ}
\newcommand{\ZX}{Z_X}
\newcommand{\ZXmirror}{Z_{\Xmirror}}
\newcommand{\OLR}{\Omega_{\mathrm{LR}}}

\newcommand{\zi}{\boldsymbol{z}}
\newcommand{\zib}{\boldsymbol{\bar{z}}}

\newcommand{\Fred}{\cF_g^{\mathrm{red}}}

\newcommand{\coh}{\mathrm{coh}}
\newcommand{\Knum}{K^{\mathrm{num}}}
\newcommand{\sH}{H}
\newcommand{\Heven}{H^{\mathrm{even}}}
\DeclareMathOperator{\ch}{ch}
\newcommand{\Td}{\mathrm{Td}}

\newcommand{\gtop}{g_s}

\newcommand{\Mcs}{\cM_{\mathrm{cs}}}
\newcommand{\PF}{\mathscr{D}}
\newcommand{\Xmirror}{X^{\circ}}
\newcommand{\piFrob}{\boldsymbol{\varpi}}
\newcommand{\piGeom}{\boldsymbol{\Pi}}
\newcommand{\periodV}{\boldsymbol{\Pi}}
\newcommand{\conifold}{\mathfrak{c}}
\newcommand{\GV}{\text{D2D0}}
\newcommand{\MSW}{\text{MSW}}

\newcommand{\oD}{\mathsf{D}}

\title{\centering Borel singularities and Stokes constants \\of the topological string free energy \\on one-parameter Calabi-Yau threefolds}

\abstract{
We study the Borel plane of the topological string free energy on all hypergeometric one-parameter Calabi-Yau models close to singular points in moduli space, focusing on the location of Borel singularities and the value of the associated Stokes constants. We find in particular that in models which exhibit massless D-branes at a singular point, the central charge of the D-brane close to the singular point coincides with the location of the leading Borel singularity, and the generalized Donaldson-Thomas invariant associated to the charge of the D-brane, in as far as its value is known, coincides with the Stokes constant associated to the Borel singularity. 
}

\author[a]{Simon Douaud}
\author[a]{Amir-Kian Kashani-Poor}

\affiliation[a]{Laboratoire de Physique de l’\'Ecole normale sup\'erieure,\\
CNRS, PSL Research University and Sorbonne Universit\'es,\\
24 rue Lhomond, 75005 Paris, France}

\emailAdd{simon.douaud@ens.fr}
\emailAdd{amir-kian.kashani-poor@ens.fr}

\DeclareFontFamily{U}{wncy}{}
\DeclareFontShape{U}{wncy}{m}{n}{<->wncyr10}{}
\DeclareSymbolFont{mcy}{U}{wncy}{m}{n}
\DeclareMathSymbol{\Sh}{\mathord}{mcy}{"58} 

\begin{document}

\maketitle
\listoftables

\section{Introduction}
A non-perturbative formulation of string theory on arbitrary backgrounds is still lacking. Topological string theory -- a remarkably rich subsector of string theory compactified on Calabi–Yau threefolds -- is more tractable than the full theory. It offers a large variety of methods to perform computations to high order in the string coupling, and often allows for a partial resummation of such contributions. However, it too still lacks a non-perturbative formulation. In the topological setting, the theory is effectively characterized by a single observable, the topological string free energy, or equivalently, its exponential, the topological string partition function. Much effort has therefore been expended on constructing a non-perturbative completion of this object, see \cite{Lockhart:2012vp,Grassi:2014zfa,Grassi:2017qee,Alim:2024dyi,Hattab:2024ewk,Bonelli:2024wha} for a sample of such works over the last decade. However, recent developments in the resurgent analysis of the topological string free energy \cite{Gu:2022sqc,Gu:2023mgf,Gu:2023wum} indicate that the situation is potentially more interesting than simply identifying a single, well-defined non-perturbative function. The resurgent data associated with the asymptotic expansion of the topological string free energy -- data encoded in the distribution of the singularities of its Borel transform in the Borel plane and the associated Stokes constants -- promises to be equally, if not more, informative. In this work, we therefore focus on studying the Borel plane of the topological string free energy in the class of one-parameter hypergeometric Calabi-Yau models.

Our work stands on two principal pillars. The pioneering work \cite{Huang:2006hq} demonstrated that by leveraging the holomorphic anomaly equations \cite{Bershadsky:1993cx}, one can compute the topological string amplitudes $F_g$ to much higher genus than what is currently feasible in the full string theory. Aside from limitations due to computer resources, the obstacle towards obtaining arbitrary high genus results lies in providing the necessary boundary conditions to resolve the holomorphic ambiguity at each genus. Substantial progress has been made on this front in the recent works \cite{Alexandrov:2023zjb, Alexandrov:2023ltz}. As a result, complete expressions for the non-holomorphic topological string amplitudes for all one-parameter hypergeometric models are now accessible via an online resource\footnote{\label{footnote:link}\href{http://www.th.physik.uni-bonn.de/Groups/Klemm/data.php}{http://www.th.physik.uni-bonn.de/Groups/Klemm/data.php}} up to genus 14 to 66, depending on the model. This data proves sufficient to study at least the leading singularity of the Borel transform of the topological string free energy. Systematically advancing beyond the leading singularity relies on a second recent breakthrough regarding the study of the holomorphic anomaly equations, now regarding their non-perturbative content  \cite{Gu_2023, Gu:2023mgf}. Building on previous work \cite{Couso-Santamaria:2013kmu, Couso-Santamaria:2014iia,Couso-Santamaria:2016vcc}, these references show how to compute instanton corrections governing the asymptotic behavior of the Borel transform of the topological string free energy to arbitrary high perturbative order.

The primary aim of our investigation into the Borel plane of hypergeometric Calabi-Yau models is to elucidate the relation between resurgence structure and BPS invariants. A plethora of works \cite{Gaiotto:2009hg,Kashani-Poor:2015pca,Gu_2022,Gu:2022sqc, Gu:2023mgf,Grassi:2022zuk,Gu_2023, Marino:2023nem,Alexandrov:2023wdj,Iwaki:2023cek,Marino:2024yme} have explored various intricate, often indirect connections between the BPS spectrum of $\cN=2$ theories and asymptotic analysis. To cite one early example in the context of Seiberg-Witten $\cN=2$ gauge theories, the construction of BPS states is reduced in \cite{Klemm:1996bj} to the computation of geodesic lines on the Seiberg-Witten curve. These same lines reappear in the exact WKB analysis of equations associated to the gauge theory: as so-called WKB curves that enter in the solution of a certain flatness condition defined in terms of the Hitchin system associated to the theory in \cite{Gaiotto:2009hg}, and, in \cite{Kashani-Poor:2015pca}, as the Stokes lines associated to the exact WKB analysis of the Mathieu equation, related to the gauge theory via the AGT correspondence \cite{Alday:2009aq}. More recently, the exact form of the Stokes automorphism acting on the topological string amplitude is computed in \cite{Iwaki:2023cek,Marino:2024tbx}; the authors of these works observe that their formula coincides with a wall-crossing formula proposed in \cite{Alexandrov:2015xir} if one identifies the Stokes constants appearing in the former with the generalized Donaldson-Thomas invariants occurring in the latter. The authors of \cite{Marino:2024yme} observe wall-crossing phenomena on the Borel plane in the context of Seiberg-Witten theory. A relation between resurgence and Donaldson-Thomas theory has also been investigated in the mathematics literature \cite{Bridgeland:2016nqw,Bridgeland:2024ecw}. Close to the spirit of this work, in \cite{Gu:2023mgf}, based on the structure of the Gopakumar-Vafa formula, Stokes constants associated to singularities close to the MUM point are identified analytically with Gopakumar-Vafa invariants at genus 0. As singularities of the Borel transform of topological string amplitudes conjecturally occur at integral periods of the Calabi-Yau variety (extending the analysis in \cite{Gu:2023mgf}, we provide further ample evidence for this conjecture in this work), we seek enumerative invariants incorporating Gopakumar-Vafa invariants that can be associated to \emph{any} integral periods of the Calabi-Yau; generalized Donaldson-Thomas invariants \cite{Thomas:1998uj,Joyce} fit the bill.

Our numerical analysis will mainly focus on the least well understood singular point of complex structure moduli space, the one lying at $z = \infty$ (in conventions in which the MUM point lies at the origin). As generalized DT invariants have not been computed around this point, we compare the Stokes constants that we can determine numerically to generalized DT invariants associated to the large radius point of the mirror, i.e. to $z=0$. When the values coincide, we interpret this as evidence that no wall-crossing occurs between the points $z=0$ and $z=\infty$; in the case of discrepancy, the simplest explanation is the occurrence of wall-crossing. An analysis of the interior conifold point of the hypergeometric models provides further evidence that wall-crossing phenomena can explain the distribution of singularities on the Borel plane of topological string amplitudes, as well as help predict the values of the associated Stokes constants.

This work is organized as follows: In Section \ref{s:topString}, we review aspects of the topological string that are central to the analysis in this work, in particular the relation between the anholomorphic amplitudes obtained via the holomorphic anomaly equations, and their holomorphic limits, from which enumerative invariants can be extracted. In Section \ref{s:hypergeometric}, we review relevant aspects of the Calabi-Yau varieties that we will be focusing on, one-parameter hypergeometric models, and study their periods close to the point $z = \infty$. In Section \ref{s:asymptoticMethods}, we summarize the basics of Borel analysis that underpin our investigation. We review the known asymptotic behavior of the topological string amplitudes close to the conifold and MUM points, and cite results from the work \cite{Gu:2023mgf} which will enter in our numerical analysis. Section \ref{s:generalizedDT} gathers essential background about generalized Donaldson-Thomas invariants. Finally, Section \ref{s:experimentalData} forms the core of this work. We explore the Borel plane of all thirteen hypergeometric Calabi-Yau models\footnote{We ignore the fourteenth model, which is not thought to correspond to a smooth variety.} close to the internal conifold point $z = \mu$, the point $z = \infty$, and also two rank 2 attractor points. The behavior close to $\mu$ is universal, and we can be brief. The behavior close to $z = \infty$ is model dependent and takes up the majority of this section. We introduce Section \ref{s:experimentalData} with a summary of our numerical observations. Finally, we outline some of the many avenues for future work in Section \ref{s:conclusions}.

\section{The topological string free energy} \label{s:topString}
\subsection{Anholomorphic topological string amplitudes}
The object of study in this paper is the topological string free energy 
\begin{equation}
    F = \sum_{g=0}^\infty F_g \gtop^{2g-2} 
\end{equation}
defined with regard to a Calabi-Yau variety $X$.
$F$ is a formal power series in $\gtop$. We will consider the topological string amplitudes $F_g$ from the B-model perspective: for $g \ge 2$, they are defined as solutions to the holomorphic anomaly equations with boundary conditions imposed to ensure the correct encoding of enumerative invariants when expanded around a point of maximum unipotent monodromy. The $F_g$ thus obtained are functions of the coordinate $\zi$ on the complex structure moduli space $\Mcs(\Xmirror)$ of $\Xmirror$, and its complex conjugate $\zib$. $\Xmirror$ here designates the mirror variety to $X$. The special geometry on $\Mcs(\Xmirror)$ is entirely determined by the holomorphic (3,0) form $\Omega \in H^{3,0}(\Xmirror)$, which is unique up to normalization, via the relations
\begin{equation}
    \e^{-K} = i \int \Omega \wedge \bar{\Omega}\,, \quad C_{ijk} = \int \Omega \wedge \partial_i \partial_j \partial_k \Omega \,, \quad i,j,k=1, \ldots, n \,,
\end{equation}
where we have set $n=b_2(X) = \frac{b_3(\Xmirror)-1}{2}$. A central role is played by the periods of $\Omega$,
\begin{equation}
    X^I=\int_{A_I}\Omega,\quad P_I=\int_{B^I}\Omega \,,
    \label{eq:Periods}
\end{equation}
defined with regard to a symplectic basis of $H_3(X,\IZ)$,
\begin{equation} 
    A_I\cap A_J=B^I\cap B^J=0\,, \quad  A_I\cap B^J=-B^J\cap A_I=\delta_I^J\,.
\label{eq:polarization}
\end{equation} 
In the following, we will organize the periods in a period vector 
\begin{equation}
    \piGeom = (P_I,X^I) \,.
\end{equation}
Given a charge vector $\boldsymbol{q} \in \IZ^{n}$, we will denote the period of $\Omega$ along $(B^0,B^1,A_0,A_1)\cdot \boldsymbol{q}$ as $\Pi_{\boldsymbol{q}}$.

It was shown in \cite{Yamaguchi:2004bt, Alim:2007qj} that the topological string amplitude $F_g$ is a polynomial in the anholomorphic functions, $S^{ij}$, $S^i$, $S$, called propagators, and derivatives of the K\"ahler potential $K$; all dependence of $F_g$ on $\zib$ is through these generators. The propagators are defined as solutions to the equations
\begin{equation} \label{eq:defPropagators}
    \partial_{\bar{k}} S^{ij} = C^{ij}_{\bar{k}} \,, \quad  \partial_{\bar{k}} S^{j} = G_{i \bar{k}} S^{ij} \,, \quad
    \partial_{\bar{k}} S = G_{i \bar{k}}S^i \,,
\end{equation}
where
\begin{equation}
    G_{i \bar{j}} = \partial_i \partial_{\bar{j}} K \,, \quad C^{ij}_{\bar{k}} = \e^{2K} G^{i \bar{m}}G^{j \bar{n}} \bar{C}_{\bar{k} \bar{m} \bar{n}} \,.
\end{equation}
Introducing shifted propagators \cite{Alim:2007qj}
\begin{equation}
    \tilde{S}^i = S^i - S^{ij}K_j \,, \quad \tilde{S} = S - S^i K_i + \frac{1}{2} S^{ij} K_i K_j 
\end{equation}
permits expressing the topological string amplitudes as polynomials in just the (shifted) propagators, $F_g(S^{ij},\tilde{S}^i, \tilde{S})$, with holomorphic coefficients.

\subsection{Taking the holomorphic limit by choosing a frame} \label{ss:holLimit}
To relate $F_g(S^{ij},\tilde{S}^i, \tilde{S})$ to quantities of physical relevance, a so-called holomorphic limit must be taken. In many cases, this amounts to fixing a point $\zib^*$ and considering the limit $\zib \rightarrow \zib^*$. This path towards introducing the holomorphic limit obscures the role played by the periods of $\Omega$. Close to a MUM point, we can introduce a canonical symplectic basis of $H_3(\Xmirror,\IZ)$, as we review briefly for one-parameter models in Section \ref{ss:singularPoints}. In the following, we will reserve the notation $(X^0,X^1, \ldots, X^n)$ and $(P_0,P_1, \ldots, P_n)$ for the A- and B-periods of $\Omega$ with regard to this choice. In these conventions, the limit $\zib \rightarrow \zib^{\mum}$ taken on the anholomorphic special geometry quantities yields
\begin{equation} \label{eq:holomorphicLimitI}
\e^{-K} \rightarrow X^0 \,, \quad K_i \rightarrow -\partial_i \log(X^0) \,, \quad \Gamma_{jk}^i \rightarrow \frac{\partial z^i}{\partial t^\ell}\frac{\partial^2 t^\ell}{\partial z^j \partial z^k} \,,
\end{equation}
where
\begin{equation} \label{eq:holomorphicLimitII}
    t^i = \frac{X^i}{X^0} \,.
\end{equation}
This induces a map
\begin{equation}
    S^{ij} \rightarrow \cS^{ij}\,, \quad \tilde{S}^i \rightarrow \tilde{\cS}^i \,, \quad \tilde{S} \rightarrow \tilde{\cS} 
\end{equation}
which permits defining a holomorphic limit of the free energies,
\begin{equation} \label{eq:holLimitMum}
    F_g(S^{ij},\tilde{S}^i, \tilde{S}) \rightarrow F_g(\cS^{ij},\tilde{\cS}^i, \tilde{\cS}) = \cF_g(X^0,X^1,\ldots,X^n)\,.
\end{equation}
The $\cF_g(X^0,X^1,\ldots,X^n)$ thus defined are homogeneous functions of degree $2-2g$ in the periods $X^0, \ldots, X^n$. By pulling out factors of $(X^0)^{2-2g}$, we obtain generating functions $\sF_g$ of Gromov-Witten invariants as functions of affine coordinates $t^i = X^i/X^0$,
\begin{equation}
    \cF_g(X^0,X^1,\ldots,X^n) = (X^0)^{2-2g}\cF_g(1,X^1/X^0,\ldots,X^n/X^0) = (X^0)^{2-2g}\sF_g(t^1, \ldots, t^n)\,.
\end{equation}
More generally, we call any choice of symplectic basis of $H_3(\Xmirror,\IZ)$ a choice of frame, and the limit \eqref{eq:holomorphicLimitI} and \eqref{eq:holomorphicLimitII} taken with regard to the A-periods that follow from this choice we call the associated holomorphic limit.

In the following, we will indicate the choice of frame which yields a particular $\cF_g$ as the holomorphic limit of $F_g$ by listing the A-periods in this frame in parentheses. Thus, $\cF_g(X^0,X^1, \ldots,X^n)$ will indicate the topological string amplitude $F_g$ denoted by $\cF_g$ in \eqref{eq:holLimitMum}. As the arguments of the formal power series $\cF_g(X^0,X^1, \ldots,X^n)$ coincide with the variables given in parentheses, we will ordinarily drop them from the notation.

\subsection{Specializing to one parameter hypergeometric models}
All of our numerics will be based on a class of one parameter models which we discuss in the next section. For one parameter models, the holomorphic limit of the propagators, in a frame which we indicate by *, can be expressed in terms of the following determinants of the $2 \times 2$ minors of the Wronskian matrix of the period vector:
\begin{equation}
  w_{i,j}=\partial^j_z X^0_* \partial_z^i X^1_*  -\partial^i_z X^0_*\partial^j_z X^1_* \,, \quad i,j = 0,1,2,3 \,.
\end{equation}
They are
\begin{equation}
  \label{propas}
 \begin{gathered}
   \CS^{zz}=-\frac{1}{C} \left( \frac{w_{0,2}}{w_{0,1}}-q^z_{zz}\right), \qquad 
 \tilde \CS^z=-\frac{1}{C} \left(  \frac{w_{1,2}}{w_{0,1}}-q_{zz} \right),\\
   \tilde \CS=-\frac{1}{2 C}\left(\frac{w_{1,3}}{w_{0,1}}-\partial_z
     q_{zz}-q_{zz} q^z_{zz}\right)+\frac{\partial_z C}{2 C^2}\left(
     \frac{w_{1,2}}{w_{0,1}}-q_{zz} \right)-\frac{q_z^z}{2}\ .
 \end{gathered}
\end{equation}
The functions $q_{zz}$, $q_z^z$ and $q^z_{zz}$ are the propagator holomorphic ambiguities that can be chosen conveniently.\footnote{A file available from the online resource linked in footnote \ref{footnote:link} contains a convenient set of choices for all hypergeometric models.}

\section{Hypergeometric Calabi-Yau models} \label{s:hypergeometric}
\subsection{Singular points on complex structure moduli space} \label{ss:singularPoints}
One approach towards systematically studying the phenomenon of mirror symmetry in one-parameter Calabi-Yau models (by this, we mean Calabi-Yau varieties with $h^{1,1}(X) = h^{2,1}(\Xmirror) = 1$) is to shift the focus from the Calabi-Yau variety to candidate Picard-Fuchs equations satisfied by the periods of the holomorphic (3,0) form $\Omega$ of the purported mirror variety, see e.g. \cite{Almkvist:2005qoo}. The operators considered are of the form
\begin{equation}
    \PF = \sum_{i=0}^k z^i P_i(\theta) \,, \quad \theta = z \frac{d}{dz} \,,
\end{equation}
with $P_i(x)$ a fourth order polynomial with integral coefficients, and they are constrained to satisfy a list of properties \cite{Almkvist:2004kj}. Restricting to $k=1$ yields the 14 so-called hypergeometric models. 13 of these describe smooth complete intersections listed in Table \ref{tab:hypergeometric}: the notation $X_{d_1,\ldots,d_r}(w_1,\ldots,w_{4+r})$ indicates the zero locus of $r$ polynomials of degree $d_1,\,\ldots,\, d_r$ in weighted projective space $\mathbb{P}^{3+r}(w_1,\ldots,w_{4+r})$.
\begin{table}[h!]
{{ 
\begin{center}
	\begin{tabular}{|c|c|c|c|c|c|c|c|}
		\hline
\#&  $X$ 	                          & $\kappa$	& $c_2 \cdot D$& $\chi(X)$    & $a_1,a_2,a_3,a_4$						        & $\mu^{-1}$ & $dT_\infty$ \\\hline
1 & $X_5(1^5)$ 	             & $5$		& $50$		& $-200$ & $\frac{1}{5},\frac{2}{5},\frac{3}{5},\frac{4}{5}$			& $5^5$	& $O_5^{\small \rm DG}$	  	 \\[1mm]
2 &  $X_{6}(1^4 2^1)$          & $3$		& $42$		& $-204$& $\frac{1}{6},\frac{1}{3},\frac{2}{3},\frac{5}{6}$			& $2^43^6$&   $O_6^{\small \rm DG}$	\\[1mm]
3& $X_{8}(1^4 4^1)$	     & $2$		& $44$		& $-296$ & $\frac{1}{8},\frac{3}{8},\frac{5}{8},\frac{7}{8}$			& $2^{16}$ & $O_8^{\small \rm DG}$	\\[1mm]
4 & $X_{10}(1^3 2^1 5^1)$  & $1$		& $34$		& $-288$ 	& $\frac{1}{10},\frac{3}{10},\frac{7}{10},\frac{9}{10}$		& $2^85^5$ & $O_{10}^{\small \rm DG}$  \\[1mm]
5 & $X_{4,3}(1^5 2^1)$        & $6$		& $48$		& $-156$  & $\frac{1}{4},\frac{1}{3},\frac{2}{3},\frac{3}{4}$			& $2^63^3$ 	& $O_{12}$	\\[1mm]
 6 & $X_{6,4}(1^3 2^2 3^1)$ & $2$		& $32$		& $-156$ & $\frac{1}{6},\frac{1}{4},\frac{3}{4},\frac{5}{6}$			& $2^{10}3^3$ & $O_{24}$	\\[1mm]
7&  $X_{4,2}(1^6)$	             & $8$		& $56$		& $-176$& $\frac{1}{4},\frac{1}{2},\frac{1}{2},\frac{3}{4}$			& $2^{10}$ & $C_{4}$	\\[1mm]
8 & $X_{6,2}(1^5 3^1)$        & $4$		& $52$		& $-256$ & $\frac{1}{6},\frac{1}{2},\frac{1}{2},\frac{5}{6}$			& $2^83^3$& $C_{6}$	\\[1mm]
9 & $X_{3,2,2}(1^7)$            &  $12$	& $60$		& $-144$  & $\frac{1}{3},\frac{1}{2},\frac{1}{2},\frac{2}{3}$			& $2^43^3$ & $C_{6}$	\\[1mm]
10 & $X_{3,3}(1^6)$               & $9$		& $54$		& $-144$ & $\frac{1}{3},\frac{1}{3},\frac{2}{3},\frac{2}{3}$			& $3^6$&  $ K_{3}$	\\[1mm]
11 & $X_{4,4}(1^4 2^2)$        &  $4$		& $40$		& $-144$ & $\frac{1}{4},\frac{1}{4},\frac{3}{4},\frac{3}{4}$			& $2^{12}$& $K_{4}$ 	\\[1mm]
12 & $X_{6,6}(1^2 2^2 3^2)$ & $1$		& $22$		& $-120$  & $\frac{1}{6},\frac{1}{6},\frac{5}{6},\frac{5}{6}$			& $2^83^6$	& $K_{6}$\\[1mm]
13 & $X_{2,2,2,2}(1^8)$	     & $16$		& $64$		& $-128$& $\frac{1}{2},\frac{1}{2},\frac{1}{2},\frac{1}{2}$			& $2^8$  	& $M_{2}$\\[1mm]\hline

	 \end{tabular}	
\end{center}}}
\caption[The 13 hypergeometric Calabi-Yau varieties]{The 13 hypergeometric Calabi-Yau varieties $X$ are smooth complete intersections of $r$ polynomials $P_j$ of degree $d_j$ in the weighted  projective ambient spaces 
 $\mathbb{P}^{3+r}(w_1,\ldots,w_{4+r})$. We denote them as  $X_{d_1, \ldots, d_r}(w_1,\ldots,w_{4+r})$. Their mirrors $\Xmirror$ have one-dimensional complex structure moduli space. Denoting  by $H$ the generator of $H^2(X,\mathbb{Z})$, we give the  triple intersection number $\kappa=H^3$, the intersection of $H$ with the second Chern class of the tangent bundle $c_2(TX)$ and the Euler constant $\chi(X)$ of $X$. We also give the data defining the operator $\PF$ and  the degeneration type $dT_{\infty}$ of the  mixed Hodge structure at $z=\infty$. }
  \label{tab:hypergeometric}
\end{table}

The operator $\PF$ at $k=1$ can be put in the normal form 
\begin{equation}  
\PF \; = \;\theta^4- \mu^{-1} z \prod_{k\; = \;1}^4(\theta+a_k) \,.
\label{eq:PFhyper}
\end{equation} 
These Picard-Fuchs operators have three singular points lying at $z \in\{0, \,\mu ,\,\infty\}$. 
The local exponents at these points are 
summarized by the Riemann symbol
\begin{equation}
{\cal  P}\left\{\begin{array}{ccc}
0& \mu& \infty\\ \hline
0& 0 & a_1\\
0& 1 & a_2\\
0& 1 & a_3\\
0& 2 & a_4 
\end{array}\right\}\ \ \,.
\label{eq:riemannsymbolgeneral}
\end{equation} 
We note that all hypergeometric models share the same local exponents around the points $z=0$ and $z=\mu$. The local exponents at a point $z \in \Mcs$ govern the behavior of a standardized set of solutions to \eqref{eq:PFhyper} in a power series solution, augmented by logarithmic solutions if necessary, around $z$; these are called the Frobenius basis $\piFrob_z$ of solutions. We will review the cases that occur at $z=\infty$ in the family of one-parameter hypergeometric models in Section \ref{ss:dominatingContributions}. At $z=0$, the four coinciding local exponents imply that the Frobenius solution is of the form
\begin{equation} \label{eq:FrobMUM}
    \piFrob_0 = 
    \begin{pmatrix}
        f_0 \\
        f_0 \log z + f_1 \\
        \frac{1}{2} f_0 \log^2 z + f_1 \log z + f_2 \\
        \frac{1}{6} f_0 \log^3 z + \frac{1}{2} f_1 \log^2 z + f_2 \log z + f_3
    \end{pmatrix} \,,
\end{equation}
with the $f_i$ holomorphic functions in the coordinate $z$. Relying on mirror symmetry, the matrix $T_0$ relating $\piFrob_0$ to a geometric basis $\piGeom(z)$ of periods,
\begin{equation} \label{eq:defT0}
    \piGeom(z) = T_0 \piFrob_0(z)\,,
\end{equation}
can be formulated in terms of topological invariants of $X$, 
\begin{equation} \label{eq:T0explicit}
    T_0 = (2 \pi \ii)^3
        \begin{pmatrix}
            \frac{\chi \zeta(3)}{(2 \pi \ii)^3} & \frac{c_2}{(2\pi \ii) 24} & 0 & \frac{\kappa}{(2 \pi \ii)^3} \\
            \frac{c_2}{24} & \frac{\sigma}{2 \pi \ii} & - \frac{\kappa}{(2\pi \ii)^2} & 0 \\
            1 & 0 & 0 & 0 \\
            0 & \frac{1}{2\pi \ii} & 0 & 0
        \end{pmatrix} \,,
\end{equation}
see e.g. \cite[Appendix B.4]{Bonisch:2022slo} for a review. We will follow the same conventions; in particular, in the following, $\{B^0, B^1, A_0,A_1\}$ will refer to the canonical choice of basis introduced with regard to the MUM point at $z=0$. Given $T_0$, analytic continuation then permits determining the transition matrices $T_z$ for any point $x\in \Mcs$.

The solutions of the Picard-Fuchs equation can undergo monodromy upon circling a singular point $z^*$. The type of monodromy they undergo depends on the local exponents at $z^*$. More specifically, 
\begin{itemize}
    \item if all four local exponents coincide, the monodromy is maximally unipotent, and $z^*$ is called a point of maximally unipotent monodromy (MUM), or an \textbf{M-point}. 

    \textbf{Occurrence:} at the origin for all hypergeometric models, at $z^* = \infty$ for $X_{2,2,2,2}$.

    \item if exactly two of the local exponents coincide, a 3-cycle shrinks to zero at the singular point $z^*$, which is called a \textbf{conifold point} or a \textbf{C-point}. The period of $\Omega$ around this shrinking cycle is the coefficient of the sole logarithm appearing in the periods in the Frobenius basis.

    \textbf{Occurrence:} at $z^* = \mu$ for all hypergeometric models, at $z^* = \infty$ for $X_{4,2}$, $X_{6,2}$ and $X_{3,2,2}$.

    \item if the local exponents consist of two pairs of coinciding local exponents, i.e. $a_1 = a_2 \neq a_3 = a_4$, the singular point $z^*$ is called a \textbf{K-point}. Two logarithmic periods occur in the Frobenius basis.

    \textbf{Occurrence:} at $z^* = \infty$ for $X_{3,3}$, $X_{4,4}$, and $X_{6,6}$.

    \item if all four local exponents at $z^*$ are pairwise different, the monodromy around $z^*$ will be of finite order, i.e. no logarithms occur in the periods. Such a point is called an \textbf{F-point}.

    \textbf{Occurrence:} at $z^* = \infty$ for $X_{5}$, $X_{6}$, $X_{8}$, $X_{10}$, $X_{4,3}$ and $X_{6,4}$.
    
\end{itemize}

\subsection{Which D-branes dominate the dynamics?} \label{ss:dominatingContributions}
In our analysis of the asymptotic behavior of the topological string amplitudes $F_g$ at a point $z^*\in \Mcs$, we will be interested in comparing the absolute value of integral periods as we approach $z^*$, as we naively would expect the periods with smallest absolute value to dominate the asymptotics. The physical intuition behind this expectation is that the asymptotics is constrained by non-perturbative physics, see Section \ref{s:asymptoticMethods}; this in turn is governed by the lightest D-branes, where the mass of a D-brane of charge $\boldsymbol{q}$ is given by the absolute value of the associated central charge,
\begin{equation} \label{eq:massDbrane}
    Z(\boldsymbol{q}) = \e^{K/2} \,\boldsymbol{q}\cdot \piGeom \,, \quad M(\boldsymbol{q}) = |Z(\boldsymbol{q})| \,, 
\end{equation}
and is hence proportional to the absolute value of the period of $\Omega$ on the cycle the brane wraps.

\paragraph{Generic points of $\Mcs$}

Note that at a generic point on moduli space, this criterion does not give us a  hierarchy of contributions which is bounded below: the set $S$ of integer linear combinations of three generic complex numbers $a, b, c \in \IC$, representing periods of $\Omega$ over integral cycles, is dense in $\IC$. To see this, find $\alpha, \beta \in \IR$ such that
\begin{equation}
    c = \alpha a + \beta b  \,.
\end{equation}
The set $S$ is hence given by
\begin{equation}
    S = \{ (r + t \alpha) a + (s + t \beta)b : r,s,t \in \IN \} \,.
\end{equation}
If $\alpha$ and $\beta$ are irrational, the set of coefficients of $a$ and of $b$ in $S$ is dense in $\IR$, hence $S$ is dense in $\IC$. Thus, at a generic point in moduli space where at least three of the four periods are linearly independent over $\IQ$, integer linear combinations of periods of arbitrarily small (but non-vanishing) norm exist. 

\paragraph{Attractor points} 

On the other hand, something remarkable happens when non-trivial linear relations among the periods exist over $\IQ$. Each such relation yields an integral period with vanishing norm. Such periods exist at point of $\Mcs$, called rank 2 attractor points, at which the elements $\mathrm{Re}\, \Omega, \,\mathrm{Im}\, \Omega \in H^3(\Xmirror,\IC)$ align with the integral lattice $H^3(\Xmirror,\IZ)$ to form a sublattice $\Lambda$ of rank 2. Elements of the dual of the orthogonal lattice $\Lambda^\perp$ will pair with $\Omega$ to yield vanishing periods. The study of such points has elicited much interest \cite{Moore:2004fg,Candelas:2019llw,Candelas:2021mwz}. However, their significance in terms of the asymptotic behavior of the topological string amplitudes $F_g$ is not clear: while precisely at an attractor point, non-vanishing integral cycles with vanishing periods exist, immediately off the point, the associated periods are no longer dominant, as the generic behavior of unboundedness of the period norm sets in. At the attractor point, the study of these periods (e.g. the extraction of the associated Stokes constant) is difficult as they lie at the origin of the Borel plane. Our analysis of the Borel plane at such points in Section \ref{s:beyondSingular} suggests their irrelevance with regard to the distribution of singularities on the Borel plane.

\paragraph{Degeneration of Hodge structure at $z^* = \infty$}

The point on $\Mcs$ that will be of primary interest to us lies at $z^* = \infty$. At this point, all periods vanish. Determining the rate at which they vanish as one approaches $z^*$ \cite{Joshi_2019, Gu:2023mgf} turns out to be a reliable criterion for predicting the location of the leading singularities in the Borel plane. By \eqref{eq:massDbrane} and the fact that $\e^{K/2} \rightarrow 0$ as $w \rightarrow 0$, this rate determines whether or not the D-brane at the singular point will be massless \cite{Joshi_2019}. 

The vanishing behavior of the periods in the Frobenius basis coincides with the local exponents of the relevant Picard-Fuchs equation. Generically, all integral periods will be a linear combination of all Frobenius periods and hence vanish as $(w)^{a_1}$, for F- and C-points, as  $(w)^{a_1}\log^3(w)$ for M-points, or $w^{a_1}\log{w}$, for K-points, as $w=\frac{1}{z}\rightarrow 0$, with $a_1$ the smallest local exponent of the system. Whether integral linear combinations of periods are possible for which this leading behavior is cancelled out is therefore determined by whether the entries in the appropriate column of the transition matrix $T_\infty$ -- recall that this matrix maps the Frobenius basis at $z^* = \infty$ to an integral basis -- satisfy an integer relation or not.\footnote{The computation of $T_\infty$ based on analytic continuation of the Meijer $G$-function has been outlined in the literature, see e.g. \cite[Section 3.2]{Bonisch:2022mgw}. For completeness, we provide a Mathematica \cite{Mathematica} readible text file containing the transition matrices for all hypergeometric models in the conventions of Section \ref{s:hypergeometric} at \url{http://www.phys.ens.fr/~kashani/}.}

\begin{itemize}
    \item F-points: 
      \begin{equation}
        \piFrob_\infty^{(F)}(w)\; = \;\left(
            \begin{array}{l} 
                w^{a_1}+O\left(w^{a_1+1}\right)\\ 
                w^{a_2}+O\left(w^{a_2+1}\right)\\ 
                w^{a_3}+O\left(w^{a_3+1}\right)\\ 
                w^{a_4}+O\left(w^{a_4+1}\right)
            \end{array}
        \right)\, .
            \label{RpointFrobenius}
        \end{equation}
    The behavior at these points is model dependent: for the three models $X_6$, $X_{4,3}$ and $X_{6,4}$, the entries of the first column of $T_\infty$ are linearly dependent over $\IZ$, permitting the definition of two periods that fall off as $z^{-a_2}$ rather than $z^{-a_1}$ as $z \rightarrow \infty$. In terms of the canonical integral symplectic basis of $H_3(\Xmirror,\IZ)$ defined below \eqref{eq:defT0}, the lattice of charge vectors with this leading behavior is spanned by
    \begin{align}
    &X_6\::\: \boldsymbol{q}_1= \begin{pmatrix}
     1\\
     0\\
     0\\
     -1
    \end{pmatrix},
    \quad  \boldsymbol{q}_2= \begin{pmatrix}
     1\\
     1\\
     -3\\
     0
    \end{pmatrix}\,;\quad X_{4,3}\::\:\boldsymbol{q}_1= \begin{pmatrix}
     1\\
     0\\
     0\\
     -2
    \end{pmatrix},
    \quad  \boldsymbol{q}_2= \begin{pmatrix}
     1\\
     1\\
     -3\\
     1
    \end{pmatrix}\,;\\
    & X_{6,4}\::\:
    \boldsymbol{q}_1= \begin{pmatrix}
     1\\
     0\\
     0\\
     -1
    \end{pmatrix},
    \quad  \boldsymbol{q}_2= \begin{pmatrix}
        -1 \\
     -1\\
     2\\
     0
    \end{pmatrix}.
     \label{eq:massless period orb}
   \end{align}
   Due to the special nature of these periods, the length of their orbits under the monodromy action of $M_\infty$ are less than the order of the monodromy group. For future reference, we record the orbits here: 
\begin{align}
      &X_{6}\::\:\{M_{\infty}\boldsymbol{q}_1,\,M_{\infty}^2\boldsymbol{q}_1,\,M_{\infty}^3\boldsymbol{q}_1\}=\{\boldsymbol{q}_2,-(\boldsymbol{q}_1+\boldsymbol{q}_2),\,\boldsymbol{q}_1\}\,,\\
       & X_{4,3}\::\:\{M_{\infty}\boldsymbol{q}_1,\,M_{\infty}^2\boldsymbol{q}_1,\,M_{\infty}^3\boldsymbol{q}_1\}=\{\boldsymbol{q}_2,-(\boldsymbol{q}_1+\boldsymbol{q}_2),\,\boldsymbol{q}_1\}\,,\\
      & 
      \begin{aligned}
          X_{6,4}\::\:\{M_{\infty}\boldsymbol{q}_1,\,\ldots,\,M_{\infty}^4\boldsymbol{q}_1\} =  \{-\bq_2,\,-\bq_1,\,\bq_2,\,\bq_1\} \,.
    \end{aligned}
   \label{monodromies}
    \end{align}

    The remaining three hypergeometric models $X_5$, $X_8$, and $X_{10}$ with F-points at $z = \infty$ do not exhibit such distinguished periods.

    The behavior of the prefactor to the period determining the central charge, see \eqref{eq:massDbrane}, has the following behavior in these models:
    \begin{equation}
        e^{\frac{K}{2}}\underset{w\rightarrow 0}{\sim}\frac{1}{\vert w\vert^{a_1}}\,.
    \end{equation}
    Hence, only the models with distinguished periods exhibit massless D-branes at $z^* = \infty$. For these, the charge lattice of massless D-branes is given by $\IZ \boldsymbol{q}_1 + \IZ \boldsymbol{q}_2$.

    Note that the existence of such massless D-branes at $z^* = \infty$ is consistent with the observation in \cite{Huang:2006hq} that the topological string amplitudes at this point for the models $X_{4,3}$ and $X_{6,4}$ are singular. The Calabi-Yau variety $X_6$ on the other hand was found to have smooth amplitudes at $z^* = \infty$. This is consistent with our observations only if the points of the charge lattice corresponding to massless states are not occupied. Our analysis of the Borel plane of this model in Section \ref{ss:numericalMethods} indeed suggests that this is the case.

    \item C-points:
        \begin{equation}
            \piFrob_{\infty}^{(C)}(w)\; = \;\left(
            \begin{array}{l} 
                w^{a_1}+O\left(w^{a_1+1}\right)\\ 
                w^{a_2}f(w)\\ 
                w^{a_2}f(w)\log(w)+O\left(w^{a_2+1}\right)\\ 
                w^{a_3}+O\left(w^{a_3+1}\right)
            \end{array}
        \right)\,,
        \label{CpointFrobenius}
    \end{equation}
    where $f(w)$ is a power series with non-vanishing constant term.
    For all three hypergeometric models $X_{4,2}$, $X_{6,2}$ and $X_{3,2,2}$ with C-points at $z=\infty$, the transition matrix $T_\infty$ takes the form
    \begin{equation}
    \setlength{\arraycolsep}{5pt}
        T_\infty = \begin{pmatrix}
                        2a & 4c & 4e & 2f \\
                        \ii \,4b  & \ii \,n \,d & 0 & \ii\,4g  \\
                        a + \ii\, b  & 2c+\ii\, d & 2e & f + \ii \,g \\
                        r \,a & c & e & r \,f
                    \end{pmatrix} \,, \quad a,b,c,d,e,f  \in \IR \,,\quad d = 2 \pi \,e \,,
    \end{equation}
    with the constants $r$ and $n$ given by
    \begin{center}
        \begin{tabular}{c|c|c|c}
             \rm{model} & $X_{4,2}$ & $X_{6,2}$ & $X_{3,2,2}$ \\ \hline
                $r$     & 1 & 2 & 2/3 \\ \hline
                $n$     & 2 & 1 & 3 
        \end{tabular}\,.
    \end{center}
    
    Hence, each of the three models exhibits two periods which vanish more rapidly than the generic $w^{a_1}$ behavior, with associated charges
    \begin{equation} \label{eq:qc}
        \boldsymbol{q}_1=\begin{pmatrix}
                    2\\
                    1\\
                    -4\\
                    0
                    \end{pmatrix}
   ,\quad \boldsymbol{q}_2=\begin{pmatrix}
                        1\\
                        0\\
                        0  \\
                        -n
                    \end{pmatrix} \,.
    \end{equation} 
    $\boldsymbol{q}_1 \cdot \piGeom$ vanishes as $w^{a_2}$ and does not contain a logarithm; it is the vanishing period characteristic of a conifold point, given by $\ii \, (n-4)d \,w^{a_2}f(w)$. As $n*r=2$, the $w^{a_1}$ contribution also vanishes in $\boldsymbol{q}_2 \cdot \piGeom$; it is a logarithmic period with contributions from only $\varpi_{\infty,2}^{(C)}$ and $\varpi_{\infty,3}^{(C)}$. The coefficient of the logarithm is $\frac{d}{2\pi}\,(4-n) w^{a_2}f(w)$, identifying it as the (unnormalized)\footnote{The intersection product of the cycles in $H_3(\Xmirror, \IZ)$ associated to these two periods is $(n-4)$. It is possible to define a correctly normalized dual period to $\boldsymbol{q}_1 \cdot \piGeom$, at the expense of including also $\varpi_{\infty,1}^{(C)}$ and $\varpi_{\infty,4}^{(C)}$ contributions.} dual period to $\boldsymbol{q}_1 \cdot \piGeom$. As $\boldsymbol{q}_1 \cdot \piGeom$ has no logarithmic contribution, its monodromy orbit around $z = \infty$ is finite. In fact, it consists of the two charges $\pm \boldsymbol{q}_1$.

    The K\"ahler potential in these three models has the following behavior:
    \begin{equation}
        e^{\frac{K}{2}}\underset{w\rightarrow 0}{\sim}\frac{1}{\vert w\vert^{a_1}} \,.
    \end{equation}
    Hence, all D-branes with charges on the lattice $\IZ \boldsymbol{q}_1 + \IZ \boldsymbol{q}_2$ are massless at $z^* = \infty$.
    
     \item K-points:
     \begin{equation}
        \piFrob_{\infty}^{(K)}(w)\; = \;\left(
         \begin{array}{l} 
                w^{a_1}f(w)\\ 
                w^{a_1}f(w) \log(w) +O\left(w^{a_1+1}\right)\\ 
                 w^{a_2}g(w)\\ 
                 w^{a_2}g(w)\log(w)+O\left(w^{a_2+1}\right)
    \end{array}
    \right)\,,
    \label{eq:KpointFrobenius}
    \end{equation}
    where $f(w)$ and $g(w)$ are power series with non-vanishing constant term. Note that following the convention that $a_1 \le a_2$, the Frobenius period which dominates as $w \rightarrow 0$ is $\varpi_{\infty,2}^{(K)} \sim w^{a_1}f(w)\log(w)$. For all three hypergeometric models $X_{3,3}$, $X_{4,4}$ and $X_{6,6}$ with K-points at $z = \infty$, the transition matrix $T_{\infty}$ is such that two integral periods exist without any logarithmic contributions. The associated charges are
    \begin{align}
    & X_{3,3}\::\:\boldsymbol{q}_1=\begin{pmatrix}
    -1\\
       0\\
        0\\
        3
    \end{pmatrix}
   ,\quad \boldsymbol{q}_2=\begin{pmatrix}
    -1\\
       -1\\
        3\\
        -1
    \end{pmatrix}
      ,\quad X_{4,4}\::\:\boldsymbol{q}_1=\begin{pmatrix}
    -1\\
        0\\
        0\\
        2
    \end{pmatrix}
    \quad \boldsymbol{q}_2=\begin{pmatrix}
    -1\\
        -1\\
        2\\
        0
    \end{pmatrix}\\
    \\
    & X_{6,6}\::\:\boldsymbol{q}_1= \begin{pmatrix}
        -1\\
        0\\
        0\\
        1
    \end{pmatrix},\quad
    \boldsymbol{q}_2= \begin{pmatrix}
        0\\
        -1\\
        1\\
        0
    \end{pmatrix}
  .\label{eq:masslessK}
\end{align}
    
    The contribution which vanishes as $w^{a_1}$ however cannot be canceled.

    Monodromy around $z = \infty$ arises from two sources: the logarithms in the periods, and the fractional local exponents. As the two periods associated to the charges $\boldsymbol{q}_1$ and $\boldsymbol{q}_2$ have no logarithmic contribution, their monodromy orbits around $\infty$, being solely due to the fractional local exponents are finite. For future reference, we record them here:
    \begin{align}
      &X_{3,3}\::\:\{M_{\infty}\boldsymbol{q}_1,\,M_{\infty}^2\boldsymbol{q}_1,\,M_{\infty}^3\boldsymbol{q}_1\}=\{\boldsymbol{q}_2,-(\boldsymbol{q}_1+\boldsymbol{q}_2),\,\boldsymbol{q}_1\}\,,\\
       & X_{4,4}\::\:\{M_{\infty}\boldsymbol{q}_1,\,\ldots,\,M_{\infty}^4\boldsymbol{q}_1\}=\{\boldsymbol{q}_2,-\boldsymbol{q}_1,\,-\boldsymbol{q}_2,\,\boldsymbol{q}_1\}\,,\\
      & 
      \begin{aligned}
          X_{6,6}\::\:\{M_{\infty}\boldsymbol{q}_1,\,\ldots,\,M_{\infty}^6\boldsymbol{q}_1\} =  \{\boldsymbol{q}_1+\boldsymbol{q}_2,\,\boldsymbol{q}_2,\,-\boldsymbol{q}_1,\,-(\boldsymbol{q}_1+\boldsymbol{q}_2),\,-\boldsymbol{q}_2,\,\boldsymbol{q}_1\} \,.
    \end{aligned}
   \label{eq:monodromies_K}
    \end{align}

    For the three hypergeometric models exhibiting K-points at $z^* = \infty$, the K\"ahler potential exhibits the behavior
    \begin{equation}
        e^{\frac{K}{2}}\underset{w\rightarrow 0}{\sim}\frac{1}{\vert w \vert^{a_1}(\log(\vert w \vert))^{\frac{1}{2}}} \,.
    \end{equation}
    Hence, all D-branes with charges on the lattice $\IZ \boldsymbol{q}_1 + \IZ \boldsymbol{q}_2$ are massless at $z^* = \infty$.

    \item M-points: 
      \begin{equation} \label{Mfrobenius}
   \piFrob_{\infty}^{(M)}(w) 
   \, = \,
    \begin{pmatrix}
        g_0(w) \\
        g_0(w)\log(w) + g_1(w) \\
        \frac{1}{2} g_0(w) \log^2(w) +g_1(w) \log(w) + g_2(w) \\
        \frac{1}{6} g_0(w) \log^3(w) + \frac{1}{2}g_1(w) \log^2(w) + g_2(w) \log(w) + g_3(w) 
    \end{pmatrix},
    \end{equation}
    where $g_0(w)=w^{a_1}+O(w^{a_1+1})$ and $g_i(w)=O(w^{a_1+1})$ for $i=1,2,3$.
\end{itemize}
Here, the Frobenius period that dominates at $w\rightarrow 0$ is $ \piFrob_{\infty,4}^{(M)}\sim w^{a_1}\log^3(w)$, followed by  $ \piFrob_{\infty,3}^{(M)}\sim w^{a_1}\log^2(w)$. Given the explicit form of the transition matrix $T_{\infty}$, we can construct integral periods without contributions from these two periods. The associated charges are
\begin{equation}
\boldsymbol{q}_1= \begin{pmatrix}
        2\\
        1\\
        -4\\
        0
    \end{pmatrix}\,,\quad
    \boldsymbol{q}_2= \begin{pmatrix}
        1\\
        0\\
        0\\
        -4
    \end{pmatrix}\,.
    \label{eq : masless M}
    \end{equation}
$\bq_1\cdot\piGeom$ has no logarithmic contribution, while $\bq_2\cdot\piGeom$ behaves as $\bq_2\cdot \piGeom \sim\bq_1\cdot\piGeom\log(w) $. This is analogous to the situation at the MUM points at $z=0$ for all hypergeometric models. We also note the monodromies
\begin{align}
 & M_{\infty}\cdot\bq_1=-\bq_1 \,\\
  &  M_{\infty}\cdot\bq_2=\bq_1-\bq_2 \,.
\end{align}
Applying $M_\infty$ repeatedly to $\bq_2$ thus produces a tower of charges of the form $\pm(\bq_2 +n\bq_1)$, $n\in\mathbb{Z}$.

The K\"ahler potential in this model has the following behavior:
\begin{equation}
    e^{\frac{K}{2}}\underset{w\rightarrow 0}{\sim}\frac{1}{\vert w \vert^{1/2}(\text{log}(\vert w \vert))^{\frac{3}{2}}}.
\end{equation}
 Hence, all D-branes with charges on the lattice $\IZ \boldsymbol{q}_1 + \IZ \boldsymbol{q}_2$ are massless at $z^* = \infty$.
We will discuss further similarities between the M-point and MUM points at $z=0$ in Section \ref{ss:BorelAndTopString}.

\section{Asymptotic methods} \label{s:asymptoticMethods}

\subsection{Basics of Borel analysis}
Perturbative series in quantum field theory tend to have vanishing radius of convergence. By an argument of Dyson \cite{Dyson:1952tj}, this is as it should be if the theory at negative coupling constant is ill-defined. Fortunately, some non-convergent, hence formal, power series can still be associated (in a unique fashion) to an underlying function.\footnote{We closely follow \cite{Sauzin} in this section.}

\begin{definition} 
Given a formal power series $\widetilde{\varphi}(z) = \sum_{n \ge 0} c_n z^{-n} \in \IC[[z^{-1}]]$, we say that $\widetilde{\varphi}$ provides a uniform asymptotic expansion of a function $\varphi: \cD \rightarrow \IC$ defined on an unbounded domain $\cD \in \IC^*$ if there exists a sequence of positive numbers $(K_N)_{N\in\IN}$ such that
    \begin{equation} \label{eq:defAsymptoticExpansion}
        | \varphi(z) - c_0 - c_1 z^{-1} - \cdots - c_{N-1}z^{-(N-1)}| \le K_N |z|^{-N} \quad \text{for all} \quad z\in \cD\,, \quad N \in \IN \,.
    \end{equation}
\end{definition}
The qualifier ``uniform'' in the definition refers to the fact that the constants $K_N$ do not depend on the value $z \in \cD$. In particular, the inequality holds no matter how one approaches infinity in $\cD$. Note that the coefficients $c_n$ are uniquely inductively determined by the relation \eqref{eq:defAsymptoticExpansion}. Indeed, by
\begin{equation}
    |\varphi(z) - c_0 | \le K_1 |z|^{-1} \,,
\end{equation}
we have that 
\begin{equation}
    c_0 = \lim_{\substack{|z|\rightarrow \infty \\ z \in \cD}} \varphi(z)  \,,
\end{equation}
where by assumption of uniformity, the limit must exist. Having defined the coefficients $c_i$, $i=0,\ldots, N-1$, define
\begin{equation}
    \rho_N(z) := \varphi(z) - \sum_{n=0}^{N-1} c_n z^{-n} \,.
\end{equation}
Then,
\begin{equation}
    |\rho_N(z) - c_N z^{-N} | \le K_{N+1} |z|^{-N-1} \,,
\end{equation}
allowing us to conclude that
\begin{equation} \label{eq:cNasymptoticExpansion}
    c_N = \lim_{\substack{|z|\rightarrow \infty \\ z \in \cD}} z^N\rho_N(z)  \,.
\end{equation}
From \eqref{eq:cNasymptoticExpansion}, we can directly read off the bound
\begin{equation}
    |c_N| \le K_N \,.
\end{equation}
When the coefficients $c_n$ of the formal power series $\widetilde{\varphi}$ satisfy the bound $|c_n| \le M L^n n!$ for constants $M,L >0$, $\widetilde{\varphi}$ can be studied by means of the Borel transform.

\begin{definition}
    A formal power series $\widetilde{\varphi} = \sum_{n\ge 0} c_n z^{-n} \in \IC[[z^{-1}]]$ is called a Gevrey-1 series if constants $M,L > 0$ exist such that $|c_n| \le M L^n n!$ for all $n\in \IN$. The vector space of all such power series is denoted by $\IC[[z^{-1}]]_{1}$.
\end{definition}
The ``1'' in the name ``Gevrey-1'' refers to the power of the factorial in the bound on the coefficients $c_n$.

\begin{definition}
    The formal Borel transform is defined as the following linear map between formal power series:
    \begin{eqnarray} \label{eq:Boreltransform}
        \cB :\quad z^{-1} \IC[[z^{-1}]]\quad &\longrightarrow& \quad \IC [[\zeta]] \\
        \widetilde{\varphi} = \sum_{n=0}^\infty c_n z^{-n-1} &\longmapsto& \sum_{n=0}^\infty c_n \frac{\zeta^n}{n!} =: \widehat{\varphi} \nonumber \,.
    \end{eqnarray}
\end{definition}
Note that the Borel transform $\widehat{\varphi}$ of a formal power series $\widetilde{\varphi}$ is a power series with finite radius of convergence if and only if $\widetilde{\varphi} \in \IC[[z^{-1}]]_1$. The relation between $\cB \widetilde{\varphi}$ and a function $\varphi$ of which $\widetilde{\varphi}$ is the asymptotic expansion is established in favorable circumstances via the Laplace transform.

\begin{definition}
    Let $\theta \in [0,2\pi)$. The Laplace transform of a function $\widehat{\varphi} :\e^{\ii \theta} \IR^+ \rightarrow \IC$ is the function $\cL^{\theta} \widehat{\varphi}$ defined by the integral
    \begin{equation} \label{eq:laplaceTransform}
        (\cL^{\theta} \widehat{\varphi})(z) = \int_0^\infty \e^{-z \xi \e^{\ii \theta}} \widehat{\varphi}(\xi \e^{\ii \theta}) \e^{\ii \theta} \dd \xi \,.
    \end{equation}
    For the integral to exist, we require that $\widehat{\varphi}$ be locally integrable along the ray $\e^{\ii \theta}(\IR^*)^+$ (i.e. integrable on any compact subset of $(\IR^*)^+$), integrable on the interval $[0,1]$ (to exclude complications in a neighborhood of the origin) and bounded by 
    \begin{equation} \label{eq:boundIntegrandLaplace}
        | \widehat{\varphi}(\zeta)| \le A \e^{c_0|\zeta|}
    \end{equation}
    for $\zeta \in \e^{\ii \theta} [1 ,\infty)$ and appropriate choice of $A>0$ and $c_0 \in \IR$. For such $\widehat\varphi$, the integral exists for all $z\in \IC: \Re(z) > c_0$.
\end{definition}
The flexibility to choose an integration ray in any direction $\theta$ along which the integral \eqref{eq:laplaceTransform} exists will play an important role in the following.

The fact that the Laplace transform of the Borel transform $\cB \widetilde{\varphi}$ may map the formal power series $\widetilde{\varphi}$ to a function $\varphi$ of which it is the asymptotic expansion is suggested by the fact that 
\begin{equation} \label{eq:LaplaceOfPower}
    \forall n\in \IN: \quad \cL^0(\frac{\zeta^n}{n!})(z) = z^{-n-1} \quad \text{for} \quad \Re z > 0 \,,
\end{equation}
i.e. term by term, the Laplace transform is the inverse of the Borel transform. In fact, the following theorem holds \cite[Theorem 5.20]{Sauzin}:
\begin{theorem}
    Let $\widehat{\varphi}$ be an analytic function in a neighborhood of the origin with analytic continuation along a half-strip containing the ray $\e^{\ii \theta} \IR^+$, and assume that $\widehat{\varphi}$ satisfies the bound \eqref{eq:boundIntegrandLaplace} for some $c_0 \ge 0$. Then the function $\varphi := \cL^{\theta} \widehat{\varphi}$ is holomorphic for $\Re z > c_0$ and has the formal series $\widetilde{\varphi} := \cB^{-1} \widehat{\varphi} \in z^{-1}\IC[[z^{-1}]]_1$ as its uniform asymptotic expansion for any $z : \Re z >c_1$, where $c_1$ is any positive constant such that $c_1 > c_0$.
\end{theorem}
We can therefore define an operator
\begin{equation} \label{eq:BorelResum}
    \sS^\theta = \cL^{\theta} \circ \cB 
\end{equation}
which under favorable circumstances and appropriate choice of $\theta$ maps a 1-Gevrey formal power series $\widetilde{\varphi}$ to a function $\varphi$ of which it is the asymptotic expansion. When this succeeds, $\varphi$ is called a Borel resummation of $\widetilde{\varphi}$.

Note that if the asymptotic expansion of the function $\varphi$ is given by the formal power series $\widetilde{\varphi}$, then for any $a\in\IC$, the asymptotic expansion of $a + \varphi$ is given by $a+\widetilde{\varphi}$. However, the definition \eqref{eq:Boreltransform} of the Borel transform does not allow for power series with a constant term, nor does \eqref{eq:LaplaceOfPower} hold for $n=-1$. To remedy the situation, we can adjoin a generator $\delta$ to the power series ring $\IC[[\zeta]]$ as the image of the constant 1 under $\cB$, and define $\cL^\theta \delta = 1$. Thus,
\begin{eqnarray} \label{eq:BoreltransformExtended}
        \cB :\quad \IC[[z^{-1}]]\quad &\longrightarrow& \quad \IC \delta \oplus \IC [[\zeta]] \\
        \widetilde{\varphi} = \sum_{n=-1}^\infty c_n z^{-n-1} &\longmapsto& c_{-1}\delta +\sum_{n=0}^\infty c_n \frac{\zeta^n}{n!} =: \widehat{\varphi} \nonumber \,,
\end{eqnarray}
and
\begin{equation}
    \cL^{\theta} (a \delta + \widehat{\varphi}) = a+\cL^{\theta} \widehat{\varphi} \,.
\end{equation}

\subsection{Singularities in the Borel plane and Stokes constants}
While the asymptotic expansion of a function, when it exists, is unique, many functions share the same asymptotic expansion. Indeed, adding a function such as $\e^{-n/x^2}$, $n\in \IN^*$, whose asymptotic expansion around infinity is identically 0, changes a function while maintaining its asymptotic expansion. This ambiguity is reflected in the $\theta$ dependence of the Borel resummation operator $\sS^\theta$. When studying an asymptotic series arising in physics, deciding which of the Borel resummations (if any) is the physically relevant one requires additional information. Interestingly, all of these choices may encode information about the non-perturbative physics. This is the point of departure of this paper.

Borel resummations along rays $\e^{\ii \theta_1}\IR^+$ and $\e^{\ii \theta_2}\IR^+$ yield different results if the rays cannot be deformed into each other without passing through a singularity of $\widehat{\varphi}$. Studying the singularities of $\widehat{\varphi}$, which are also referred to as the Borel singularities of $\widetilde{\varphi}$, is hence tantamount to studying the ambiguity in assigning a function to the formal power series $\widetilde{\varphi}$. 

\paragraph{A point on conventions:} From this point forth, it will be more convenient to consider asymptotic expansions around the origin rather than infinity, which amounts to the variable redefinitions $z \rightarrow \frac{1}{z}$, $\zeta \rightarrow \frac{1}{\zeta}$.

\vspace{0.5cm}

We will be interested in a simple pole or logarithmic singularities of $\widehat{\varphi}$, labelled by a set $\Omega$, of the form
\begin{equation} \label{eq:logSing}
    \widehat{\varphi}(\zeta_\omega + \xi) = - \frac{S_\omega}{2\pi} \left( \frac{a_{\omega,-1}}{\xi} +  \log \xi \,\widehat{\varphi}_\omega(\xi)\right) + \text{reg.}\,, \quad \omega \in \Omega\,,\,\,\zeta_\omega \in \IC \,,
\end{equation}
with $\widehat{\varphi}_\omega(\xi)$ an analytic function in a neighborhood of the origin with series expansion
\begin{equation}
    \widehat{\varphi}_\omega(\xi) = \sum_{n=0}^\infty \widehat{a}_{\omega,n} \xi^n \,.
\end{equation}
The constant $S_\omega$ is called the Stokes coefficient and will play an important role in our analysis. It is well-defined if a distinguished normalization of $a_{\omega,-1}$ and $\widehat{\varphi}_\omega$ exists. 

By analyticity, the coefficients $\frac{c_n}{n!}$ of the power series expansion of $\widehat{\varphi}$ in a neighborhood of the origin, as given in \eqref{eq:Boreltransform}, are given by the contour integral
\begin{equation} \label{eq:CoeffCauchy}
    \frac{c_n}{n!} = \frac{1}{2\pi \ii} \oint_0 \frac{\widehat{\varphi}(\zeta)}{\zeta^{n+1}}\dd \zeta \,.
\end{equation}
Again by analyticity, they are sensitive to all singularities of $\widehat{\varphi}(\zeta)$ in the $\zeta$-plane, as these contribute to the integral as the integration contour in \eqref{eq:CoeffCauchy} is pushed out to infinity. When the singularity at $\zeta_\omega$ is of the form \eqref{eq:logSing}, this yields the following leading contribution from this singularity to the coefficients $c_k$ at large $k$:
\begin{equation} \label{eq:CoeffAsymp}
    c_k \sim  \frac{S_\omega}{2\pi} \sum_{n \ge -1} \frac{\Gamma(k-n)}{\zeta_{\omega}^{k-n}} a_{\omega,n} \,, \quad k \gg 1 \,,
\end{equation}
where we have defined $a_{\omega,n} = n!\,\widehat{a}_{\omega,n}$ for $n\ge0$. The singularities of $\widehat{\varphi}$ also are responsible for the ambiguities associated to the Borel resummation. Let us introduce the notation $\Omega_\theta$ as the index set for all singularities of $\widehat{\varphi}$ lying along a ray $\e^{\ii \theta}\IR^+$ in the complex plane. The Laplace transform along this ray is ill-defined. If we can avoid all singularities by performing the Laplace transform slightly below or slightly above the ray, the difference between the two transforms is given by
\begin{eqnarray} 
    \left(\cL^{\theta_+} - \cL^{\theta_-}\right)(\widehat{\varphi})(z) &=& \ii \sum_{\omega \in \Omega_\theta} S_\omega \e^{-\zeta_\omega/z} \left( a_{\omega,-1} + \left(\cL^{\theta_-} \widehat{\varphi}_\omega\right)(z) \right) \label{eq:discRaw}\\
    &=& \ii \sum_{\omega \in \Omega_\theta} S_\omega \e^{-\zeta_\omega/z} \big(\cL^{\theta_-} \left(a_{\omega,-1}\delta +  \widehat{\varphi}_\omega\right)\big)(z) \,.\nonumber
\end{eqnarray}
This motivates the definition of the formal power series
\begin{equation}
    \widetilde{\varphi}_\omega(z) = \cB^{-1}\left(a_{\omega,-1}\delta +  \widehat{\varphi}_\omega \right)(z) = \sum_{n=-1}^\infty a_{\omega,n} z^{n+1} \,. 
\end{equation}
The expression $\e^{-\zeta_\omega/z} \widetilde{\varphi}(z)$ is an example of a trans-series in $z$. To express \eqref{eq:discRaw} as a relation between trans-series, we extend the definition of the Borel resummation operator $\sS^\theta$ introduced in \eqref{eq:BorelResum} to act on trans-series via
\begin{equation}
    \sS^\theta \left(\e^{-\zeta_\omega/z} \widetilde{\varphi}\right)(z) := \e^{-\zeta_\omega/z} \sS^\theta \left(\widetilde{\varphi}\right)(z) \,.
\end{equation}
We can then rewrite \eqref{eq:discRaw} as
\begin{equation}
\sS^{\theta_+}  (\widetilde{\varphi}) = \sS^{\theta_-} \left( \widetilde{\varphi} + \ii \sum_{\omega \in \Omega_\theta} S_\omega  \e^{-\zeta_\omega/z} \widetilde{\varphi}_\omega \right)\,,
\end{equation}
or
\begin{equation} \label{eq:disc}
    \disc_\theta(\widetilde{\varphi}) := \left(\sS^{\theta_+} - \sS^{\theta_-}\right) (\widetilde{\varphi}) = \ii \sum_{\omega \in \Omega_\theta} S_\omega  \e^{-\zeta_\omega/z}  \sS^{\theta_-}(\widetilde{\varphi}_\omega) \,.
\end{equation}

\subsection{Periodic structures on the Borel plane} \label{ss:BorelInQuantum}

The Borel planes of many asymptotic series that we encounter in physics exhibit a periodic structure of singularities, distributed along a ray extending from the origin to infinity. Such a periodic structure arises e.g. when considering solutions to linear differential equations with periodic potentials (such as the Matthieu equation, see e.g. \cite{Kashani-Poor:2015pca} for an analysis of this equation in a context similar to the one in this paper) or non-linear differential equations (such as the Painlevé equation, see e.g. \cite{Garoufalidis:2010ya}). 

In such situations, we call any contribution at $\omega \in \Omega$ which is uncorrelated, in a sense which must be made precise depending on context, to any $\omega' \in \Omega$ such that $\zeta_\omega/\zeta_\omega' \in \IN^{>1}$ a primitive or 1-instanton contribution. We denote the set of singularities at which such primitive contributions occur $\Omega^{(1)}$. For  $\omega \in \Omega^{(1)}$, we introduce the notation
\begin{equation} \label{eq:nInstanton}
    \tilde{\varphi}^{(n\zeta_\omega)}(z) := \e^{-n\zeta_\omega/z} \tilde{\varphi}^{(n)}_\omega(z) 
\end{equation}
for the contribution at $n\zeta_\omega$, $n \in \IN$ which is correlated to the contribution at $\zeta_\omega$, and refer to it as an $n$-instanton correction to $\widetilde{\varphi}$. At $n=1$, we may also drop the super index ${}^{(1)}$ on $\tilde{\varphi}^{(1)}_\omega(z)$. A singular point $\zeta_\omega$ may receive contributions from all $m\zeta_{\omega'} = \zeta_\omega$ for some $\omega' \in \Omega^{(1)}$, $m\in\IN$.

\subsection{Borel analysis and the topological string} \label{ss:BorelAndTopString}
In the classic work \cite{Shenker:1990uf}, the growth of string theory amplitudes with the genus $g$ is estimated to be~$\sim (2g)!$ based on an analysis of the moduli space of Riemann surfaces of genus $g$. This asymptotic behavior, in the setting of topological string theory, has been confirmed in many subsequent works \cite{Marino:2007te,Mari_o_2008,Drukker_2011,Couso-Santamaria:2014iia,  Couso-Santamaria:2016vcc, Gu_2023, Gu:2023mgf}.

In the topological string setting, two structural results allow us to go further: the large radius behavior as governed by the Gopakumar-Vafa formula, and the gap condition at the conifold. Note that we always study the asymptotic behavior of the topological string amplitudes in a holomorphic limit. 

\paragraph{The topological string at large radius} Near the large radius point of $X$, or the associated MUM point of $\Xmirror$, the structure of $\cF_g(X^0,X^1)$ is governed by the Gopakumar-Vafa formula \cite{Gopakumar:1998jq}, from which one can extract the asymptotic behavior of the amplitudes \cite{Gu:2023mgf}. Specialized to one-parameter models, this behavior is given by 
\begin{equation} \label{eq:MumAsymptotics}
    \cF_g(X^0,X^1) \sim \sum_{(d,m) \in \tilde{\Omega}_0^{(1)}} \sum_{\ell =1}^\infty \frac{c_{0,d}}{2\pi^2 \ell^2} \left( \frac{\Gamma(2g-1)}{(\ell\cA_{d,m})^{2g-1}} \ell \cA_{d,m} +\frac{\Gamma(2g-2)}{(\ell\cA_{d,m})^{2g-2}}  \right)
\end{equation}
with
\begin{equation} 
    \cA_{d,m} = \aleph (d X^1 + m X^0) \,, \quad \tilde{\Omega}_0^{(1)} = \big(\IN^* \times \IZ \cup \{0\} \times \IN^* \big)  
\end{equation}
and
\begin{equation}
    c_{0,d} = \left\{
    \begin{array}{ll}
        n_{0, d} & \text{if}\, d\ne0 \\
        -\chi & \text{if}\, d=0\,.\end{array}\right.
\end{equation}
More precisely, the $d=0$ contribution stems from the constant map contribution to $\cF_g(X^0,X^1)$, with both the $\ell$ dependence and the contributions from $(0,m)$ for $m>1$ stemming from the factor $\zeta(2g)\zeta(2g-2)$, while the $\ell$ dependence for $d>0$ can be traced back to a $\zeta(2g)$ factor in the expansion of the $\sin^{-2}g_s$ contribution to the Gopakumar-Vafa formula. This gives a concrete realization of the concept of ``correlated" singularities introduced in Section \ref{ss:BorelInQuantum}: a given curve of degree $d$ gives rise to a doubly infinite lattice of singularities lying at $\ell \cA_{d,m}$; $\ell$ corresponds to the multi-instanton degree. At non-primitive degree $d = m d'$, instantons of different degree can contribute.

$\aleph$ is a normalization constant,\footnote{How $\aleph$ depends on the normalization conventions chosen to define $F_g$ is explained in \cite{Gu:2023mgf} in the discussion around Table 5.}
\begin{equation}
    \aleph = \ii \sqrt{2\pi \ii} \,,
\end{equation}
and the $n_{0,d}$ are genus 0 Gopakumar-Vafa invariants at degree $d$, and $\chi$ is the Euler characteristic of the Calabi-Yau variety $\X$. Comparing to \eqref{eq:CoeffAsymp}, we can identify
\begin{equation} \label{eq:cIdMUM}
    c_{2g-2} = \cF_g(X^0,X^1)  \,.
\end{equation}
This matches the conventions introduced in \eqref{eq:Boreltransform} if we define
\begin{equation} \label{eq:cFIdMUM}
    \widetilde{\cF}(X^0,X^1) = \sum_{g=1}^\infty \cF_g(X^0,X^1) g_s^{2g-1} \,;
\end{equation}
in particular, we set $c_{2k+1} = 0$ for $k\in \IN$. As the associated Borel transform
\begin{equation} \label{eq:borelTransformFgMUM}
    \widehat{\cF}(X^0,X^1) =\sum_{g=1}^\infty \cF_g(X^0,X^1) \frac{\zeta^{2g-2}}{(2g-2)!}
\end{equation}
is a function of $\zeta^2$, the set of singularities in the $\zeta$ Borel plane must be closed under $\zeta_\omega \mapsto -\zeta_\omega$, $\omega \in \Omega$. We indeed observe this reflection symmetry of the Borel plane in our numerical analysis in Section \ref{s:experimentalData}. We therefore identify each contribution to \eqref{eq:MumAsymptotics} proportional to $n_{0,d}$ as being the sum of two singularities situated at $\pm \zeta_\omega$ which contribute identically. The set of Borel singularities contributing to \eqref{eq:MumAsymptotics} is hence parametrized by
\begin{equation}
    \Omega_0^{(1)} = \IZ \times \IZ -\{(0,0)\}\,.
\end{equation}
Once again comparing \eqref{eq:MumAsymptotics} to \eqref{eq:CoeffAsymp} now lets us further identify the one-instanton contributions at $\zeta_\omega = \cA_{d,m}$ with
\begin{equation} \label{eq:StokesDataMUM}
   S_\omega= \left\{
    \begin{array}{ll}
        n_{0,\vert d\vert} & \text{if}\, d\ne0 \\
        -\chi & \text{if}\, d=0
    \end{array}
\right.\,, \quad a_{\omega,-1} = \frac{\cA_{d,m}}{2\pi} \,,\quad a_{\omega,0} = \frac{1}{2\pi} \,, \quad a_{\omega,n>0} = 0  \quad \text{for} \quad \omega = (d,m) \in \Omega_0^{(1)} \,.
\end{equation}
Fixing $\zeta_\omega$, it is natural to interpret all contributions written in terms of $\ell \cA_{d,m}$, as multi-instanton contributions, leading to the identification
\begin{equation}
    S_{\omega,\ell} =  n_{0,|d|}\,, \quad a_{\omega,-1}^{(\ell)} = \frac{\cA_{d,m}}{2\pi \ell} \,,\quad a_{\omega,0}^{(\ell)} = \frac{1}{2\pi \ell^2} \,, \quad a_{\omega,n>0}^{(\ell)} = 0  \quad \text{for} \quad \omega = (d,m) \in \Omega_0^{(1)} \,.
\end{equation}
In particular, all correlated contributions, in the sense of Section \ref{ss:BorelInQuantum}, exhibit the same Stokes constant.\footnote{Note that to arrive at this conclusion, we have organized the sum in \eqref{eq:MumAsymptotics} in the $d=0$ sector differently compared to \cite{Gu:2023mgf}, interpreting the sum over $k$ in $\sum_{k \ge 1}\sum_{m\ge1} k^{-2g+2} m^{-2g} = \sum_{\ell \ge 1} \sigma_2(\ell) \ell^{-2g}$ as contributions from primitive instantons.}
Note that the choice of overall normalization of the coefficients $a_{\omega,n}$, $a_{\omega,n}^{(\ell)}$ leads to integral Stokes constants, and is ultimately motivated by the third of the conjectures put forth at the end of this section. 

In the notation of \eqref{eq:nInstanton}, we have found
\begin{equation} \label{eq:instTransMUM}
    \tilde{\cF}^{(\ell \cA_{d,m})}(X_0,X_1) = \e^{-\ell \cA_{d,m}/g_s} \frac{1}{2\pi} \left(\frac{\cA_{d,m}}{\ell}\frac{1}{g_s} + \frac{1}{\ell^2} \right) \,.
\end{equation}
Note that the Gopakumar-Vafa form of the topological string amplitude does not bring all singularities in the Borel plane to light; this is why we have written the singularity set which is accessible as $\Omega_0 \in \Omega$.

\paragraph{The topological string near the conifold point at $z=\mu$} All one-parameter hypergeometric models yield topological string amplitudes that satisfy, in an appropriate frame, the gap condition near the conifold point which lies in the interior of moduli space \cite{Huang:2006hq}.\footnote{The behavior near conifold points at infinity is more nuanced and will be discussed below.} These conifold points are characterized by a unique primitive vanishing cycle; we denote the associated period by $X^1_\conifold$. Any non-logarithmic period $X^0_{\conifold}$ over a cycle which does not intersect the vanishing cycle can be chosen to define a frame exhibiting the gap behavior. We refer to such frames as conifold frames. For all hypergeometric models, $(X^0_\conifold, X^1_\conifold) = (X^1,P_0)$ defines a conifold frame.

The gap condition implies that the only negative power of the vanishing period $X^1_\conifold$, and thus the leading contribution to $\cF_g(X^0_\conifold,X^1_\conifold)$ as we approach the conifold point, is of the form
\begin{equation} \label{eq:interiorGap}
    \cF_g(X^0_\conifold,X^1_\conifold) = \frac{(-1)^{g-1}B_{2g}}{2g(2g-2)}\left(\frac{2\pi}{\aleph X^1_\conifold}\right)^{2g-2} + \text{regular} \,.
\end{equation}
The asymptotic behavior of this contribution can be extracted from the asymptotic expansion of the Bernoulli numbers \cite{Gu:2023mgf},
\begin{equation}
    B_{2g}=\frac{2(2g)!}{(2\pi \ii)^{2g}}\zeta(2g) \,.
\end{equation}
This analysis yields
\begin{equation} \label{eq:asymptoticsConifold}
    \cF_g(X^0_\conifold,X^1_\conifold) \sim \frac{1}{2\pi^2} \sum_{\ell=1}^\infty \frac{1}{\ell^2}\left(\frac{\Gamma(2g-1)}{(\ell \aleph X^1_\conifold)^{2g-1}} \ell \aleph X^1_\conifold + \frac{\Gamma(2g-2)}{(\ell \aleph X^1_\conifold)^{2g-2}} \right) \,.
\end{equation}
Once again comparing to \eqref{eq:CoeffAsymp} and \eqref{eq:Boreltransform}, we arrive at the same identifications as in \eqref{eq:cIdMUM} and \eqref{eq:cFIdMUM} with $(X^0,X^1)$ replaced by $(X^0_\conifold, X^1_\conifold)$.

The asymptotic behavior of the topological string amplitudes near the MUM and the conifold point, in appropriately chosen frames, is thus consistent with the singularities of the Borel transform
\begin{equation} \label{eq:borelTransformFg}
    \widehat{\cF} =\sum_{g=1}^\infty \cF_g \frac{\zeta^{2g-2}}{(2g-2)!} 
\end{equation}
of the formal power series
\begin{equation}
    \widetilde{\cF} = \sum_{g=1}^\infty \cF_g g_s^{2g-1}
\end{equation}
being of the form \eqref{eq:logSing}. The conventional definition of the topological string partition amplitude $\cF$ is related to this via
\begin{equation}
    \cF = \frac{1}{g_s} \left( \frac{\cF_0}{g_s} + \widetilde{\cF} \right) =\sum_{g=0}^\infty \cF_g g_s^{2g-2} \,. 
\end{equation}

Once again identifying the contributions to the asymptotics in \eqref{eq:asymptoticsConifold} as stemming from two singularities in the Borel plane, at $\aleph X^1_\conifold$ and $-\aleph X^1_\conifold$ respectively, and comparing to \eqref{eq:CoeffAsymp}, yields the $\ell$-instanton contributions at $\ell \zeta_\conifold = \ell \aleph X^1_\conifold$
\begin{equation} \label{eq:conifoldResurgenceData}
     S_{\conifold,\ell} = 1\,, \quad a_{\conifold,-1}^{(\ell)} = \frac{\aleph X^1_\conifold}{2\pi \ell} \,,\quad a_{\conifold,0}^{(\ell)} = \frac{1}{2\pi \ell^2} \,, \quad a_{\conifold,n>0}^{(\ell)} = 0  \,.
\end{equation}
In the notation of \eqref{eq:nInstanton}, we have found
\begin{equation} \label{eq:instTransConifold}
    \tilde{\cF}^{(\ell \aleph X^1_\conifold)}(X_0,X_1) = \e^{-\ell \aleph X^1_\conifold/g_s} \frac{1}{2\pi} \left(\frac{\aleph X^1_\conifold}{\ell}\frac{1}{g_s} + \frac{1}{\ell^2} \right) \,.
\end{equation}

\paragraph{The topological string near a conifold point at $z=\infty$} The three hypergeometric models $X_{4,2}$, $X_{6,2}$ and $X_{3,2,2}$ each exhibit a conifold point at $z = \infty$. The vanishing cycles for all three models have the same coefficients, denoted by $\bq_1$ in \eqref{eq:qc}, in terms of the canonical basis of cycles introduced at the MUM point. Completing $X^1_\conifold = \bq_1 \cdot \piGeom$ to a frame by choosing a period $X^0_\conifold$ without logarithmic contributions which is associated to a symplectically orthogonal cycle (a canonical choice would be $P^0$), the topological string amplitudes for $X_{4,2}$ and $X_{6,2}$ exhibit the gap behavior 
\begin{equation}
    \cF_g(X^0_\conifold,X^1_\conifold)=S\frac{(-1)^{g-1}B_{2g}}{2g(2g-2)}\left(\frac{2\pi }{\aleph X^1_\conifold}\right)^{2g-2}+\,\text{regular}\,,  
    \label{eq:gapX42X62}
\end{equation}
with $S=2$ for $X_{4,2}$ and $S=1$ for $X_{6,2}$.
Since we can construct two massless particles, we might have expected another singular contribution to the $\cF_g$. The occurrence of a simple gap behavior suggests that the particle associated to the charge $\bq_2$ is unstable. Indeed, as we will see in Section \ref{s:X42} and \ref{s:X62}, no singularity occurs in the Borel plane of the geometries $X_{4,2}$ and $X_{6,2}$ at $\pm \bq_2 \cdot \piGeom$, consistent also with the observation that the generalized DT invariant associated to this charge vanishes for both models.

The same frame does not give rise to a gap behavior for the Calabi-Yau threefold $X_{3,2,2}$. The leading singularity however does exhibit a structure reminiscent of all other conifold points of hypergeometric models, given by 
\begin{equation}
    \mathcal{F}_g(\periodV^{\bot},\bq_1\cdot\periodV)= \frac{(-1)^{g-1}B_{2g}}{2g(2g-2)}\left(14\left(\frac{(2\pi \text{i})^{1/2}}{\bq_1\cdot\periodV}\right)^{2g-2}-2\left(\frac{(2\pi \text{i})^{1/2}}{2\bq_1\cdot\periodV}\right)^{2g-2}\right)+ \,\text{singular}\,.
    \label{eq:leadingX322}
\end{equation}
This structure implies that the asymptotics is governed by two singularities with associated Stokes constants 
\begin{align}
   & \ell\aleph \bq_1\cdot\periodV,\quad S_{\ell}=14,\quad \ell\in \mathbb{N}^* \,,\\
    & 2\ell\aleph \bq_1\cdot\periodV,\quad S_{\ell} =-2,\quad \ell\in \mathbb{N}^* \,.
\end{align}
The singularity at $2\bq_1\cdot\periodV$ hence receives both a primitive and a two-instanton contribution.

It is noteworthy that for this geometry, the Borel plane does exhibit a singularity at the location of the second lightest period $\pm \aleph \bq_2\cdot \Pi$ (although our numerics here has very low precision); also, both the generalized DT invariant associated to the charge $\bq_2$ and the one associated to the charge $-\bq_2$ are non-vanishing. It is tempting to conjecture that it is the presence of these additional massless particles that give rise to the subleading singular terms in \eqref{eq:leadingX322}.

\paragraph{The topological string near the M-point of $X_{2,2,2,2}$} This point can be interpreted as the mirror to the large radius limit of a non-commutative resolution of a degeneration of the $X_8$ model \cite{Katz:2022lyl}. From this geometrical point of view, one can derive a constant map contribution in the frame $(X^0_{\fm},X^1_{\fm})=(\bq_1\cdot\periodV,\bq_2\cdot\periodV)$ given by \begin{equation}
\label{eq:constantmapM}
    (X^0_\fm)^{2g-2}\mathcal{F}_g(X^0_{\fm},X^1_{\fm})=\frac{B_{2g}B_{2g-2}\left((2\pi\ii)^{3/2}\right)^{2g-2}(-1)^{g-1}}{2g(2g-2)(2g-2)!}\left (\frac{20}{2^{2g-2}}-84\right)+O(q_\fm) \,, 
\end{equation}
where $q_\fm=\text{exp}(2\pi\ii X^1_{\fm}/2X^0_{\fm})$. Just as for MUM points at $z=0$, we can determine the leading Borel singularities and the associated Stokes constants from this formula. We find two towers of multi-instanton contributions of the form $\pm\ell \cA_{1_n}$ and $\pm \ell \cA_{2,n}$
where \begin{equation}
    \cA_{1,n}= \aleph\,n X^0_\fm,\quad \cA_{2,n}= \aleph \,2n X^0_\fm,\quad n\in\IN^*\,,
\end{equation}
with
\begin{equation}
 a_{i,n,-1}^{(\ell)} = \frac{\cA_{i,n}}{2\pi \ell} \,,\quad a_{i,n,0}^{(\ell)} = \frac{1}{2\pi \ell^2} \,, \quad a_{i,n,k>0}^{(\ell)} = 0  \,,\quad i=1,2 \,,
\end{equation}
and associated Stokes constants 
\begin{equation}
   S_{1,n}=-84 \,,\quad S_{2,n}=20 \,.
\end{equation}
The BPS spectrum for this theory is governed by a torsion refined version of the Gopakumar-Vafa formula \cite{Katz:2022lyl}, involving invariants  $n^{l}_{g,d}$ labeled by a $\IZ_2$ charge $l=0,1$, 
\begin{equation}
  \label{eq:gvX2222} 
  \mathcal{F}(X^1_\fm,X^0_\fm) =\sum_{g\geq 0}\sum_{d \in \mathbb{N}}\sum_{k\geq 1}\sum_{l=0,1}
  \frac{n^l_{g,d}}{k}
  \left(2\sin\frac{k \lambda_\fm }{2}\right)^{2g-2} (q_\fm^de^{\ii\pi l})^k,
\end{equation}
with $\lambda_\fm=\frac{(2\pi\ii)^{3/2}g_s}{X^0_\fm}$. This leads to a formula analogous to \eqref{eq:MumAsymptotics} for the subleading asymptotics,
\begin{equation} \label{eq: M point asymptotics}
    \cF_g(X^0_\fm,X^1_\fm) \sim \sum_{(d,m,l) \in \tilde{\Omega}_{0,\fm}^{(1)}} \sum_{\ell =1}^\infty \frac{n_{0,d}^l}{2\pi^2 \ell^2} \left( \frac{\Gamma(2g-1)}{(\ell\cA_{d,m,l})^{2g-1}} \ell \cA_{d,m,l} +\frac{\Gamma(2g-2)}{(\ell\cA_{d,m,l})^{2g-2}}  \right)
\end{equation}
with 
\begin{equation} 
    \cA_{d,m,l} = \aleph \ell (d X^1_\fm +(2 m-l)X^0_\fm),\quad \tilde{\Omega}_{0,\fm}^{(1)}=\IN^*\times\IZ\times\{0,1\}.
   \label{eq: sing X222}
\end{equation}
Taking into account the $\IZ_2$ symmetry of the Borel plane, one is led to identify the position of Borel singularities at $\ell\zeta_{\omega}=\ell\epsilon \cA_{d,m,l}$, where $(\epsilon,d,m,l)\in \IZ_2\times\tilde{\Omega}_{0,\fm}^{(1)}=\Omega_{0,\fm}^{(1)}$.

The trans-series data associated to these singularities is 
\begin{equation}
    S_{\omega,\ell} =  n_{0,d}^{l}\,, \quad a_{\omega,-1}^{(\ell)} = \frac{\epsilon\cA_{d,m,l}}{2\pi \ell} \,,\quad a_{\omega,0}^{(\ell)} = \frac{1}{2\pi \ell^2} \,, \quad a_{\omega,n>0}^{(\ell)} = 0  \quad \text{for} \quad \omega = (\epsilon,d,m,l) \in \Omega_{0,\fm}^{(1)},
\end{equation}
which is again of the same form as \eqref{eq:instTransMUM} and \eqref{eq:instTransConifold}.

\paragraph{Conjectures regarding the asymptotics of $\cF_g$}
The generic asymptotic behavior of the holomorphic topological string amplitudes at MUM and conifold points in appropriate frames that we have just reviewed motivates the following conjectures \cite{Gu:2023mgf,Grassi_2020, 
Pasquetti_2010,Marino:2024yme,Marino:2023gxy}. 
\begin{enumerate}
    \item The Borel singularities of the holomorphic topological string amplitude are of the form \eqref{eq:logSing}.
    \item They lie at integral periods of $\Omega$ (up to an overall constant determined by choices of normalization).
    \item When the period $\cA$ determining the location $\zeta_\omega = \aleph \cA$ of the singularity $\omega \in \Omega$ figures among the A-periods specifying the frame defining $\cF_g$, the trans-series encoding the contribution of such singularities takes the form
    \begin{equation}
     \tilde{\cF}^{(\ell \cA)} = \e^{-\ell \cA/g_s} \frac{1}{2\pi} \left(\frac{\cA}{\ell}\frac{1}{g_s} + \frac{1}{\ell^2} \right) \,, \quad \ell \in \IN^* \,.
    \end{equation}
\end{enumerate}

\subsection{All order results for the instanton trans-series}
In \cite{Couso-Santamaria:2013kmu, Couso-Santamaria:2016vcc}, a strategy is proposed and implemented to low order to compute the parent object $\tilde{F}^{(\zeta_\omega)}$ to the instanton trans-series $\tilde{\cF}^{(\zeta_\omega)}$ (in the notation introduced in general in \eqref{eq:nInstanton} and applied to the study of topological string amplitudes in \eqref{eq:instTransMUM} and \eqref{eq:instTransConifold}). By parent object, we designate the analogue of the non-holomorphic topological string amplitude $F_g(S^{ij},\tilde{S}_i,\tilde{S})$, from which the holomorphic topological string amplitudes $\cF_g$ in any frame can be obtained upon taking an appropriate holomorphic limit. The strategy consists in finding trans-series solutions to the holomorphic anomaly equations. Just as in the case of the computation of $F_g(S^{ij},\tilde{S}_i,\tilde{S})$, boundary conditions are needed to uniquely determine the relevant solution. The papers \cite{Couso-Santamaria:2013kmu,Couso-Santamaria:2014iia} propose to impose the third conjecture listed at the end of Section \ref{ss:BorelAndTopString} as the required boundary condition. Numerical analysis, comparing the results thus obtained to the asymptotic behavior of the topological string amplitudes, must then provide evidence for the validity of this strategy \cite{Couso-Santamaria:2014iia, Gu_2023,Gu:2023mgf}. The authors of \cite{Gu_2023} succeeded in obtaining a closed form expression for $\tilde{F}^{(\zeta_\omega)}$ in the case of the topological string on non-compact geometries. This result was generalized to compact geometries in \cite{Gu:2023mgf}.

We will here review the result of \cite{Gu:2023mgf} for the one-instanton trans-series, as this will play a key role in our numerical analysis in Section \ref{s:experimentalData}. For a one-instanton singularity at $\zeta_\omega = \cA$ in the Borel plane, it is given by
\begin{equation}
    F^{(\cA)} = \frac{1}{2\pi}\mathfrak{a}\:\exp(\Sigma)\,,
\end{equation}
where 
\begin{align}
   & \mathfrak{a}=1+G+\oD\Sigma,\quad\quad \Sigma=\sum_{k\geq1}\frac{\oD^{k-1}}{k!}G\,, \\
   & G=\frac{\mathcal{A}}{g_s}+\sum_{g\geq 1} \oD(g_s^{2g-2} F_g)\,.
   \label{eq:oneInstanton}
\end{align}
Here, $\oD$ is a differential operator with regard to the complex moduli and $g_s$. The necessity to introduce derivatives with regard to $g_s$ stems from the technical requirement of avoiding sections of $\cL^n$, with $\cL$ denoting the K\"ahler line bundle. Given a section of $s \in \Gamma(\cL^n)$ which is independent of $g_s$, $\oD$ acts upon mapping $s$ to $\cL^0$ by dividing by $g_s^n$. With regard to a holomorphic frame $(X_*^I,P_I^*)$ in which
\begin{equation}
    \cA = c^I P_I^* + d_I X^I_*
\end{equation}
(recall that $\cA/\aleph$ is an integral period of $\Omega$, hence $c^I, \, d_I \in \aleph \IZ$), the holomorphic limit of $\oD$ acting on a homogeneous function $s$ of degree $n$ in the variables $X^I_*$ acts as
\begin{equation} \label{eq:holLimitDI}
    \oD \left(g_s^{-n}s \right) \rightarrow g_s^{-n+1}c^I \frac{\partial s}{\partial X^I_*} \,. 
\end{equation}
The derivative operator can also be expressed in terms of global coordinates on complex structure moduli space \cite[equation (5.51)]{Gu:2023mgf}. For one-parameter models, its holomorphic limit is
\begin{equation} \label{eq:holLimitDII}
  \oD \rightarrow  g_s\left(\mathfrak{f}_1(z)g_s\frac{\partial}{\partial g_s} + \mathfrak{f}_2(z)\frac{\partial}{\partial z}\right).
\end{equation}
Here \begin{align}
    \mathfrak{f}_1(z)=\mathcal{A}'(\tilde{\mathcal{S}}^z-\tilde{\mathcal{S}}^z_{\mathcal{A}})-2\mathcal{A}(\tilde{\mathcal{S}}-\tilde{\mathcal{S}}_{\mathcal{A}}) \,,\\
      \mathfrak{f}_2(z)=\mathcal{A}'(\mathcal{S}^{zz}-\mathcal{S}^{zz}_{\mathcal{A}})-\mathcal{A}(\tilde{\mathcal{S}}^z-\tilde{\mathcal{S}}^z_{\mathcal{A}})\,,
\end{align}
where a subscript $\cA$ on a propagator indicates the holomorphic limit of the propagator in any frame containing the period $\cA/\aleph$ as an A-period. As the expression \eqref{eq:holLimitDI} shows, the limit \eqref{eq:holLimitDII} does not depend on the periods which complete $\cA/\aleph$ to a full set of A-periods.

For explicit evaluations, it is convenient to note that acting on a function $f$ that is independent of $g_s$,
\begin{equation}
    D^k(g_s^{-n} f(z))=g_s^{k-n}B_{k,n}(z)(f(z))\,,
\end{equation}
where 
\begin{equation}
    B_{k,n}(z)=\prod_{i=0}^{k-1}\left((i-n)\mathfrak{f}_1(z) + \mathfrak{f}_2(z)\frac{d}{dz}\right)\,.
\end{equation}
The operator $B_{k,n}$ can be computed recursively as a polynomial in $\frac{d}{dz}$. Using the operator $B_{k,n}(z)$ for specific value of $(k,n)$, the derivatives of the amplitudes, and the derivatives of $\mathcal{A}$ one can easily compute the one instanton amplitudes.

\section{Generalized Donaldson-Thomas invariants} \label{s:generalizedDT}
By the conjectures presented at the end of section \ref{ss:BorelAndTopString}, Borel singularities occur at periods of $\Omega$ over integral 3-cycles of $\Xmirror$. Given the close relation between D-branes and non-perturbative physics already discussed in Section \ref{ss:dominatingContributions}, it is natural to look for a manifestation of the D-brane degeneracy associated to such cycles in the non-perturbative data accessible via Borel analysis. This degeneracy is associated to enumerative invariants of $\X$. In section \ref{ss:BorelAndTopString}, we have already argued for the occurrence of genus 0 Gopakumar-Vafa invariants $n_{0,d}$ as Stokes constants near large radius, occurring at Borel singularities at $\sim d X^1+mX^0$, see \eqref{eq:StokesDataMUM}. A fuller understanding of the non-perturbative structure of the theory would allow us to predict the Stokes constant, given the position of the singularity. A first step towards this ambitious goal is to seek out enumerative invariants of $\X$ which are associated to integral periods of $\Xmirror$, as this, by the second conjecture stated at the end of section \ref{ss:BorelAndTopString}, is where Borel singularities lie. Generalized Donaldson-Thomas (DT) invariants \cite{Joyce} fit this bill. In this section, we review selective aspects of the theory of these invariants that will be relevant in our numerical study in Section \ref{s:experimentalData}.

\subsection{Mapping between $\Knum(\coh(X))$ and $H_3(X,\IZ)$ via mirror symmetry}
Generalized DT invariants are rational invariants associated to elements of the numerical Grothendieck group $\Knum(\coh(X))$ of the coherent sheaves on a Calabi-Yau threefold $X$. The isomorphic image of this space in $\Heven(X, \IQ)$ is shown in \cite[Theorem 4.19]{Joyce} to be given by the lattice $\Lambda_X$ defined via
\begin{align} \label{eq:lambdaX}
    \Lambda_X = &\Big\{(\lambda_0, \lambda_1, \lambda_2, \lambda_3) \in \Heven(X,\IQ) \, | \, \lambda_0 \in \sH^0(X,\IZ)\,,\,\, \lambda_1 \in \sH^2(X,\IZ)\,, \,\,\\ &\lambda_2 - \frac{1}{2} \lambda_1^2 \in \sH^4(X,\IZ)\,,\,\, \lambda_3 + \frac{1}{12} \lambda_1 c_2(TX) \in \sH^6(X,\IZ)\Big\} \,,\nonumber
\end{align}
with the isomorphism provided by taking the Chern character,
\begin{equation}
    \ch \, : \, \Knum(\coh(X)) \xrightarrow{\sim} \Lambda_X \,.
\end{equation}
Defining stability on the derived category of coherent sheaves on $X$ requires the choice of a bounded t-structure on the category, as well as a central charge function
\begin{equation}
    Z: \Lambda_X \rightarrow \IC 
\end{equation}
which maps charges associated to elements in the heart of the bounded t-structure to a fixed half-plane $\mathrm{H} \subset \IC$. The D-branes whose stability is captured by the formalism reside in the heart. To introduce the family of central charge functions relevant here, we introduce the Mukai vector
\begin{eqnarray}
     \gamma \,: \, \Knum(\coh(X)) &\rightarrow& \Heven(X,\IQ)\\
    E &\mapsto& \ch(E)\sqrt{\Td(T\X)} \,.
\end{eqnarray}
For the one-parameter models we will be considering, $b_2(X)=1$. We denote the generator of $H^2(X,\IZ)$ by $H$. Introducing a complex parameter $t \in \IC$ associated to this generator permits us to define the family of central charges
\begin{equation} \label{eq:centralChargeX}
    \ZX = \int_X \e^{-t H} \gamma(E) \,.
\end{equation}
We can introduce generators of $\Heven(X,\IZ)$ in terms of $H$ via $\Heven(X,\IZ) = \langle 1, H, \frac{1}{\kappa}H^2, \frac{1}{\kappa}H^3\rangle$, where the factor involving the triple intersection number $\kappa = \int H^3$ follows by imposing integrality of the pairing between $H^{i}(X,\IZ)$ and $H^{6-i}(X, \IZ)$. We map this generating set to rational cohomology via the inclusion $\Heven(X,\IZ) \hookrightarrow \Heven(X,\IQ)$. Writing
\begin{equation} \label{eq:rationalCoordinates}
    \gamma = p^0 +p^1 H - \frac{q_1}{\kappa} H^2 + \frac{q_0}{\kappa}H^3
\end{equation} 
introduces the rational coordinates $(p^0, p^1, q_1,q_0)$ used in \cite[equation (2.8)]{Alexandrov:2023zjb} with the intricate integrality structure which follows from \eqref{eq:lambdaX}. By mirror symmetry, the same invariants can be associated to the mirror variety $\Xmirror$, where they are associated to elements of $\sH_3(\Xmirror,\IZ)$, which clearly has a more straightforward integrality structure than $\Lambda_X$.

$\Lambda_X$ is referred to as the charge lattice of BPS states from the A-model perspective, $\sH_3(\Xmirror,\IZ)$ as the charge lattice from the B-model perspective.

We will arrive at the map between the charge lattices by comparing the expression for the central charge in $X$ and in $\Xmirror$. On the mirror variety, the central charge associated to an element $\gammaMirror \in H_3(\Xmirror,\IZ)$ is given by
\begin{equation} \label{eq:centralChargeMirror}
    \ZXmirror = \int_{\gammaMirror} \Omega \,,
\end{equation}
with $\Omega \in H^{(3,0)}(\Xmirror)$ unique up to normalization. In fact, equation \eqref{eq:centralChargeMirror} provides a more general family of central charges than $\ZX$, parametrized by the complex structure $\cM_{\mathrm{cmplx}}(\Xmirror)$ of $\Xmirror$. The compactification of this space has the topology of $\IP^1$. We have called $z$ the affine coordinate on it which situates the MUM point with local exponents $(0,0,0,0)$ at the origin and the singular point with fractional local exponents at infinity. $\ZX$ and $\ZXmirror$ are meant to coincide at leading order in $z \rightarrow 0$, up to a small discrepancy we shall discuss presently.

The variable $t$ parametrizing the central charge \eqref{eq:centralChargeX} of $X$ is related to the variable $z$ parametrizing the central charge \eqref{eq:centralChargeMirror} of $\Xmirror$ via the mirror map 
\begin{equation}
    t = \frac{X^1}{X^0} \,.
\end{equation}
To leading order in $z$, this yields
\begin{equation}
    \e^{2\pi \ii t} = z + \cO(z^2) \,.
\end{equation}
Hence, in terms of the notation introduced in \eqref{eq:FrobMUM},
\begin{equation}
    \lim_{z \rightarrow \infty} \piFrob_0 = f_0\begin{pmatrix}
        1 \\
        2\pi \ii t \\
        \frac{1}{2}(2 \pi \ii)^2 t^2 \\
        \frac{1}{6}(2 \pi \ii)^3 t^3
    \end{pmatrix} \,.
\end{equation}
Expanding $\gammaMirror$ in the generating set of integral cycles corresponding to the explicit form of $T_0$ given in \eqref{eq:T0explicit}, the central charge associated to $\gammaMirror$ is given to leading order in $z$ by
\begin{eqnarray}
    \ZXmirror &\sim& \frac{1}{(2 \pi \ii)^3}
    \begin{pmatrix}
        m_0 & m_1 & n_0 & n_1
    \end{pmatrix}
    T_0 
    \begin{pmatrix}
        1 \\
        2\pi \ii t \\
        \frac{1}{2}(2 \pi \ii)^2 t^2 \\
        \frac{1}{6}(2 \pi \ii)^3 t^3
    \end{pmatrix}  \nonumber \\
    &=&
    \begin{pmatrix}
        m_0 & m_1 & n_0 & n_1
    \end{pmatrix}
    \begin{pmatrix} 
        \frac{\chi \zeta(3)}{(2\pi \ii)^3} + \frac{c_2}{24}t + \frac{\kappa}{6}t^3 \\
        \frac{c_2}{24} +\sigma t - \frac{\kappa}{2}t^2 \\
        1 \\
        t
    \end{pmatrix} \,,\label{eq:ZX0charges}
\end{eqnarray}
where we have normalized $\Omega$ by setting the holomorphic period $X^0 = 1$. On the other hand, in terms of the expansion of $\gamma$ introduced in \eqref{eq:rationalCoordinates}, the central charge \eqref{eq:centralChargeX} has the expression
\begin{equation} \label{eq:ZXcharges}
    \ZX = \int_X \e^{-  t H} \gamma(E) = 
    \begin{pmatrix}
        p^0 & p^1 & q_1 & q_0
    \end{pmatrix}
    \begin{pmatrix}
        0 & 0 & 0 & -\frac{\kappa}{6} \\
        0 & 0 &  \frac{\kappa}{2}&0 \\
        0 & 1 & 0 & 0 \\
        1 & 0 & 0 & 0
    \end{pmatrix}
    \begin{pmatrix}
        1 \\
        t \\
         t^2 \\
         t^3
    \end{pmatrix} \,.
    \end{equation}
$\ZX$ and the leading expression for $\ZXmirror$ can almost be identified. The small discrepancy lies in the coefficient of the $t^0$ term proportional to the $\zeta$ function. To include this, the expression for $\ZX$ must be modified by replacing the square root of the Todd class by the $\Gamma$-class \cite{Iritani2009}. Dropping this term from \eqref{eq:ZX0charges}, we can obtain the explicit map between the coordinates $(p^0,p^1,q_1,q_0)$ and $(m_0,m_1,n_0,n_1)$ by equating to \eqref{eq:ZXcharges}. We obtain
\begin{equation}
    (p^0,p^1,q_1,q_0)^T = T (m_0,m_1,n_0,n_1)^T
\end{equation}
with 
\begin{equation} \label{eq:mapBtoA}
    T = \begin{pmatrix}
        -1 & 0 & 0 & 0 \\
        0 & -1 & 0 & 0 \\
        \frac{c_2}{24} & \sigma & 0 & 1  \\
        0 &  \frac{c_2}{24} & 1 & 0 
    \end{pmatrix} \,.   
\end{equation}

\subsection{Rank 0 and rank 1 generalized DT invariants}
The value of $p_0$ is referred to as the rank of a generalized DT invariant.
Building on the works \cite{Feyzbakhsh:2020wvm,Feyzbakhsh:2021nds,Feyzbakhsh:2021rcv,Feyzbakhsh:2022ydn}, the paper \cite{Alexandrov:2023zjb} uses techniques relying on the holomorphic anomaly equations, wall-crossing and modularity to compute rank 0 and rank 1 generalized DT invariants for all one-parameter hypergeometric Calabi-Yau varieties at large radius. We will label these as $\OLR(\gamma)$, $\gamma \in \Heven(Z,\IQ)$. The values computed in \cite{Alexandrov:2023zjb} are available via the online resource linked in footnote \ref{footnote:link}, using a parametrization that we now review.

\paragraph{Rank -1: PT invariants} Generalized DT invariants at rank $p_0 = -1$ can be identified with Pandharipande-Thomas (PT) invariants. It turns out that computing the invariants at $p_1= 0$ is sufficient, as they only depend on the following combination of charges:
\begin{equation}
    Q_-=-q_1+\frac{1}{2}\kappa(p^1)^2+\frac{c_2}{24},\quad n_-=-q_0+p^1q_1-\frac{1}{3}\kappa(p^1)^3 \,.
\end{equation}

\paragraph{Rank 1: DT invariants} Generalized DT invariants at rank $p_0 = 1$ can be identified with the original DT invariants defined in \cite{DTGauge,Thomas:1998uj}. Again, computing the invariants at $p_1= 0$ is sufficient; the invariants depend on the following combination of charges:
\begin{equation}
    Q_+=q_1+\frac{1}{2}\kappa(p^1)^2+\frac{c_2}{24},\quad n_+=-q_0-p^1q_1-\frac{1}{3}\kappa(p^1)^3 \,.
\end{equation}

\paragraph{Rank 0: MSW and GV invariants} The rank 0 invariants\footnote{When it is clear from context that one is working with $p^0$ invariants, the value of $p^1$ is somewhat unfortunately also referred to as the rank.} at $p^1 =0$ have been conjectured (\cite[conjecture 6.20]{Joyce} and \cite{Alexandrov:2013yva}) to be equal to genus zero Gopakumar-Vafa invariants, 
\begin{equation}\label{eq:DTGV}
    \OLR(0,0,q_1,q_0)=n^0_{\vert q_1\vert} \,.
\end{equation}
The invariants at $p^1 =1$ have been identified with rank 1 MSW invariants \cite{Alexandrov:2013yva}, i.e. D4-D2-D0 invariants with one unit of D4 charge. In the notation of \cite{Alexandrov:2023zjb}, 
\begin{equation}
    \Omega_{LR}(0,1,q_1,q_0)=\Omega_{\mu}(\hat{q}_0) \,,
\end{equation}
with
\begin{equation}
    \mu=q_1-\frac{1}{2}\kappa r^2 \,\,\mathrm{mod}\, \kappa r\,,\quad \hat{q}_0=q_0-\frac{q_1^2}{2\kappa r} \quad \text{at} \,\,r=1\,.
    \label{eq:MSWmu}
\end{equation}
It is also possible to replace $\hat{q}_0$ by the integral parameter
\begin{equation}
    n=\frac{\kappa r^3+c_2 r}{24}-\frac{\mu^2}{2\kappa r}-\frac{\mu r}{2}-\hat{q}_0 \,.
    \label{eq:MSWn}
\end{equation}
MSW invariants of rank 2 have recently been computed in \cite{Alexandrov:2023ltz} for the Calabi-Yau threefolds $X_8$ and $X_{10}$. Periods associated to such charges do not arise among the leading singularities in the Borel plane of these models that we identify in Section \ref{s:experimentalData}.

\section{Experimental data} \label{s:experimentalData}
Much evidence has already accumulated in previous work that the setup reviewed in Section \ref{s:asymptoticMethods} applies to topological string amplitudes, both in the local \cite{Alexandrov:2023wdj,Couso-Santamaria:2013kmu,Couso-Santamaria:2014iia,Gu_2023,
Pasquetti_2010,Marino:2024yme,Marino:2023gxy} and the compact setting \cite{Gu:2023mgf}. Our focus in this work is taking a step towards understanding the systematics of where Borel singularities occur, and which values the associated Stokes constants take. As we have reviewed in Section \ref{ss:BorelAndTopString}, these questions have appealing answers which can be argued for analytically close to MUM points, and conifold points which lie in the interior of moduli space. After briefly reconsidering such conifold points in Section \ref{ss:conifoldNumerics}, we will turn our attention to the remaining singular point of the moduli space of one-parameter models. As we lack the same level of analytic control for most of these points, we must resort to numerics. We will describe the numerical methods we use in Section \ref{ss:numericalMethods} and then exhibit results for all 13 hypergeometric models in Section \ref{ss:numericalResults}. Here is a summary of the structure we find:

\begin{enumerate}
    \item All Borel singularities lie at integral periods of $\Omega$, determined up to an overall normalization constant which we call $\aleph$ (this constant is discussed in some detail in \cite{Gu:2023mgf}).
    \item Except in the case of conifold points (both at $z=\mu$ and $z=\infty$), we find that the set of singularities in the Borel plane are invariant under the action of the monodromy matrix $M_\infty$, and that the Stokes constants for all elements in an $M_\infty$ orbit coincide. Note that for K-points, our numerics only allows us to study the leading singularities in the Borel plane which are linear combinations of the charges $\bq_1$ and $\bq_2$, introduced in Section \ref{ss:dominatingContributions}. The invariance under $M_\infty$ may very well be restricted to this class of charges (indeed, at conifold points, $M_\infty$ acts with a sign on $\bq_1$ and $\bq_2$, which is indeed a symmetry of the Borel plane).   
    
    \item With the exception of the model $X_{10}$, whenever a Borel singularity at a period $\Pi$ is associated to a rank $\pm 1$ charge, such that its partner singularity at $-\Pi$ is associated to a rank $\mp 1$ charge, then the large radius generalized DT invariant associated to either one or the other charge (or both, when they coincide) coincides with the Stokes constant of the singularity. When $\Pi$ carries an MSW charge, on the other hand, the MSW invariant never coincides with the associated Stokes constant. 

    \item If the Calabi-Yau variety exhibits massless periods, then these coincide with the leading Borel singularities, except if the associated generalized DT invariants vanish.
\end{enumerate}

\subsection{Numerical methods} \label{ss:numericalMethods}
Let us collect and slightly enhance the notation we have introduced so far, as well as the results that will underlie the numerical analysis. We consider the formal power series $\widetilde{\cF}$ and its Borel transform $\widehat{\cF}$,
\begin{equation}
    \widetilde{\cF} = \sum_{g=1}^\infty \cF_g g_s^{2g-1}  \quad \Rightarrow \quad
    \widehat{\cF} =\sum_{g=1}^\infty \cF_g \frac{\zeta^{2g-2}}{(2g-2)!} \,.
\end{equation}
We have good evidence that $\widehat{\cF}$ exhibits singularities of the form
\begin{equation} 
    \widehat{\cF}(\zeta_\omega + \xi) = - \frac{S_\omega}{2\pi} \left( \frac{\cF_{\omega,-1}}{\xi} +  \log \xi \,\widehat{\cF}_\omega(\xi)\right) + \text{reg.}\,, \quad \omega \in \Omega\,,\,\,\zeta_\omega \in \IC \,,
\end{equation}
with $\widehat{\cF}_\omega(\xi)$ an analytic function in a neighborhood of the origin which we consider as the Borel transform of the formal power series
\begin{equation}
    \widetilde{\cF}_\omega(g_s) = \sum_{n=-1}^\infty \cF_{\omega,n} g_s^n \,.
\end{equation}
The coefficients $\cF_{\omega,n}$ determine the coefficients $\cF_g$ asymptotically via
\begin{equation} \label{eq:FgAsymp}
    \cF_{g} \sim  \frac{S_\omega}{2\pi} \sum_{n \ge -1} \frac{\Gamma(2g-2-n)}{\zeta_{\omega}^{2g-2-n}} \cF_{\omega,n} \,, \quad g \gg 1 \,.
\end{equation}
Finally, we have also introduced notation to take into account correlated singularities:
\begin{equation} \label{eq:nInstantonFg}
    \tilde{\cF}^{(n\zeta_\omega)}(z) := \e^{-n\zeta_\omega/z} \tilde{\cF}^{(n)}_\omega(z) 
\end{equation}

\subsubsection{Identifying Borel singularities via Padé approximation}
We only have access to finitely many of the topological string amplitudes $\cF_g$ required to compute the Borel transform $\widehat{\cF}_g$ as a power series in $\zeta$, see \ref{eq:borelTransformFg}. Approximating the series via truncation will yield a polynomial in $\zeta$, thus erasing any trace of possible singularities of $\widehat{\cF}_g$. A more adapted approximation scheme is provided by the Padé transform.

\begin{definition}
    The Padé approximation of order $(m,n)$ to a function $f(z)$ which is analytic around the origin is provided by two polynomials $p(z)$, $q(z)$ such that $\ord p(z) \le m$, $\ord q(z) \le n$ and the Taylor expansion of $f(z)$ and $p(z)/q(z)$ coincide up to order $m+n$ in $z$.
\end{definition}
The poles of Padé approximants to an analytic function $f$ provide an approximation to the poles of $f$ near the origin. By computing Padé approximants to the truncated Taylor expansion of the principal branch of the logarithm $\log(z_0 - z)$ around the origin, one can also verify that the poles of the Padé approximants occur on a ray extending from a vicinity of the point $z_0$ along the positive $x$-axis to infinity. An affine linear change of variables then shows that the poles of Padé approximants indicate the location of branch cuts in the Borel plane, in the convention that these extend from the branch point to infinity along a ray whose extension runs through the origin.

We typically work with the diagonal Padé approximant at $m=n$.

\subsubsection{Identifying Stokes constants}
\paragraph{Leading singularity is dominant} When the singularities dominating the asymptotics share the same Stokes constant, it can be extracted from the relation \eqref{eq:FgAsymp}. In the case of the topological string, singularities always come in pairs at $\pm \zeta_\omega$, both contributing equally to the asymptotics. If such a pair constitutes the leading singularity, we obtain
\begin{equation} \label{eq:SomegaFromAsymp}
    S_{\omega} \sim \frac{2\pi}{2} \left(\frac{(2g-2)!\cF_{\omega,-1}}{\zeta_{\omega}^{2g-1}}+\frac{(2g-3)!\cF_{\omega,0}}{\zeta_{\omega}^{2g-2}}+ \ldots \right)^{-1}\cF_g   \,.
\end{equation}
By the third of our conjectures presented at the end of Section \ref{ss:BorelAndTopString}, the form of the instanton correction simplifies if we choose a frame $(X^0_\omega, X^1_\omega)$ in which $\aleph^{-1}\zeta_\omega$ figures as an A-period; indeed, only the two leading contributions are non-zero, yielding 
\begin{equation} \label{eq:SomegaFromAsympSimple}
    S_{\omega} \sim \frac{(2\pi)^2}{2} \frac{ \zeta_{\omega}^{2g-2}}{(2g-3)!(2g-1)}\cF_g(X^0_\omega, X^1_\omega)   \,.
\end{equation}
When the leading singularities in the Borel plane constitute an $M_\infty$ monodromy orbit, then the Stokes constant associated to this orbit (assuming that all Stokes constants coincide), can be obtained via \eqref{eq:SomegaFromAsymp}, by summing in the parentheses over all elements in the orbit,\footnote{When the orbit contains the negative of each of its elements, the sum must extend over only one representative per pair, to avoid overcounting. This is indicated by the prime in the summation region in \eqref{eq:SomegaFromAsympOrbit}.}
\begin{equation} \label{eq:SomegaFromAsympOrbit}
    S_{\omega} \sim \frac{2\pi}{2} \left(\sum_{\omega' \in (M_\infty \text{ orbit of } \omega)'}\left( \frac{(2g-2)!\cF_{\omega',-1}}{\zeta_{\omega'}^{2g-1}}+\frac{(2g-3)!\cF_{\omega',0}}{\zeta_{\omega'}^{2g-2}}+\ldots \right)\right)^{-1}\cF_g   \,.
\end{equation}

To check the assumption that the Stokes constants are $M_\infty$ invariant, we can use the method of the following paragraph. 

\paragraph{Leading singularity on a ray} When one singularity $\zeta$ lying on a radial ray dominates the contribution of singularities on this ray to the asymptotics, the corresponding Stokes constant can be determined by computing the discontinuity \eqref{eq:disc}, in terms of which
\begin{equation} \label{eq:SomegaFromDisc}
    S_\omega \sim -\ii \,\e^{\zeta_\omega/g_s}\frac{\disc_\theta \widetilde{\cF}(g_s)}{\sS^{\theta_-}(\widetilde{\cF}_\omega)(g_s)} \,.
\end{equation}
Numerically, we approximate $\sS^{\theta_-}(\widetilde{\cF}_\omega)(g_s)$ by its asymptotic expansion. Note that this method of determining the Stokes constant requires choosing a value of $g_s$. Numerically, we choose the phase of $g_s$ to cancel that of $\zeta_\omega$ to improve the dominance of the contribution from this singularity to $S_\omega$. Due to the approximation $\sS^{\theta_-}(\widetilde{\cF}_\omega)(g_s) \sim \widetilde{\cF}_\omega(g_s)$ that we use, only a finite range of values for $|g_s|$ yields a good approximation of $S_\omega$. To compute the discontinuity numerically, we consider the Padé approximant to a truncation of $\widetilde{\cF}$ and then either perform the required integrals slightly above and below the $\theta$ direction numerically, or evaluate the difference directly via a residue calculation (recall that in the Padé approximant, a logarithm is approximated by a series of poles).

Ideally, we choose a frame in which $\aleph^{-1}\zeta_\omega$ figures as an $A$-period, to simplify the denominator in equation \eqref{eq:SomegaFromDisc}.

\paragraph{Removing contributions from leading singularities} It is often possible to identify subleading singularities in the Borel plane by subtracting the known asymptotic contribution of leading singularities to the topological string amplitude. To this end, we introduce the reduced topological string amplitudes $\cF_{g,\Omega_{\mathrm{lead}}}^{\mathrm{red},n}$ (or $\cF_g^{\mathrm{red}}$ for brevity) via
\begin{equation}
    \cF_{g,\Omega_{\mathrm{lead}}}^{\mathrm{red},n}=\mathcal{F}_g- \sum_{\omega \in \Omega_{\mathrm{lead}}} \frac{S_{\omega}}{2\pi}\frac{(2g-2)!}{\zeta_\omega^{2g-2}}\left(\cF_{\omega,-1}+\frac{\cF_{\omega,0}\,\zeta_\omega}{2g-2}+ \ldots + \frac{\cF_{\omega,n}\zeta_\omega^{n+1}}{(2g-2)\cdots(2g-2-n)}\right)\,,
  \label{eq:red}
\end{equation}
with the sum over $\omega$ ranging over the singularities we wish to remove, and $n$ being chosen as large as computational viable (typically $n\sim 10$).

Depending on the precision of the numerics, the poles associated to $\omega \in \Omega_{\mathrm{lead}}$ completely or partially disappear in the Borel plane of the reduced amplitudes $\cF_g^{\mathrm{red}}$ (recall that logarithmic singularities manifest themselves in Padé approximants as poles). In fact, the disappearance of these poles proves very sensitive to having identified the correct value for the $S_\omega$ and can therefore serve as a check on the determination of these constants.

At M- and C-points, we have exact results for the leading singularities, given by the constant map contribution for MUM points and the leading coefficient in the expansion \eqref{eq:interiorGap} at conifold points. We can hence completely subtract the leading singularity and all associated higher instanton corrections in these cases. For conifold points, the reduced topological string amplitudes take the form
\begin{equation}
    \cF^{\mathrm{red}}_g(X^0_\conifold,X^1_\conifold)=\cF_g(X^0_\conifold,X^1_\conifold)-\frac{(-1)^{g-1}B_{2g}}{2g(2g-2)}\left(\frac{2\pi}{\aleph \X^1_\conifold}\right)^{2g-2} \,. \label{eq:fredconifold}
\end{equation}

\paragraph{Choosing the point $\boldsymbol{z}$ on moduli space at which to study the Borel plane} To study the Borel plane close to the point $z = \infty$, we typically consider points between $z_0 = 10^4 \mu$ and $z_0 = 10^{10} \mu$, with $\mu$ as always denoting the position of the conifold point in the interior of moduli space. Working at larger $z_0$ improves the convergence of the periods which are represented as power series around $\infty$, but makes it more difficult to study subleading singularities.

\subsection{Subleading singularities at the conifold point in the interior of moduli space} \label{ss:conifoldNumerics}

Extending and refining the analysis in \cite{Gu:2023mgf} to all hypergeometric models, we find, by studying $\cF^{\mathrm{red}}_g$ as introduced in \eqref{eq:fredconifold},  that the location of the Borel singularities depends on the relative position of $z$ with regard to $\mu$. 
\begin{figure}
    \centering
    \includegraphics[]{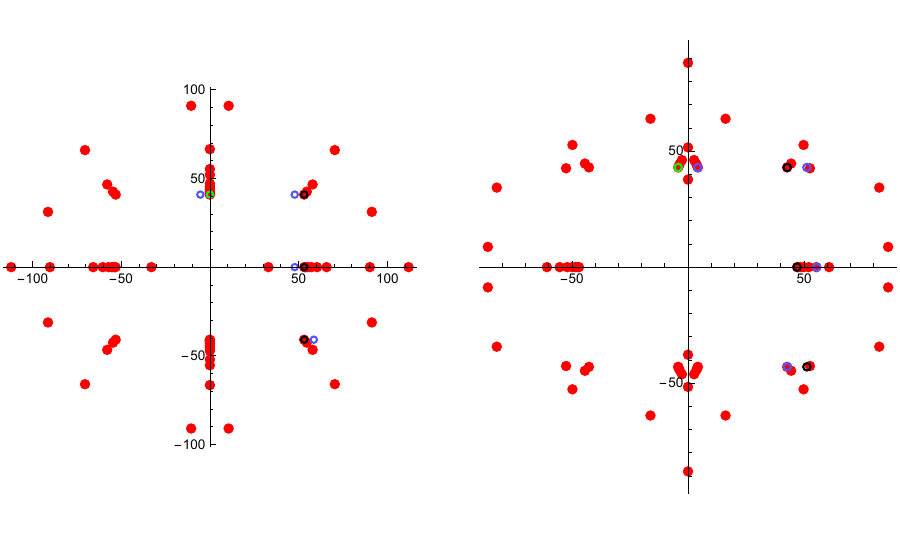}
    \caption{Borel plane of $X_5$ in the frame $(P_0,X^1)$ upon removing the leading gap singularity, at $z=0.7\mu$ on the left, $z=2\mu$ on the right. The green dot indicates the position of the point $-\aleph X^0$. Note that it moves off of the imaginary axis as we move from inside to outside the interval $(0,\mu)$. The blue circle next to the point $-\aleph X^0$ indicates the position of the point $\aleph (P_0 - X^0)$. No Borel singularity is visible at this point on the left. Upon crossing the conifold point, a Borel singularity appears. The black disks indicate the positions of the points $\aleph (X^1 + n X^0)$, with $n=-1,0,1$ above, on, below the real axis. The blue disks next to these points are at $\aleph (-P_0 + X^1 + n X^0)$ or $\aleph (P_0 + X^1 + n X^0)$; $P_0$ is real and negative on the left, real and positive on the right.}
    \label{fig : Conifold point X5}
\end{figure}
For all hypergeometric models (except for $X_{3,2,2}$, for which the $F_g$ are currently computed only up to genus 14), we can observe the disappearance of a Borel singularity at the points $\pm \aleph (X^0-P_0)$ as we decrease $z$ from the right to the left of the point $\mu$ on the real axis. This is plotted in Figure \ref{fig : Conifold point X5}  for the case of the quintic. We can interpret this phenomenon as the decay of BPS states  $(X^0-P_0)\rightarrow (X^0)+(-P_0)$ upon crossing a line of marginal stability passing through the conifold point (recall that by \eqref{eq:conifoldResurgenceData}, the leading singularity at all internal conifold points lies at $\pm \aleph P^0$, with associated Stokes constant 1). The simplest form of the wall crossing formula of \cite{Denef:2007vg} then predicts the relation
\begin{equation}
    \Omega(\gamma_1+\gamma_2,\mu^+)-\Omega(\gamma_1+\gamma_2,\mu^-)=(-1)^{\langle\gamma_1,\gamma_2\rangle-1}\vert\langle\gamma_1,\gamma_2\rangle\vert\Omega(\gamma_1,\mu)\Omega(\gamma_2,\mu)
    \label{eq:wallcrossingCpoint}
\end{equation}
between the associated BPS invariants, with $ \gamma_1=X^0$ and $\gamma_2=-P_0$. Setting $\Omega(\gamma_1+\gamma_2,\mu^-)=0$, we obtain the prediction 
\begin{equation}
    \Omega(X^0-P_0,\mu^+) = \Omega(X^0,\mu) \,.
\end{equation}
And indeed, we numerically find that the Stokes constants at $\pm \aleph X^0$ and $\pm \aleph (X^0-P_0)$ coincide for all hypergeometric models. For the quintic, we also have some numerical evidence for similar decays $(\mp P_0 + X^1 \pm X^0) \rightarrow (X^1 \pm X^0)$, see Figure \ref{fig : Conifold point X5}. However, we do not have enough numerical precision to determine the associated Stokes constants.

\subsection{Organizing the data} \label{ss:numericalResults}
In this following subsections, we consider the 13 hypergeometric Calabi-Yau models close to $z= \infty$ in turn.

We organize the discussion as follows. We first give a map of the Borel plane in which the visible branch points are identified with integral periods of the holomorphic (3,0) form $\Omega$. Next, we summarize the Stokes data in a table in the following manner.

\begin{itemize}
    \item The first column indicates a choice of period as an $M_\infty$ orbit generator for the observed Borel singularities. By the reflection symmetry of the Borel plane, both a period and its negative appear. The generators are listed in order of proximity to the origin, but note that this ordering does not extend to all elements of the orbits.
    \item The following columns corresponds to each element in the orbit in turn. Note that in some cases, $M_{\infty}^k=-1$ for some $k$. In such cases, we will only list half of the orbit elements, as to not be redundant with the orbit elements of the negative generator.
    \item If the charge associated to the period, via the discussion in Section \ref{s:generalizedDT}, is a rank $\pm 1$ or rank $0$ charge, then we indicate the charge and the associated large radius invariant. More precisely, to each rank $\pm 1$ charge, we indicate, for each element in its $M_\infty$ orbit, the normalized charge $(Q_\pm,n_\pm)_{\mathrm{DT},\mathrm{PT}}$ introduced in Section \ref{s:generalizedDT}, as well as the associated generalized DT invariant at $z=0$. To each rank 0 charge with one unit of D4 charge, we indicate the normalized charge $(\mu,n)_{\text{MSW} }$ introduced in equation (\ref{eq:MSWmu}) as well as the associated generalized DT invariant at $z=0$. To each D2-D0 charge, we indicate the corresponding D0 and D2 charge $(q_0,q_1)_{\GV}$ and the corresponding large radius BPS invariant given, by \eqref{eq:DTGV}, by $n^{0}_{|q_1|}$. We indicate a charge that does not belong to the set just described as well as the associated generalized DT invariant by the symbol $\ast$. 
    \item To every element in the orbit, we indicate, numerical precision permitting, the numerically determined Stokes constant. The symbol $\ast$ indicates a lack of the required numerical precision.
\end{itemize}
As anticipated in Section \ref{ss:BorelAndTopString}, the Borel plane exhibits a reflection symmetry through the origin, and this symmetry is also reflected in the equality of the Stokes constants associated to pairs of singularities related by sign change. The generalized Donaldson-Thomas invariants do not respect this symmetry. In fact, if a charge belongs to an element in the heart of the t-structure, its negative will not. On the other hand, a phase rotation of the central charge will shift which charges are mapped into the appropriate half plane $\mathrm{H}$, without changing the invariants associated to the charges that map to $\mathrm{H}$ for both choices of phase. Whether a distinguished choice of phase for the central charge exists, allowing us to determine before the fact whether the Stokes constant associated to a pair of charges $\pm \gamma$ coincides with the generalized DT invariant $\Omega(\gamma)$ or $\Omega(-\gamma)$ is a question which merits further study, and on which we hope to report elsewhere. In this section, to each pair of singularities in the Borel plane, we include both associated charges $\pm \gamma$. When the charges are of rank 1, the associated invariants are known. As we mentioned above, in all but one case (the $X_{10}$ model), either $\Omega_{\mathrm{LR}}(\gamma)$ or $\Omega_{\mathrm{LR}}(-\gamma)$ (or both, when they coincide) correctly predicts the value of the Stokes constant.

\newpage
\subsection{F-points with massless periods}
\subsubsection{$X_6$}
\begin{figure}[H]
    \centering
    \includegraphics[scale=1]{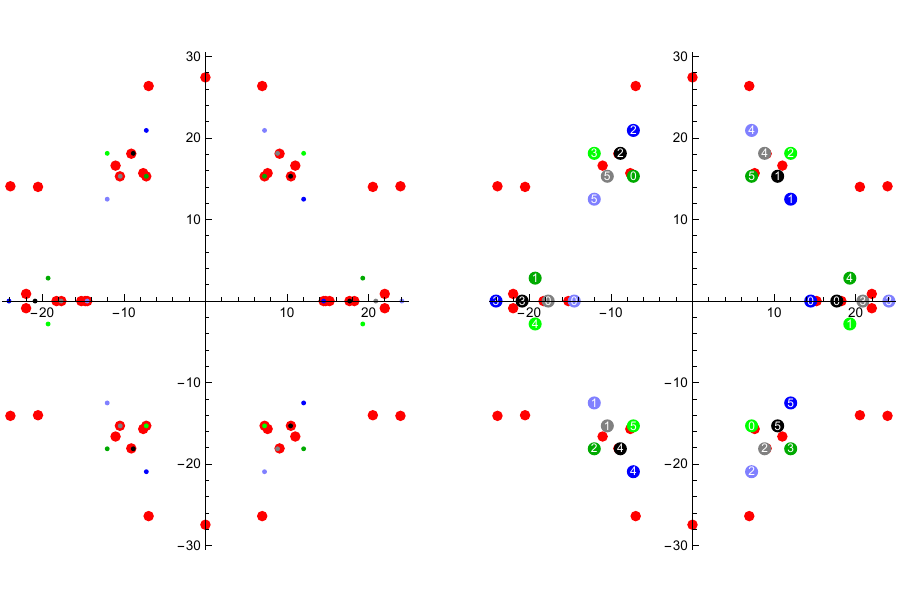}
    \caption{Borel plane of $X_{6}$ at $z=10^{4}\mu$ in the frame $\big(\bq_1,(0,1,-1,0)\big)$. The (light) blue points are orbits of $(-)\aleph P_0$ under $M_{\infty}$, the (light) green points of $(-) \aleph X^0$, the (gray) black points of $(-)\aleph X^1$. We draw the diagram twice: on the right, we indicate the position $n$ of the point in the orbit $\bq\cdot M_{\infty}^n$ inside the colored dots, at the expense of partially covering up the underlying red dots indicating the position of Borel singularities, completely visible on the left. In the following, we will choose the representation on the right, indicating in the caption when the match between period and singularity is sufficiently accurate to cover up the red dot indicating the position of the singularity.}
    \label{fig:X6}
\end{figure}

\begin{table}[h!]
    \begin{center}
        \begin{tabular}{|c| c | c | c | c | c | c | c|}
        \hline
            && $\pi$ & $\pi \cdot M_\infty$ & $\pi \cdot M_\infty^2$ & $\pi \cdot M_\infty^3$ & $\pi \cdot M_\infty^4$ & $\pi \cdot M_\infty^5$  \\ \hline
                $\bq_1$ & charge  & $(1,0)_{\PT}$ &$(1,0)_{\PT}$ &$*$ & $(1,0)_{\PT}$ & $(1,0)_{\PT}$ & $*$ \\
                     &  gen. DT invariant & 0    & 0&  $*$ &   0    & 0 & $*$ \\
                     &  Stokes constant &  0    & 0 &  0    &  0    & 0 & 0 \\\hline
             $-\bq_1$ & charge & $(1,0)_{\DT}$ &$(1,0)_{\DT}$ &$*$ & $(1,0)_{\DT}$ & $(1,0)_{\DT}$ & $*$ \\
                     &  gen. DT invariant & 0    & 0&  $*$ &   0    & 0 & $*$ \\
                     &  Stokes constant &  0    & 0 &  0    &  0    & 0 & 0 \\\hline

             $P_0$ & charge & $(0,0)_{\PT}$ & $(0,0)_{\PT}$ & $*$& $*$ & $*$ & $*$\\
                     &  gen. DT invariant & 1    & 1 &  $*$   &   $*$   & $*$ &$*$\\
                     &  Stokes constant &  1    & 1 &  1   &  1    & 1 &  1\\\hline
                       $-P_0$ & charge & $(0,0)_{\DT}$ & $(0,0)_{\DT}$ & $*$& $*$ & $*$ & $*$\\
                     &  gen. DT invariant & 1    & 1 &  $*$   &   $*$   & $*$ &$*$\\
                     &  Stokes constant &  1    & 1 &  1   &  1    & 1 &  1\\\hline
             $X_0$  & charge & $(0,1)_{\GV}$&$(0,-1)_{\PT}$ & $*$&$(0,2)_{\text{MSW}}$ & $*$&   $(0,-1)_{\DT}$\\
                     &  gen. DT invariant &$-\chi=204$   & 0&  $*$   &   -40392   & $*$ &0\\
                     &  Stokes constant &  $-\chi$    &   $-\chi$     &    $-\chi$       &    $-\chi$        &   $-\chi$     &    $-\chi$    \\\hline
            $-X_0$  & charge & $(0,-1)_{\GV}$&$(0,1)_{\DT}$ & $*$&$*$ & $*$&   $(0,1)_{\PT}$\\
                     &  gen. DT invariant & $-\chi$    &   $-\chi$&  $*$   &   $*$   & $*$ &  $-\chi$\\
                     &  Stokes constant &  $-\chi$    &   $-\chi$     &    $-\chi$       &    $-\chi$        &   $-\chi$     &    $-\chi$    \\\hline
             $X_1$ & charge & $(1,0)_{\GV}$ & $(1,-1)_{\GV}$&$(1,1)_{\DT}$ & (2,0)$_{\PT}$ & (2,0)$_{\PT}$  & (1,-1)$_{\DT} $\\
              &  gen. DT invariant & $n_{0,1}=7884$  & $n_{0,1}$&  $n_{0,1}$    &   $n_{0,1}$   & $n_{0,1}$ &0\\
                     &  Stokes constant &  $n_{0,1}$   &   $n_{0,1}$      &    $n_{0,1}$        &   $n_{0,1}$        &   $n_{0,1}$       &   $n_{0,1}$     \\\hline

              $-X_1$ & charge & $(-1,0)_{\GV}$ & $(-1,1)_{\GV}$&$(1,-1)_{\PT}$ & (2,0)$_{\DT}$ & (2,0)$_{\DT}$  & (1,1)$_{\PT} $\\
              &  gen. DT invariant & $n_{0,1}=7884$    & $n_{0,1}$&  $0$    &   $n_{0,1}$   & $n_{0,1}$ &$n_{0,1}$\\
                     &  Stokes constant &  $n_{0,1}$   &   $n_{0,1}$      &    $n_{0,1}$        &   $n_{0,1}$        &   $n_{0,1}$       &   $n_{0,1}$     \\\hline

        \end{tabular}
    \end{center}
     \caption[Borel singularities of $X_6$]{Borel singularities of $X_6$  at $z=\infty$  and their corresponding invariants. Note that ${\bq_2 = \bq_1 \cdot M_\infty}$, and $\bq_1 \cdot M_\infty^3 = \bq_1$.}
    \end{table}
\paragraph{Comments}
\begin{itemize}
    \item Deviating from all other models, we have also listed the massless period $\bq_1$ in the table, even though it does not appear as a singularity, consistent with the fact that the associated generalized DT invariant vanishes. Note that the $\bq_2$ row in the table is redundant, as $\bq_2 = \bq_1\cdot M_{\infty}^2$ lies in the orbit of $\bq_1$. 
    \item $\aleph X^1$ and $\aleph P_0$ lie on the real axis and are close in value. Thus, both contribute to the RHS of \eqref{eq:disc}, but with different Stokes constants. To separate the contributions, 
    \begin{enumerate}
        \item We conjecture that the Stokes constant associated to $X^1$ is $n_{0,1}= 7884$.
        \item We check this conjecture by computing the reduced amplitudes $\cF_g^{\mathrm{red}}$ with regard to the Borel singularities at $\pm X^1$ via \eqref{eq:red} and verifying that the poles associated to these singularities are removed.
        \item We compute the Stokes constant associated to $P_0$ via \eqref{eq:SomegaFromDisc} applied to $\cF_g^{\mathrm{red}}$.
        \item To check the procedure, we compute the reduced amplitudes now with regard to the singularities at $\pm P^0$ using the Stokes constant determined, and compute the Stokes constant associated to $\pm X^1$ via \eqref{eq:SomegaFromDisc}.
    \end{enumerate}
    As the cycles underlying $X^1$ and $P_0$ are symplectically orthogonal, we can choose a frame in which $X^1, P_0$ are both A-periods, simplifying the instanton corrections required for the computation of $\cF_g^{\mathrm{red}}$ in both cases.
\end{itemize}

\newpage

\subsubsection{$X_{4,3}$}
\begin{figure}[H]
\centering\includegraphics[scale=0.8]{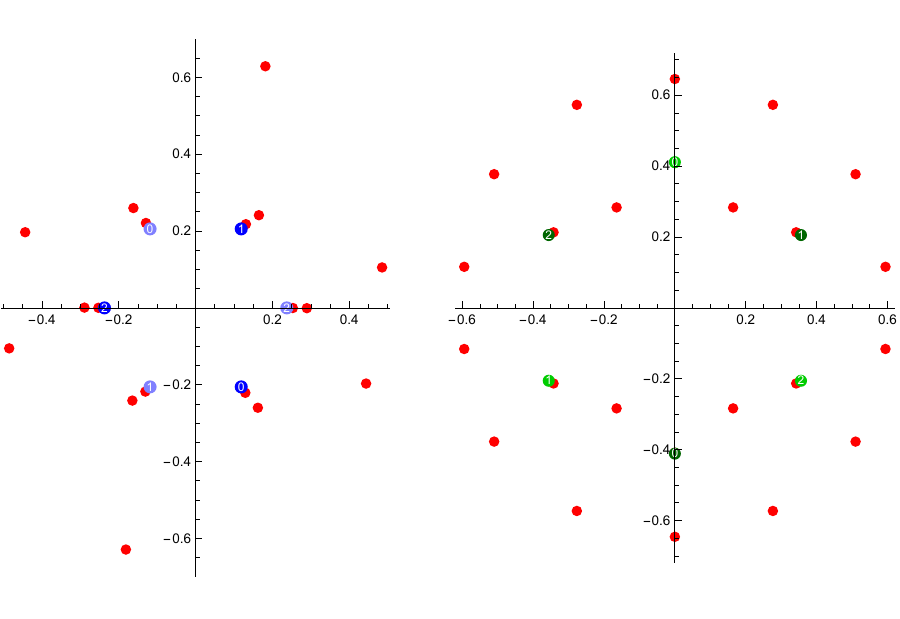}
\caption{Borel plane of $X_{4,3}$ at $z=10^{8}\mu$ in the frame $\big(\bq_1,(0,1,-2,0)\big)$. On the right, the leading singularities are subtracted. The (light) blue dots indicate the orbit elements of $\aleph\bq_1\cdot\periodV$ ($-\aleph\bq_1\cdot\periodV$) under the monodromy $M_\infty$, the (light) green dots the orbit elements of $\aleph(\bq_1 - \bq_2)\cdot\periodV$ ($\aleph(\bq_2 - \bq_1)\cdot\periodV)$. The blue dots on the left and the two green dots on the imaginary axis on the right completely cover red dots.}
\label{fig:X43}
\end{figure}
\begin{table}[H]
    \begin{center}
        \begin{tabular}{|c| c | c | c | c | c | c| c| }
           \hline && $\pi$ & $\pi \cdot M_\infty$ & $\pi \cdot M_\infty^2$  \\ \hline
          $\bq_1$ &charge& $(2,0)_{\PT}$&$(2,0)_{\PT}$&$*$\\ 
          &gen. DT invariant&27&27&$*$\\
          &Stokes constant &27&27&27\\\hline
           $-\bq_1$ &charge& $(2,0)_{\DT}$&$(2,0)_{\DT}$&$*$\\ 
          &gen. DT invariant&27&27&$*$\\
          &Stokes constant &27&27&27\\\hline
           $\bq_1-\bq_2$ &charge&$(0,2)_{\MSW}$ &$*$&$*$\\
          &gen. DT invariant&$*$&$*$&$*$\\
          &Stokes constant &$*$&$*$&$*$\\\hline
           $\bq_2-\bq_1$ &charge&$(0,-2)_{\MSW}$ &$*$&$*$\\
          &gen. DT invariant&$*$&$*$&$*$\\
          &Stokes constant &$*$&$*$&$*$\\\hline
        \end{tabular}
         \caption[Borel singularities of $X_{4,3}$]{Leading Borel singularities of $X_{4,3}$ at $z=\infty$ and their corresponding invariants. Note that $\bq_1 \cdot M_\infty =\bq_2$ and $\bq_2 \cdot M_\infty^2=-\bq_1$.}
    \end{center}
\end{table}
\paragraph{Comments}
\begin{itemize}
    \item Removing the leading singularities associated to the $M_\infty$ orbit elements of $\pm\bq_1$ reveals singularities at $M_\infty$ orbit elements of $\pm(\bq_1 - \bq_2)$. Our numerical precision is not sufficient to determine the Stokes constant at these points.
\end{itemize}

 \subsubsection{$X_{6,4}$}

\begin{figure}[H]
\centering\includegraphics[scale=1]{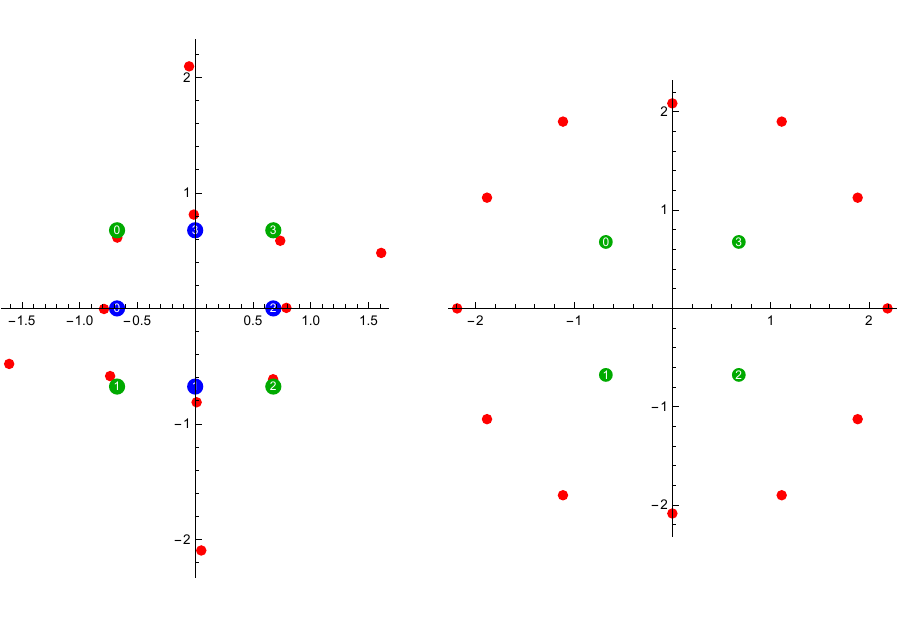}
\caption{Borel plane of $X_{6,4}$ at $z=10^{8}\mu$ in the frame $\big(\bq_1,(0,1,-1,0)\big)$. On the right, the leading singularities are subtracted. The blue dots indicate the orbit elements of $\aleph\bq_1\cdot\periodV$ under the monodromy $M_\infty$, the green dots the orbit elements of $\aleph(\bq_1 + \bq_2)\cdot\periodV$. The blue dots on the left and the green dots on the right completely cover red dots.}
\label{fig:X64}
\end{figure}

\begin{table}[h!]
    \begin{center}
        \begin{tabular}{|c| c | c | c | c | c | }
            \hline&& $\pi$ & $\pi \cdot M_\infty$   \\ \hline
          $\bq_1$ &charge& $(1,0)_{\PT}$&$(1,0)_{\PT}$\\ 
                     &gen. DT invariant&8&8\\
                  &Stokes constant &8&8\\\hline
           $-\bq_1$ &charge&  $(1,0)_{\DT}$  & $(1,0)_{\DT}$\\ 
                     &gen. DT invariant&8&8\\
                     &Stokes constant &8&8\\\hline
    $\bq_1+\bq_2$   &charge& $(0,-1)_{\text{MSW}}$&$*$\\ 
                    &gen. DT invariant&$-304$&$*$\\
                    &Stokes constant &8&8\\\hline
     $-(\bq_1+\bq_2)$ &charge& $*$&$*$\\ 
                    &gen. DT invariant&$*$&$*$\\
                    &Stokes constant &8&8\\\hline
        \end{tabular}
    \end{center}
    \caption[Borel singularities of $X_{6,4}$]{Leading Borel singularities of $X_{6,4}$ at $z=\infty$ and their corresponding invariants. Note that $\bq_1 \cdot M_\infty^2=-\bq_1$ and $\bq_2 \cdot M_\infty^2=-\bq_2$.}
    \end{table}

\paragraph{Comments}
\begin{itemize}
    \item The subleading singularity is uncovered by studying $\cF_g^{\mathrm{red}}$ as defined in \eqref{eq:red}, with regard to the contributions from the four singularities in the $M_{\infty}$ orbit of $\bq_1$. We computed the subleading Stokes constant in the frame $(\bq_1+\bq_2, X^0)$ at $z=10^8\mu$ via the formula \eqref{eq:SomegaFromAsympOrbit}. Despite the relatively low genus, $g_{\mathrm{max}}=17$, to which the $\cF_g$ are available, the value $S=8$ emerges reliably.
\end{itemize}

\subsection{F-point without massless period}
\subsubsection{$X_5$}
\begin{figure}[H]
    \centering
    \includegraphics[trim = 0 20 0 20, clip, scale=1]{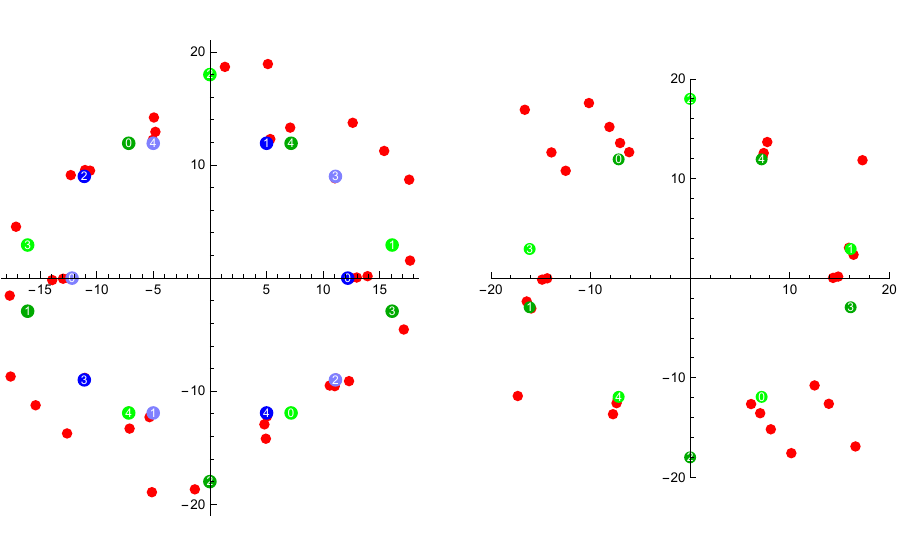}
    \caption{Borel plane of $X_{5}$ at $z=10^{8}\mu$ in the frame $\big(X^0,X^1\big)$. On the right, the leading singularities are subtracted. The (light) blue dots indicate the orbit elements of $\aleph X^1$ ($-\aleph X^1$) under the monodromy $M_\infty$, the (dark) green dots the orbit elements of $\aleph X^0$ ($-\aleph X^0$). The blue dots on the left completely cover red dots, the green dots with numbers $0$ and $4$ cover red dots both on the left and the right.}
\label{fig:X5}
\end{figure}

\begin{table}[H]
    \begin{center}
        \begin{tabular}{|c| c | c | c | c | c | c| }
        \hline
            && $\pi$ & $\pi \cdot M_\infty$ & $\pi \cdot M_\infty^2$ & $\pi \cdot M_\infty^3$&$\pi \cdot M_\infty^4$  \\ \hline
          $X^1$ &charge& $(1,0)_{\GV}$&$(1,-1)_{\GV}$&$(1,1)_{\DT}$&$*$&$(1,-1)_{\DT}$\\
                   &gen. DT invariant&$n_{0,1}=2875$&$n_{0,1}$&$n_{0,1}$&$*$&0\\
                  &Stokes constant&$n_{0,1}$&$n_{0,1}$&$n_{0,1}$&$n_{0,1}$&$n_{0,1}$\\\hline
        $-X^1$ &charge&$(-1,0)_{\GV}$&$(-1,1)_{\GV}$&$(1,-1)_{\PT}$&$*$&$(1,1)_{\PT}$\\
                 &gen. DT invariant&$n_{0,1}$&$n_{0,1}$&0&$*$&$n_{0,1}$\\
                   &Stokes constant&$n_{0,1}$&$n_{0,1}$&$n_{0,1}$&$n_{0,1}$&$n_{0,1}$\\\hline
          $X^0$ &charge& $(0,1)_{\GV}$&$(0,-1)_{\PT}$&$*$&$*$&$(0,-1)_{\DT}$\\
                   &gen. DT invariant&$-\chi=200$&0&$$*$$&$*$&0\\
                   &Stokes constant&$-\chi$&$-\chi$&$-\chi$&$-\chi$&$-\chi$\\\hline
          $-X^0$ &charge& $(0,-1)_{\GV}$&$(0,1)_{\DT}$&$*$&$*$&$(0,1)_{\PT}$\\
                   &gen. DT invariant&$-\chi$&$-\chi$&$$*$$&$*$&$-\chi$\\
                   &Stokes constant&$-\chi$&$-\chi$&$-\chi$&$-\chi$&$-\chi$\\\hline

        \end{tabular}
    \end{center}
    \caption[Borel singularities of $X_5$]{Leading Borel singularities of $X_5$  at $z=\infty$  and their corresponding invariants.}
    \label{tab:X5}
\end{table}
\paragraph{Comments}
\begin{itemize}
    \item Unlike the two other models in this class, $X_8$ and $X_{10}$, $X_5$ does not exhibit $P_0$ as a leading Borel singularity, consistent with the observation that at large $z$, $|P_0| > |X_0|$ for this model.
\end{itemize}

\subsubsection{$X_8$}
\begin{figure}[H]
    \centering
     \includegraphics[scale=1]{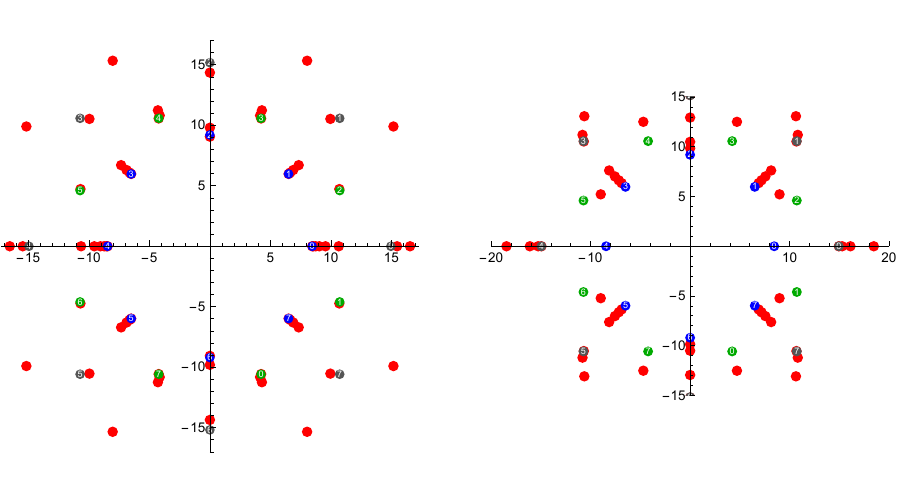}
   \caption{Borel plane of $X_{8}$ at $z=10^{6}\mu$ in the frame $\big(P^0,X^1\big)$. On the right, the leading singularities are subtracted. The blue dots indicate the orbit elements of $\aleph P_0$ under the monodromy $M_\infty$, the green dots the orbit elements of $\aleph X^0$, the gray dots the orbit elements of $\aleph X^1$. The blue and green dots on the left and the gray dots on the right completely cover red dots.}
\label{fig:X8}
\end{figure}

   \begin{table}[H]
    \begin{center}
    \scalebox{0.85}{
        \begin{tabular}{|c| c | c | c | c | c | c | c | c | c |}\hline
            && $\pi$ & $\pi \cdot M_\infty$ & $\pi \cdot M_\infty^2$ & $\pi \cdot M_\infty^3$ \\ \hline
          $P_0$ &charge& $(0,0)_{\PT}$&$(0,0)_{\PT}$&$*$&$*$\\
              &gen. DT invariant&1&1&$*$&$*$\\
               &Stokes constant&1&1&1&1\\\hline
        $-P_0$ &charge& $(0,0)_{\DT}$&$(0,0)_{\DT}$&$*$&$*$\\
                &gen. DT invariant&1&1&$*$&$*$\\
                 &Stokes constant&1&1&1&1\\\hline
        $X_0$ &charge& $(0,1)_{\GV}$&$(0,-1)_{\PT}$&$*$&$(0,1)_{\PT}$\\
                &gen. DT invariant&$-\chi=296$&$0$&$*$&$-\chi$\\
                &Stokes constant&$-\chi$&$-\chi$&$-\chi$&$-\chi$\\
          \hline
             $-X_0$ &charge& $(0,-1)_{\GV}$&$(0,1)_{\DT}$&$*$&$(0,-1)_{\DT}$\\
                     &gen. DT invariant&$-\chi$&$-\chi$&$*$&0\\
                     &Stokes constant&$-\chi$&$-\chi$&$-\chi$&$-\chi$\\
          \hline
             $X_1$ &charge& $(1,0)_{\GV}$&$(1,-1)_{\GV}$&$(1,1)_{\DT}$&$(1,1)_{\PT}$\\
                    &gen. DT invariant&$n_{0,1}=29504$&$n_{0,1}$&$n_{0,1}$&$n_{0,1}$\\
                    &Stokes constant&$n_{0,1}$&$n_{0,1}$&$n_{0,1}$&$n_{0,1}$\\ \hline
           $-X_1$ &charge& $(-1,0)_{\GV}$&$(-1,1)_{\GV}$&$(1,-1)_{\PT}$&$(1,-1)_{\DT}$\\
                     &gen. DT invariant&$n_{0,1}$&$n_{0,1}$&0&0\\
                 &Stokes constant&$n_{0,1}$&$n_{0,1}$&$n_{0,1}$&$n_{0,1}$\\ \hline
        \end{tabular}}
    \end{center}
    \caption[Borel singularities of $X_8$]{Leading Borel singularities of $X_{8}$ and their corresponding invariants. Note that $M_\infty^4 = -1$.}
\end{table}
\paragraph{Comments}
\begin{itemize}
    \item To determine the Stokes constant associated to $\pm X^1$, we choose the frame $(X^1,P_0)$ at $z=10^6\mu$ and apply \eqref{eq:SomegaFromDisc}. We compute the Stokes discontinuity along the ray in the $\aleph X^1$ direction with regard to $\Fred$ with the asymptotic contributions of $M_\infty$ orbit elements of $P_0$ removed. The value $S\sim 29504$ emerges for values of $g_s$ around 1/3. We check this value by using it to remove $\pm X^1$ from the Borel plane.
\end{itemize}

\subsubsection{$X_{10}$}
\begin{figure}[H]
    \centering
    \includegraphics[trim = 0 60 0 60, clip, scale=0.4]{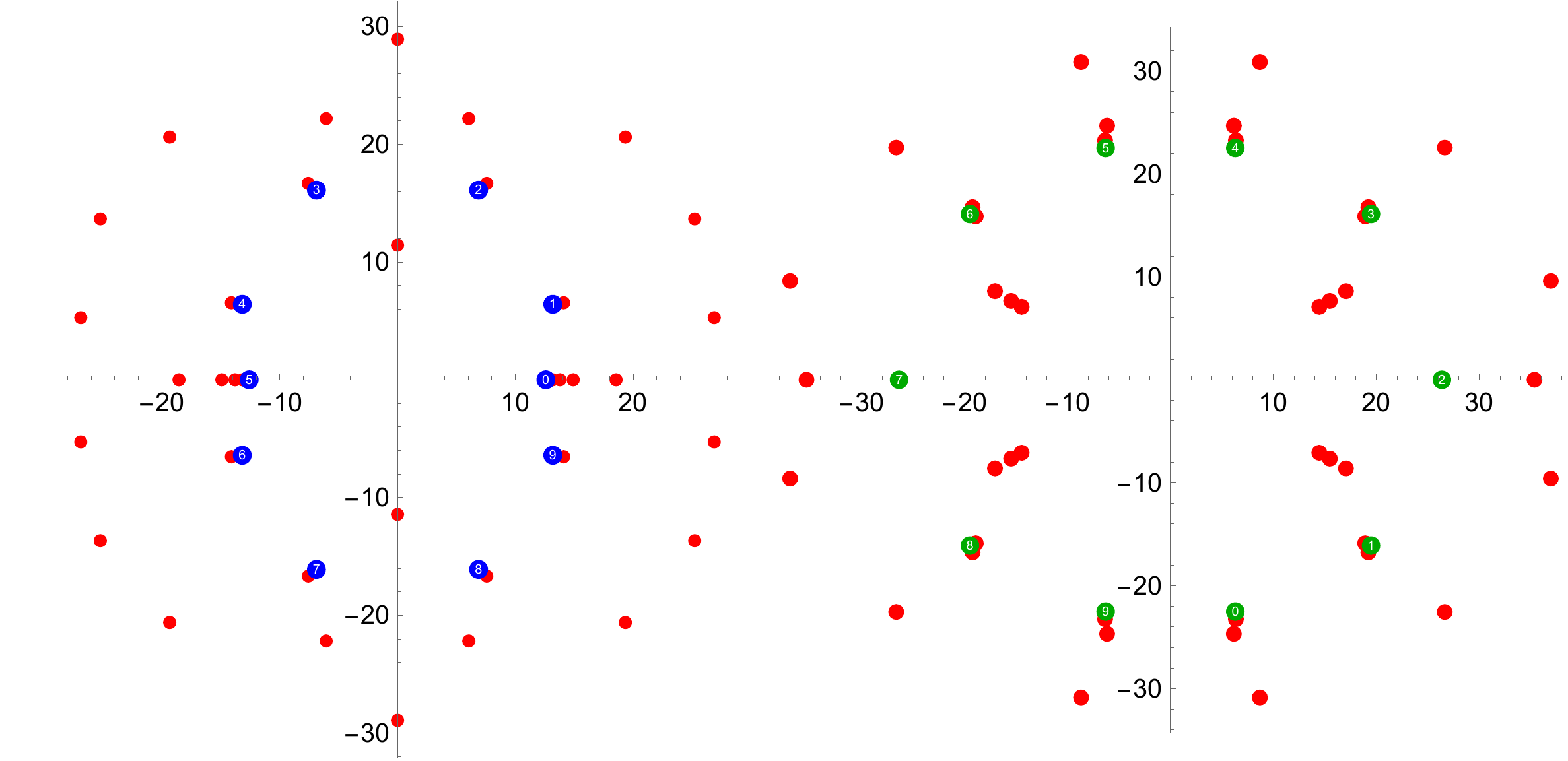}
    \caption{Borel plane of $X_{10}$ in the frame $(P_0,X^1)$ at $z=10^4\mu$. On the right, the leading singularities are subtracted. Blue dots indicate orbit elements of $\aleph P_0$ under $M_\infty$, green dots  orbit elements of $\aleph X^0$. All blue and green dots completely cover red dots.}
    \label{fig:BPX10}
\end{figure}

\begin{table}[H]
    \begin{center}
        \scalebox{0.8}{
       \begin{tabular}{|c| c | c |c| c | c | c |  }
        \hline
            && $\pi$ & $\pi \cdot M_\infty$ & $\pi \cdot M_\infty^2$ & $\pi \cdot M_\infty^3$&$\pi \cdot M_\infty^4$ \\ \hline
          $P_0$ &charge& $(0,0)_{\PT}$&$(0,0)_{\PT}$&$*$&$(0,1)_{\MSW}$&$*$\\
          &gen. DT invariant&1&1&$*$&-575&$*$\\
          &Stokes constant&1&1&1&1&1\\\hline
          $-P_0$ &charge&$(0,0)_{\DT}$&$(0,0)_{\DT}$&$*$&$*$&$*$\\
          &gen. DT invariant&1&1&$*$&$*$&$*$\\
          &Stokes constant&1&1&1&1&1\\\hline
            $X_0$ &charge&$(0,1)_{\GV}$& $(0,-1)_{\PT}$&$(1,0)_{\DT}$&$(1,0)_{\DT}$&$(0,1)_{\PT}$\\
          &gen. DT invariant&$ -\chi = 288 $&0&1150&1150&0\\
          &Stokes constant&$-\chi$&$-\chi$&$-\chi$&$-\chi$&$-\chi$\\\hline  
           $-X_0$ &charge&$(0,-1)_{\GV}$&$(0,1)_{\DT}$&$(1,0)_{\PT}$&$(1,0)_{\PT}$&$(0,-1)_{\DT}$\\
          &gen. DT invariant&$-\chi$&$-\chi$&286&286&0\\
          &Stokes constant&$-\chi$&$-\chi$&$-\chi$&$-\chi$&$-\chi$\\\hline  
        \end{tabular}}
    \caption[Borel singularities of $X_{10}$]{Leading Borel singularities of $X_{10}$ at $z=\infty$ and their corresponding invariants. Note that $M_\infty^5 = -1$.}
    \end{center}
\end{table}
\paragraph{Comments}
\begin{itemize}
    \item Removing the leading singularities associated to the $M_\infty$ orbit elements of $\aleph P_0$ reveals singularities at $M_\infty$ orbit elements of $\aleph X^0$. To compute the Stokes constant associated to the subleading singularities, our numerical precision proves insufficient for the use of formula \eqref{eq:SomegaFromDisc}. We therefore seek the constant which allows for the removal of the leading singularities. For all singularities other than the one associated to $M_{\infty}^2 X^0$, this procedure yields the value $S=-\chi$ reliably in the frame $(P_0,X^1)$ at $z=10^4\mu$. For the remaining singularity, the result in this frame is inconclusive, but the frame $(P_0,M_{\infty}^2 X^0)$ at $z=10^6\mu$ permits us to identify the Stokes constant also here as $S=-\chi$.
\end{itemize}

\newpage

\subsection{K-points}
For all three models with K-points, all visible Borel singularities lie at the lightest periods $\bq_1$ and $\bq_2$ and their images under the repeated action of $M_\infty$, which are linear combinations of these two periods. Unfortunately, choosing a frame defined by $\bq_1$ and $\bq_2$ -- a distinguished frame for studying the instanton corrections associated to all of these singularities -- proves to be numerically unfeasible, probably due to the smallness of both periods $\aleph \bq_i \cdot \piGeom$, $i=1,2$. 

\subsubsection{$X_{3,3}$}
\begin{figure}[H]
   \center \includegraphics[scale=1]{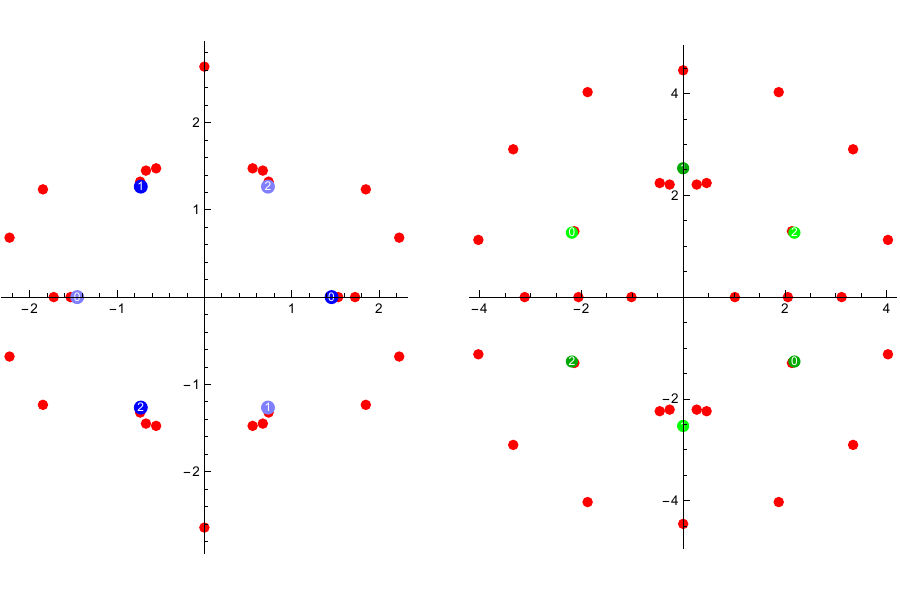}     \caption{Borel plane of $X_{3,3}$ at $z=10^{6}\mu$ in the frame $(\bq_1\cdot\periodV,P_0)$. On the right, the leading singularities are subtracted. The (light) blue dots indicate the orbit elements of $\aleph\bq_1\cdot\periodV$ (-$\aleph\bq_1\cdot\periodV$) under the monodromy $M_\infty$, the (light) green dots the orbit elements of $\aleph(\bq_1 - \bq_2)\cdot\periodV$ ($\aleph(\bq_2 - \bq_1)\cdot\periodV$). All blue dots completely cover red dots.}
    \label{fig:X33}
\end{figure}

\begin{table}[h!]
    \begin{center}
    \scalebox{0.9}{
        \begin{tabular}{|c| c | c | c | c |}
        \hline
            && $\pi$ & $\pi \cdot M_\infty$ & $\pi \cdot M_\infty^2$    \\ \hline
          $\bq_1$ &charge& $(3,0)_{\DT}$&$(3,0)_{\DT}$&$*$\\ 
          &gen. DT invariant&3402&3402&$*$\\
          &Stokes constant &3402&3402&3402\\\hline
   $-\bq_1$ &charge& $(3,0)_{\PT}$&$(3,0)_{\PT}$&$*$\\ 
          &gen. DT invariant&3402&3402&$*$\\
          &Stokes constant &3402&3402&3402\\\hline
      $\bq_1 - \bq_2$ &charge&$*$ & $*$& $*$\\ 
          & gen. DT invariant& $*$&$*$&$*$\\
          & Stokes constant & $*$ & $*$ & $*$ \\ \hline
          $\bq_2 - \bq_1$ &charge&$(0,-3)_{\MSW}$ & $*$& $*$\\ 
          & gen. DT invariant&-678474&$*$&$*$\\
          & Stokes constant & $*$ & $*$ & $*$ \\ \hline
        \end{tabular}}
    \end{center}
    \caption[Borel singularities of $X_{3,3}$]{Leading Borel singularities of $X_{3,3}$ and their corresponding invariants.}
\end{table}

\paragraph{Comments}
\begin{itemize}
    \item Removing the leading singularities associated to the $M_\infty$ orbit elements of $\pm\bq_1$ reveals singularities at $M_\infty$ orbit elements of $\pm(\bq_1 - \bq_2)$. Our numerical precision is not sufficient to determine the Stokes constant at these points.
\end{itemize}

\subsubsection{$X_{4,4}$}
\begin{figure}[H]
  \center \includegraphics[trim=0 20 0 20 ,clip,scale=1]{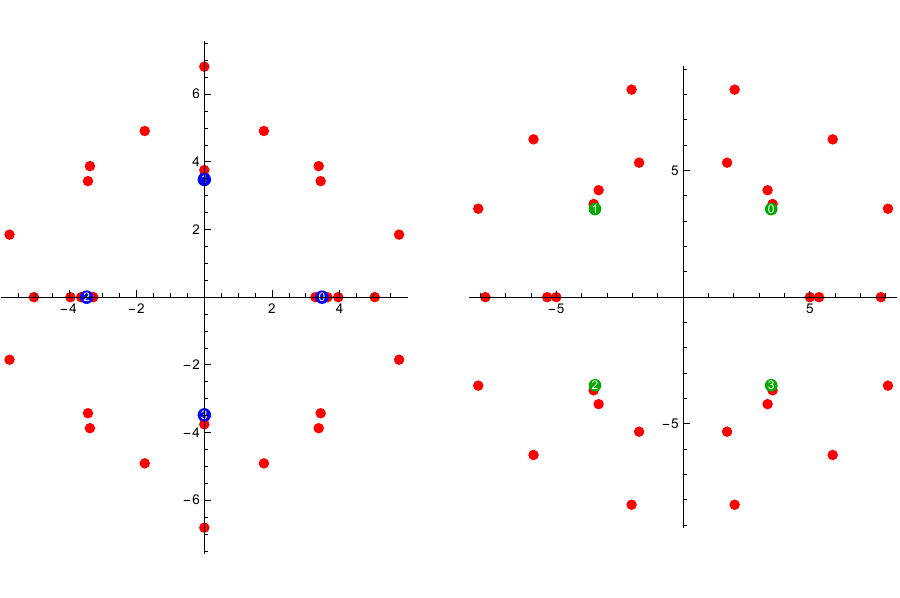}
    \caption{Borel plane of $X_{4,4}$ at $z=10^{6}\mu$ in the frame $(\bq_1\cdot\periodV,P_0)$. On the right, the leading singularities are subtracted. The blue dots indicate the orbit elements of $\aleph\bq_1\cdot\periodV$ ($-\bq_1 = \bq_1\cdot M_\infty^2$) under the monodromy $M_\infty$, the  green dots the orbit elements of $\aleph(\bq_1 - \bq_2)\cdot\periodV$. The blue and the green dots completely cover red dots.}
    \label{fig:X44}
\end{figure}
\begin{table}[H]
    \begin{center}
    \scalebox{0.8}{
        \begin{tabular}{|c| c | c | c | c | c | c| }
           \hline && $\pi$ & $\pi \cdot M_\infty$   \\ \hline
          $\bq_1$ &charge& $(2,0)_{\DT}$&$(2,0)_{\DT}$\\ 
                 &gen. DT invariant&1408&1408\\
                &Stokes constant &1408&1408\\\hline
        $-\bq_1$ &charge& $(2,0)_{\PT}$&$(2,0)_{\PT}$\\ 
                 &gen. DT invariant&1408&1408\\
                &Stokes constant &1408&1408\\\hline
        $\bq_1+\bq_2$ &charge&$*$&$(0,2)_{\text{MSW}}$\\
          &gen. DT invariant&$*$&-10032\\
          &Stokes constant&9984&9984\\\hline
         $-(\bq_1+\bq_2)$ &charge&$*$&$*$\\
          &gen. DT invariant&$*$&$*$\\
          &Stokes constant&9984&9984\\\hline
        \end{tabular}}
    \end{center}
    \caption[Borel singularities of $X_{4,4}$]{Leading Borel singularities of $X_{4,4}$ at $z=\infty$ and their corresponding invariants. Note that $\bq_1 \cdot M_\infty = \bq_2$ and $\bq_1 \cdot M_\infty^2 = -\bq_1$.}
    \label{X44 table}
\end{table}
\paragraph{Comments}
\begin{itemize}
    \item Removing the leading singularities lying at the $M_\infty$ orbit elements of $\bq_1$ reveals the singularities at the $M_\infty$ orbit elements of $\bq_1 + \bq_2$. In the frame $((\bq_1+\bq_2)\cdot\periodV,X^0)$ at $z=10^6\mu$, we apply \eqref{eq:SomegaFromDisc} with regard to the reduced amplitude $\Fred$ with the asymptotic contributions from $M_\infty$ orbit elements of $\bq_1$ removed stabilizes around $9984\pm1$ for values of around $\vert g_s\vert\sim \frac{1}{10}$. We check this value by using it to remove the singularities associated to $\bq_1 + \bq_2$ in the Borel plane. We thus verify that the value of the Stokes constant that makes the singularities disappear is indeed $9984$. 
\end{itemize}

\subsubsection{$X_{6,6}$}
\begin{figure}[H]
   \center  \includegraphics[trim= 0 55 0 55, clip,scale=0.9]{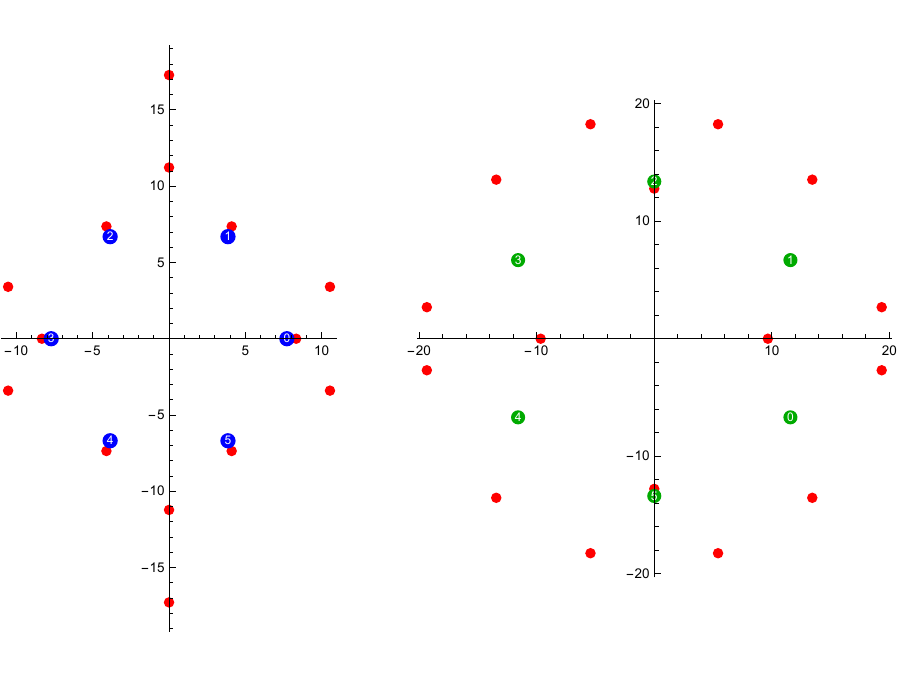}
    \caption{Borel plane of $X_{6,6}$ at $z=10^{6}\mu$ in the frame $(\bq_1\cdot\periodV,P_0)$. On the right, the leading singularities are subtracted. The blue dots indicate the orbit elements of $\aleph\bq_1\cdot\periodV$ ($-\bq_1 = \bq_1\cdot M_\infty^3$) under the monodromy $M_\infty$, the  green dots the orbit elements of $\aleph(\bq_1 - \bq_2)\cdot\periodV$. The blue dots, and the green dots other than dot number 2 and 5 completely cover red dots.}
    \label{fig:X66}
\end{figure}

\begin{table}[H]
    \begin{center}
    \scalebox{0.8}{
        \begin{tabular}{| c| c | c | c | c | c | c| c | }
        \hline
            && $\pi$ & $\pi \cdot M_\infty$ & $\pi \cdot M_\infty^2$  \\ \hline
          $\bq_1$ &charge& $(1,0)_{\DT}$&$(1,0)_{\DT}$&$(0,1)_{\MSW}$\\ 
          &gen. DT invariant&482&482&482\\
          &Stokes constant &362&362&362\\\hline
                  $-\bq_1$ &charge& $(1,0)_{\PT}$&$(1,0)_{\PT}$&$*$\\ 
          &gen. DT invariant&362&362&$*$\\
          &Stokes constant &362&362&362\\\hline

        \end{tabular}}
    \end{center}
    \caption[Borel singularities of $X_{6,6}$]{Leading Borel singularities of $X_{6,6}$  at $z=\infty$ and their corresponding invariants. Note that $\bq_1 \cdot M_\infty = \bq_2$ and $\bq_2 \cdot M_\infty^2 = -\bq_1$.}
\end{table}

\paragraph{Comments:} 
\begin{itemize}
    \item We can propose a naive wall-sacrossing for this model to correct the generalized DT invariant for the charge $\bq_1 \cdot M_{\infty}=(-1,-1,1,1)$ at $z=0$, $\Omega_{\mathrm{LR}}(\bq_1 \cdot M_{\infty}) = 482$ to the numerically determined value of the Stokes constant, $362$. This proceeds via the decay
    \begin{equation}
        (-1,1,1,1) \rightarrow \underbrace{(-1,-1,2,1)}_{=:\gamma_1}  + \underbrace{(0,0,-1,0)}_{=: \gamma_2} \,.
    \end{equation}
    If we conjecture that the Stokes constants $S_{\gamma_1}=1$ and $S_{\gamma_2} = -\chi = 120$ associated to the charges $\gamma_1$ and $\gamma_2$ at $z=-10 \mu$ as determined in Section \ref{s:beyondSingular} (see Table \ref{tab:X66beyond}) coincide with the invariants $\Omega(\gamma_i,z_0)$, $i=1,2$  at the location $z_0$ of the wall, then an application of the wall-crossing formula \eqref{eq:wallcrossingCpoint} yields
    \begin{equation}
    \Delta \Omega((-1,-1,1,1))=120 \,.
\end{equation}
\end{itemize} 

\newpage
\subsection{C-points}
By the discussion in Section \ref{ss:BorelAndTopString}, we have analytical control over the leading singularities in the case of C-points. This predicts the location and the associated Stokes constants of the leading singularity at $\aleph \bq_1 \cdot \Pi$. In addition, we can subtract the leading behavior more precisely than in other models (subtracting the contributions in \eqref{eq:gapX42X62} and \eqref{eq:leadingX322} amounts to subtracting the instanton contributions to all multi-instanton order).

\subsubsection{$X_{4,2}$} \label{s:X42}
\begin{figure}[H]
    \centering
    \includegraphics[trim= 0 50 0 50, clip, scale=1]{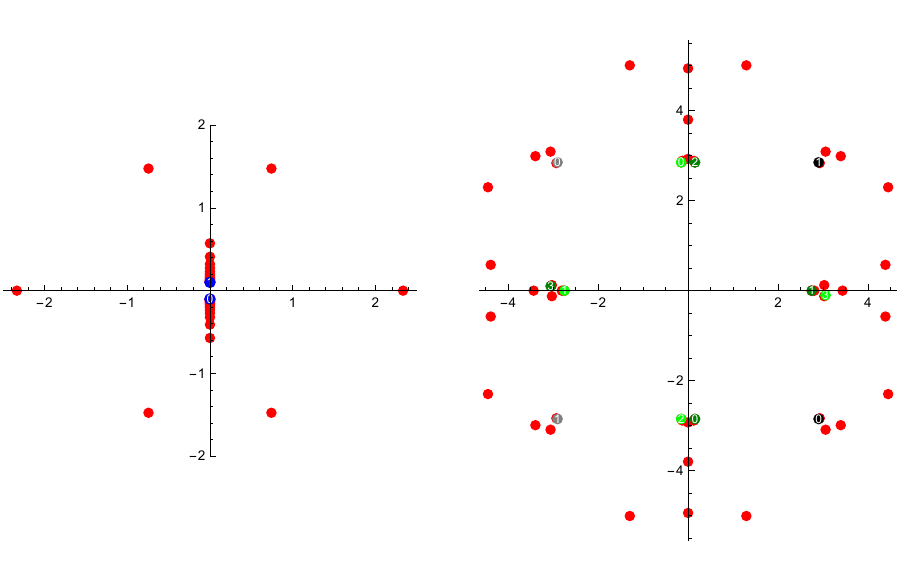}
    \caption{Borel plane of  $X_{4,2}$  at $z=10^6\mu$ in the frame $(\bq_1\cdot\periodV,P_1)$. On the right, the leading gap singularity is subtracted. The blue dots indicate the points $\bq_1 \cdot M_\infty^i$, $i=0,1$, the (light) green dots  the points $(-)(-1,0,1,1) \cdot M_\infty^i$, $i=0,\ldots,3$, the (gray) black dots the points $(-)(-1,0,1,2) \cdot M_\infty^i$, $i=0,1$.}
    \label{X42 X62 leading}
\end{figure}

\begin{table}[h!]
    \begin{center}
    \scalebox{0.8}{
        \begin{tabular}{|c| c | c |c| c |    }
          \hline  && $\pi$& $\pi\cdot M_{\infty}$&$\pi \cdot M^2_{\infty}$ \\ \hline
          $\bq_1$ &charge& $*$&$*$&$*$\\
               &gen. DT invariant&$*$&$*$&$*$\\ 
               &Stokes constant &2&2&2\\ \hline
         $\bq_2$ &charge& $(2,0)_{\text{PT}}$&$(2,0)_{\text{PT}}$&$*$\\ 
                 &gen. DT invariant&0&0&$*$\\
                 &Stokes constant &0&0&0\\\hline
        $-\bq_2$ &charge& $(2,0)_{\text{DT}}$&$(2,0)_{\text{DT}}$&$*$\\ 
                 &gen. DT invariant&0&0&$*$\\
                 &Stokes constant &0&0&0\\\hline
    
          $(-1,0,1,1)$ &charge& $(1,-1)_{\text{DT}}$&$(1,0)_{\text{D2D0}}$&$(1,-1)_{\text{D2D0}}$\\ 
          &gen. DT invariant&$0$&$n_{0,1}=1280$&$n_{0,1}$\\
          &Stokes constant &$n_{0,1}$&$n_{0,1}$&$n_{0,1}$\\\hline
           $-(-1,0,1,1)$ &charge& $(1,1)_{\text{PT}}$&$(-1,0)_{\text{D2D0}}$&$(-1,1)_{\text{D2D0}}$\\ 
          &gen. DT invariant&$n_{0,1}$&$n_{0,1}$&$n_{0,1}$\\
          &Stokes constant &$n_{0,1}$&$n_{0,1}$&$n_{0,1}$\\\hline
          $(-1,0,1,2)$ &charge& $(2,-1)_{\text{DT}}$&$(2,-1)_{\text{D2D0}}$& $(2,1)_{\text{DT}}$\\
                  &gen. DT invariant&0&$n_{0,2}=92288$& 92288\\
                   &Stokes constant&$*$&$*$& 0 \\\hline
          $-(-1,0,1,2)$ &charge& $(2,1)_{\text{PT}}$&$(-2,1)_{\text{D2D0}}$& $(2,-1)_{\text{PT}}$\\
                &gen. DT invariant&92288&$n_{0,2}$& 0 \\
                &Stokes constant &$*$&$*$& 0\\\hline
        \end{tabular}}
    \end{center}
    \caption[Borel singularities of $X_{4,2}$]{Leading Borel singularities of $X_{4,2}$ at $z=\infty$ and their corresponding invariants. As for all conifold points, $\bq_1 \cdot M_\infty = - \bq_1$.}
    \label{X42 table}
    \end{table}
\paragraph{Comments}
    \begin{itemize}
        \item The asymptotic behavior of $\cF_g$ given in \eqref{eq:gapX42X62} implies that the Stokes constant $S=2$ at $\aleph \bq_1 \cdot \Pi$ coincides for all associated multi-instanton singularities, located at $\ell \aleph \bq_1 \cdot \Pi$, $\ell \in \IN$.
        \item Even though the infinite order monodromy $M_\infty$ is not a symmetry of the Borel plane, the subleading singularities that we identify do lie in an $M_\infty$ orbit.
        \item We find no singularity in the Borel plane at the location of the second lightest period $\aleph \bq_2 \cdot \Pi$, consistent with the vanishing of the associated generalized DT invariant.
    \end{itemize}

\subsubsection{$X_{6,2}$} \label{s:X62}
\begin{figure}[H]
    \centering
    \includegraphics[trim = 0 50 0 45, clip,scale=1]{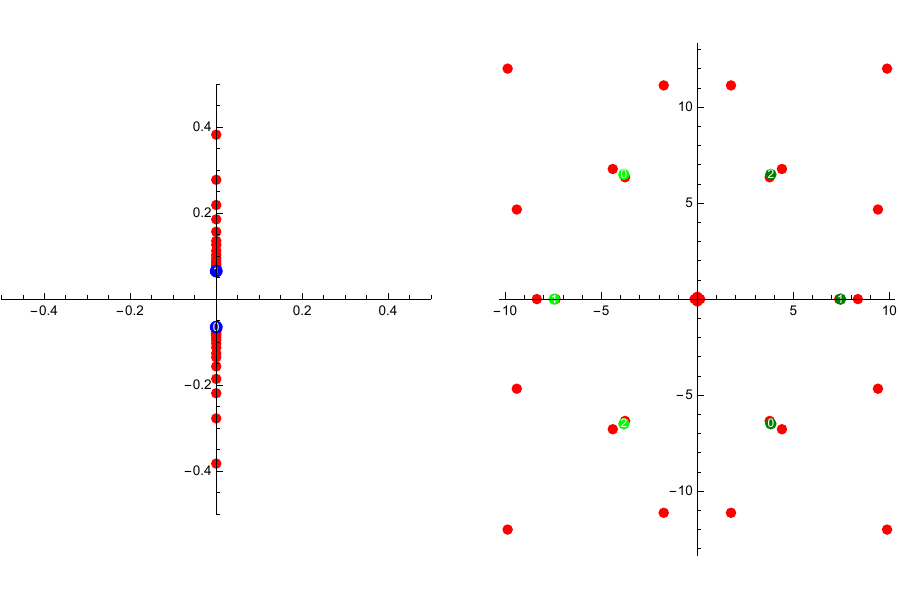}
    \caption{Borel plane of  $X_{6,2}$  at $z=10^6\mu$ in the frame $(\bq_1\cdot\periodV,P_1)$. On the right, the leading gap singularity is subtracted. The blue dots indicate the points $\bq_1 \cdot M_\infty^i$, $i=0,1$, the (light) green dots  the points $(-)(-1,0,1,1) \cdot M_\infty^i$, $i=0,\ldots,2$.}
    \label{fig:X62}
\end{figure}

 \begin{table}[H]
    \label{X62_table}
    \begin{center}
    \scalebox{0.9}{
        \begin{tabular}{|c| c | c | c | c|  }
          \hline  && $\pi$  & $\pi \cdot M_\infty$ & $\pi \cdot M_\infty^2$\\ \hline
          $\bq_1$ &charge&$*$ & $*$ & $*$\\
          &gen. DT invariant&$*$ & $*$ & $*$\\ 
          &Stokes constant &1 & 1 & 1\\ \hline
    $\bq_2$ &charge& $(1,0)_{\text{PT}}$ & $(1,0)_\PT $ & $*$\\ 
          &gen. DT invariant&0 & 0  & $*$\\
          &Stokes constant &0 & 0 & 0\\\hline
      $-\bq_2$ &charge& $(1,0)_{\text{DT}}$ & $(1,0)_{\text{DT}}$ & $*$\\ 
          &gen. DT invariant&0 & 0 & $*$\\
          &Stokes constant &0 & 0 & 0\\\hline
          $(-1,0,1,1)$ &charge& $(1,-1)_{\text{DT}}$ & $(1,0)_\GV$ & $(1,-1)_\GV$\\ 
               &gen. DT invariant& 0 & $n_{0,1}=4992$ & $n_{0,1}$ \\
              &Stokes constant &$*$ & $*$ & $*$\\\hline
          $-(-1,0,1,1)$ &charge& $(1,1)_{\text{PT}}$ & $(-1,0)_\GV$ & $(-1,1)_\GV$\\ 
                &gen. DT invariant& 4992 & $n_{0,1}$ & $n_{0,1}$ \\
                &Stokes constant &$*$ & $*$ & $*$\\\hline
        \end{tabular}}
    \end{center}
    \caption[Borel singularities of $X_{6,2}$]{Leading Borel singularities of $X_{6,2}$ at $z=\infty$ and their corresponding invariants. As for all conifold points, $\bq_1 \cdot M_\infty = - \bq_1$.}
    \label{tab:X62}
    \end{table}

\paragraph{Comments}
    \begin{itemize}
        \item The asymptotic behavior of $\cF_g$ given in \eqref{eq:gapX42X62} implies that the Stokes constant $S=1$ at $\aleph \bq_1 \cdot \Pi$ coincides for all associated multi-instanton singularities, located at $\ell \aleph \bq_1 \cdot \Pi$, $\ell \in \IN$.
        \item Even though the infinite order monodromy $M_\infty$ is not a symmetry of the Borel plane, the subleading singularities that we identify do lie in an $M_\infty$ orbit.
        \item We find no singularity in the Borel plane at the location of the second lightest period $\aleph \bq_2 \cdot \Pi$, consistent with the vanishing of the associated generalized DT invariant.
    \end{itemize}

\subsection{$X_{3,2,2}$}
\begin{figure}[H]
    \centering
 \includegraphics[trim=0 0 0 0, clip,scale=0.4]{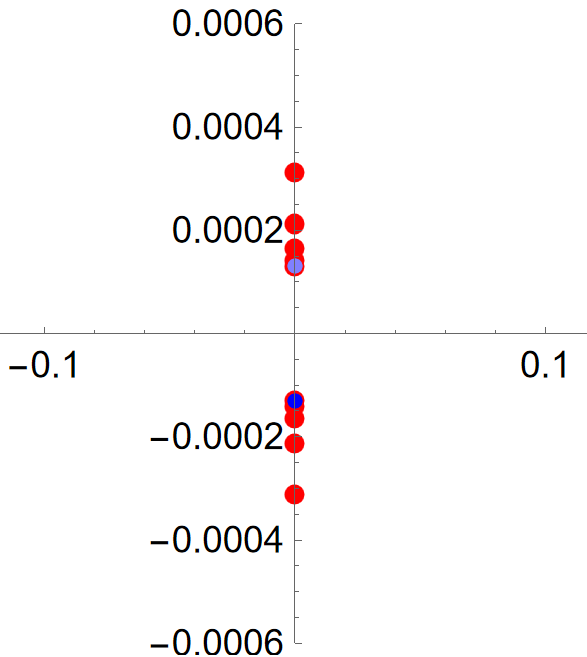}
  \includegraphics[trim=0 15 0 15, clip,scale=0.4]{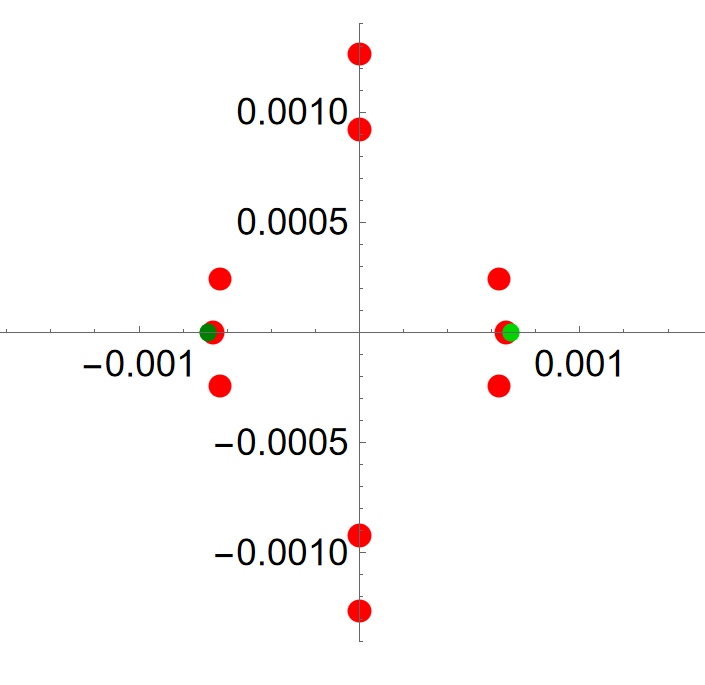}
    \caption{Borel plane of $X_{322}$  at $z=10^{12}\mu$ in the frame $(\periodV^{\bot},\bq_1\cdot\periodV)$. On the right, the contributions from $\pm\ell\aleph \bq_1\cdot\periodV$ are subtracted up to $\ell=6$. The (light) blue dot indicates the point at $(-)\aleph \bq_1\cdot\periodV$, the (light) green dot the point at $(-)\aleph \bq_2\cdot\periodV$.}
    \label{sub sing X322}
\end{figure}
   
\begin{table}[H]
    \label{X322 table}
    \begin{center}
        \begin{tabular}{|c| c | c |  }
          \hline  && $\pi$  \\ \hline
          $\pm\bq_1$ &charge& $*$\\
          &gen. DT invariant&$*$\\ 
          &Stokes constant &14\\ \hline
    $\pm2\bq_1$ &charge&$*$\\ 
          &gen. DT invariant&$*$\\
          &Stokes constant &$-2$\\\hline
        $\bq_2$ &charge& $(3,0)_{\PT}$\\ 
          &gen. DT invariant& 64\\
          &Stokes constant &$*$\\\hline
          $-\bq_2$ &charge& $(3,0)_{\DT}$\\ 
          &gen. DT invariant& 1238016\\
          &Stokes constant &$*$\\\hline
        \end{tabular}
    \end{center}
    \caption[Borel singularities of $X_{3,2,2}$]{Leading Borel singularities of $X_{3,2,2}$ at $z=\infty$ and their corresponding invariants}
    
\end{table}

\paragraph{Comments}
    \begin{itemize}
    \item For this model, the topological string amplitudes are only known up to $g_{\mathrm{max}}=14$, so our numerics are not stable. The singularity around the point $\aleph \bq_2\cdot\periodV$ appears when we remove only the first few of the multi-instanton contributions (up to $\ell=6$). 
    \end{itemize}

\subsection{M-point: $X_{2,2,2,2}$}
\begin{figure}[H]
    \centering
    \includegraphics[trim=20 100 20 100,clip,scale=0.35]{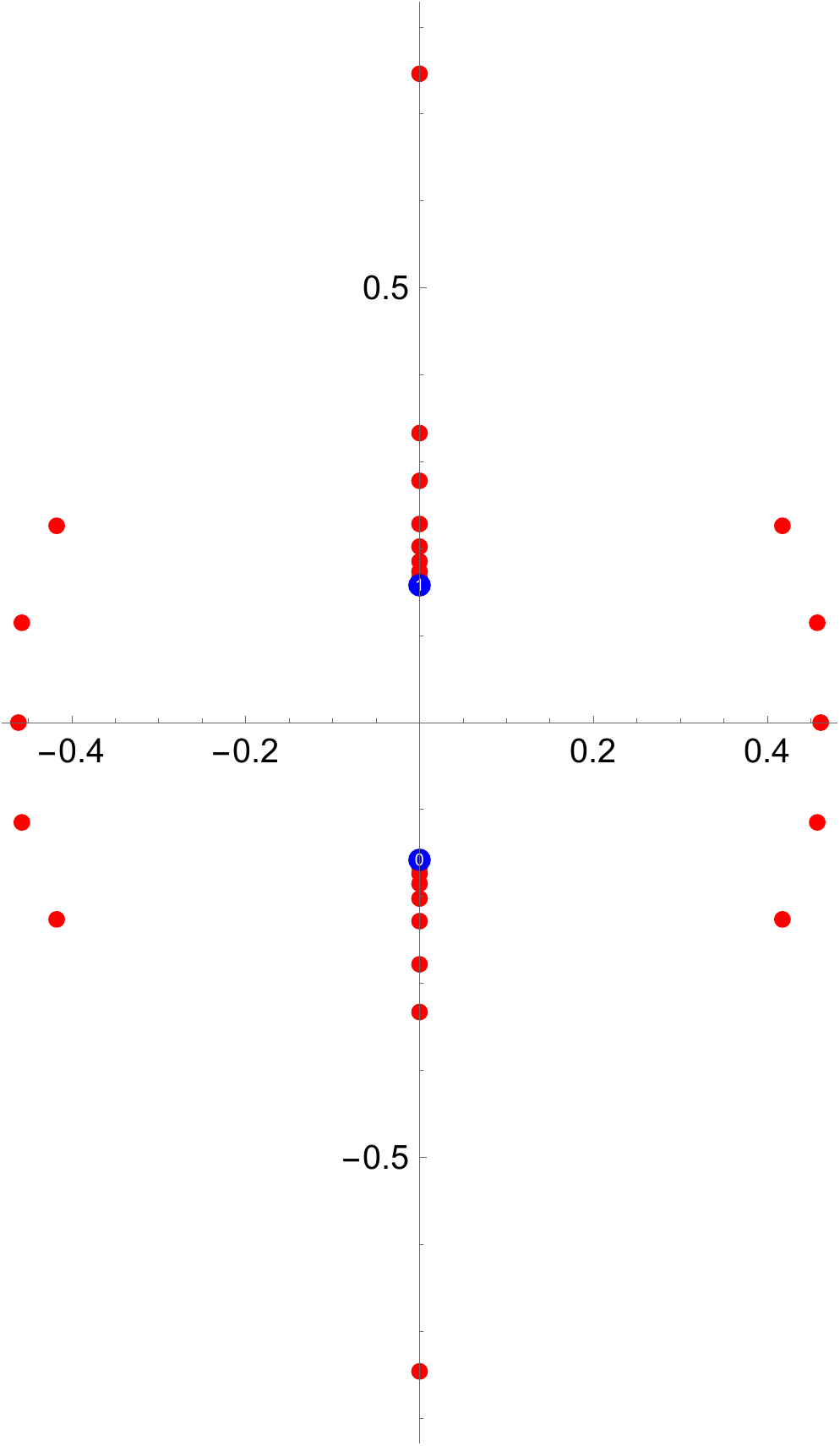}
    \includegraphics[scale=0.35]{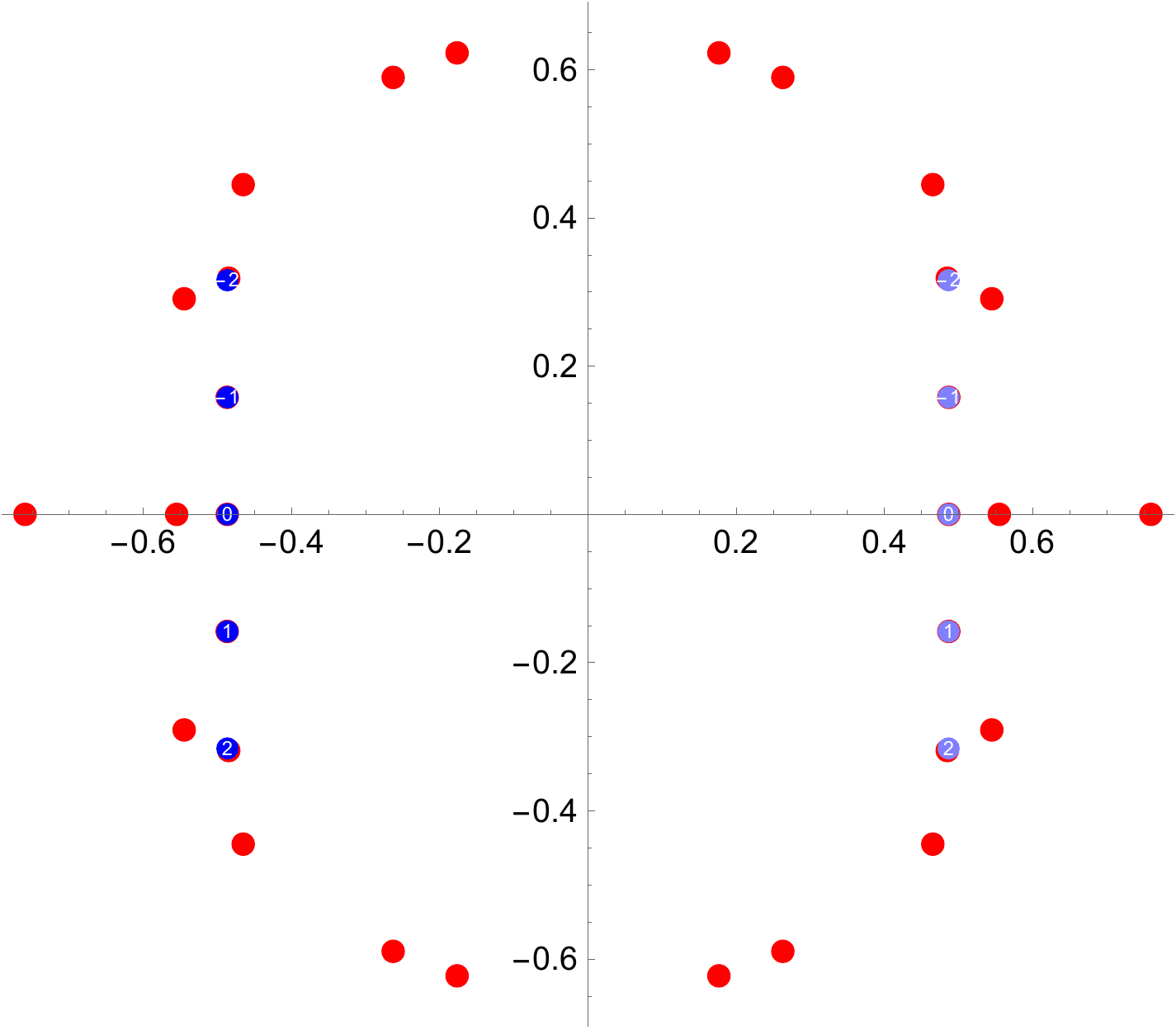}
    \caption{Borel plane of $X_{2,2,2,2}$ in the frame $(X^0_\fm,X^1_\fm$) at $z=10^6\mu$. On the right, the constant map contribution is subtracted. The blue dots on the left indicate the orbit elements of $\aleph X^0_\fm$. On the right, (light) blue dots indicate the points $\aleph((-)X^1_\fm+nX^0_\fm)$, with the factor $n$ indicated within the dot.}
    \label{Borel plane X2222}
\end{figure}
\begin{table}[H]
   \begin{center}
        \begin{tabular}{|c| c | c |  c | c |  c |  c | }
           \hline &$N$& $-2$ & $-1$ &$0$ & $1$ & $2$  \\ \hline
          $\bq_2+N\bq_1$ &charge& $*$&$(4,0)_{\text{DT}}$&$(4,0)_{\text{PT}}$&$*$&$*$\\
          &gen. DT invariant&$*$&$14752$&14752&$*$&$*$\\ 
          &Stokes constant &14752&14752&14752&14752&14752\\ \hline
     $-\bq_2+N\bq_1$ &charge&$*$&$*$& $(4,0)_{\text{DT}}$&$(4,0)_{\text{PT}}$&$*$\\ 
          &gen. DT invariant&$*$&$*$&14752&14752&$*$\\
          &Stokes constant &14752&14752&14752&14752&14752\\\hline
        \end{tabular}
            \caption[Borel singularities of $X_{2,2,2,2}$]{Subleading Borel singularities of $X_{2,2,2,2}$ at $z=\infty$ and their corresponding invariants.}
    \label{tab:X2222}
    \end{center}
    \end{table}

\paragraph{Comments}
\begin{itemize}
    \item As discussed in Section \ref{ss:BorelAndTopString}, for this model, we can predict the location of Borel singularities and the associated Stokes constant at $z=\infty$ using the same reasoning that applies to all MUM points at $z=0$. The role of the periods $X^0$ and $X^1$ is here played by $X^0_\fm = \bq_1 \cdot \piGeom$ and $X^1_\fm = \bq_2 \cdot \piGeom$.
    \item In Table \ref{tab:X2222}, we cite, as always, the generalized DT invariants of the model near $z=0$. This coincides with the torsion refined GV invariants $n_{0,1}^0$ and $n_{0,1}^1$ of the non-commutative resolution associated to this point in \cite{Katz:2022lyl}.
\end{itemize}

\subsection{Searching in-between singular points }\label{s:beyondSingular}
Up to this point, we have studied the asymptotics of topological string amplitudes close to singular points in the moduli space. We have seen that periods that fall off to zero faster than generically play a special role: as long as the generalized DT invariant associated to their charge is non-vanishing, they contribute the leading singularities in the Borel plane. As discussed in Section \ref{ss:dominatingContributions}, certain geometries exhibit rank 2 attractor points outside of the set $(0,\mu,\infty)$. 

Such rank 2 attractor points are identified in the two hypergeometric models $X_{4,3}$ and $X_{3,3}$ in \cite[Table 1, operators 4 and 11]{Bonisch:2022slo}.

\begin{figure}[h!!]
    \centering
    \includegraphics[scale=0.4]{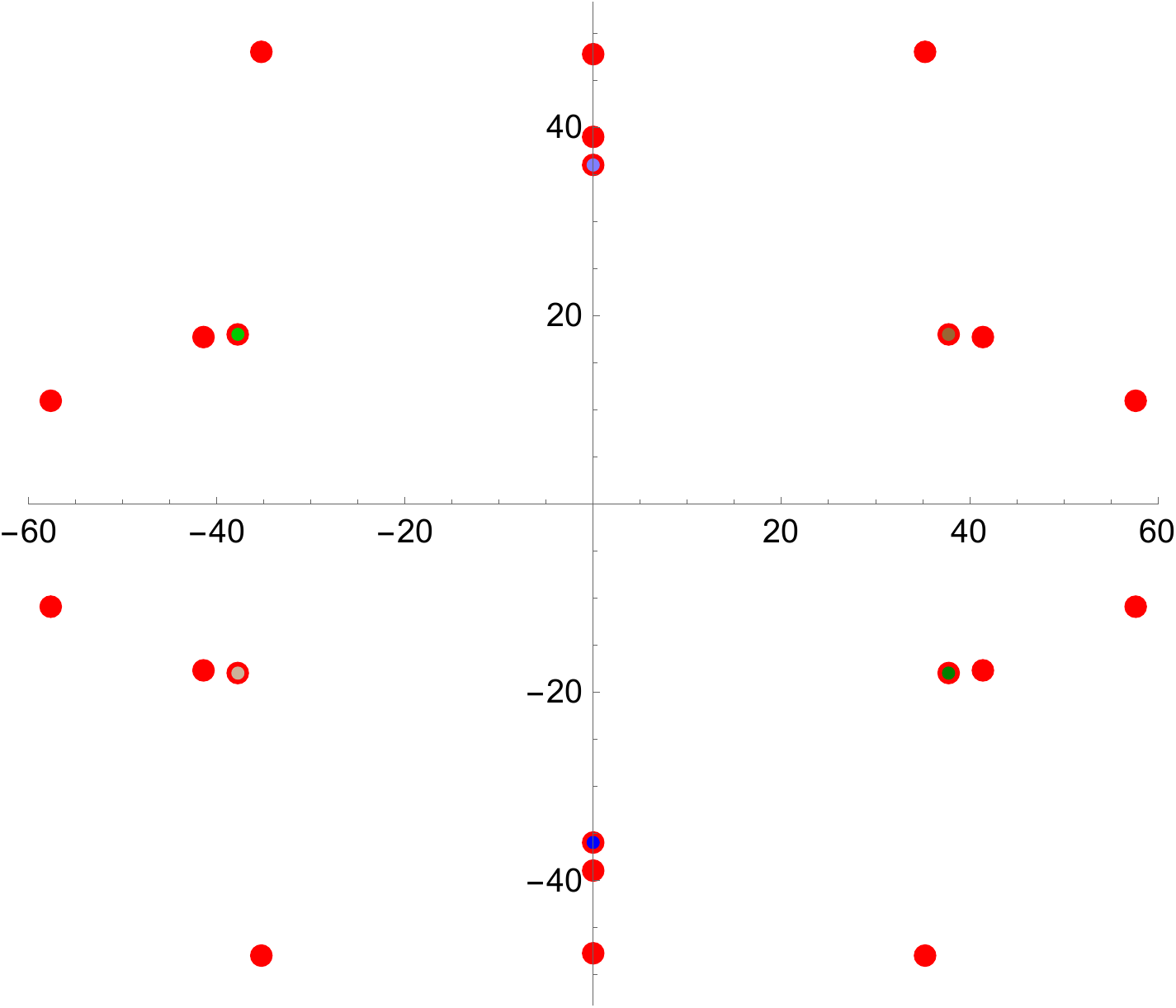}
    \caption{Borel plane of $X_{4,3}$ at $z=-4\mu$ in the frame $(P_0-X^0,X^1)$. The (light) blue dot indicates the point $(-)\aleph(P_0-X^0)$, the 
    (light) green dot the point $(-)\aleph(-P_0+X^0+X^1)$, and the (light) brown dot the point $(-)\aleph X^1$.}
    \label{fig: X43 attractor}
\end{figure}

\begin{figure}
     \centering
     \includegraphics[scale=0.4]{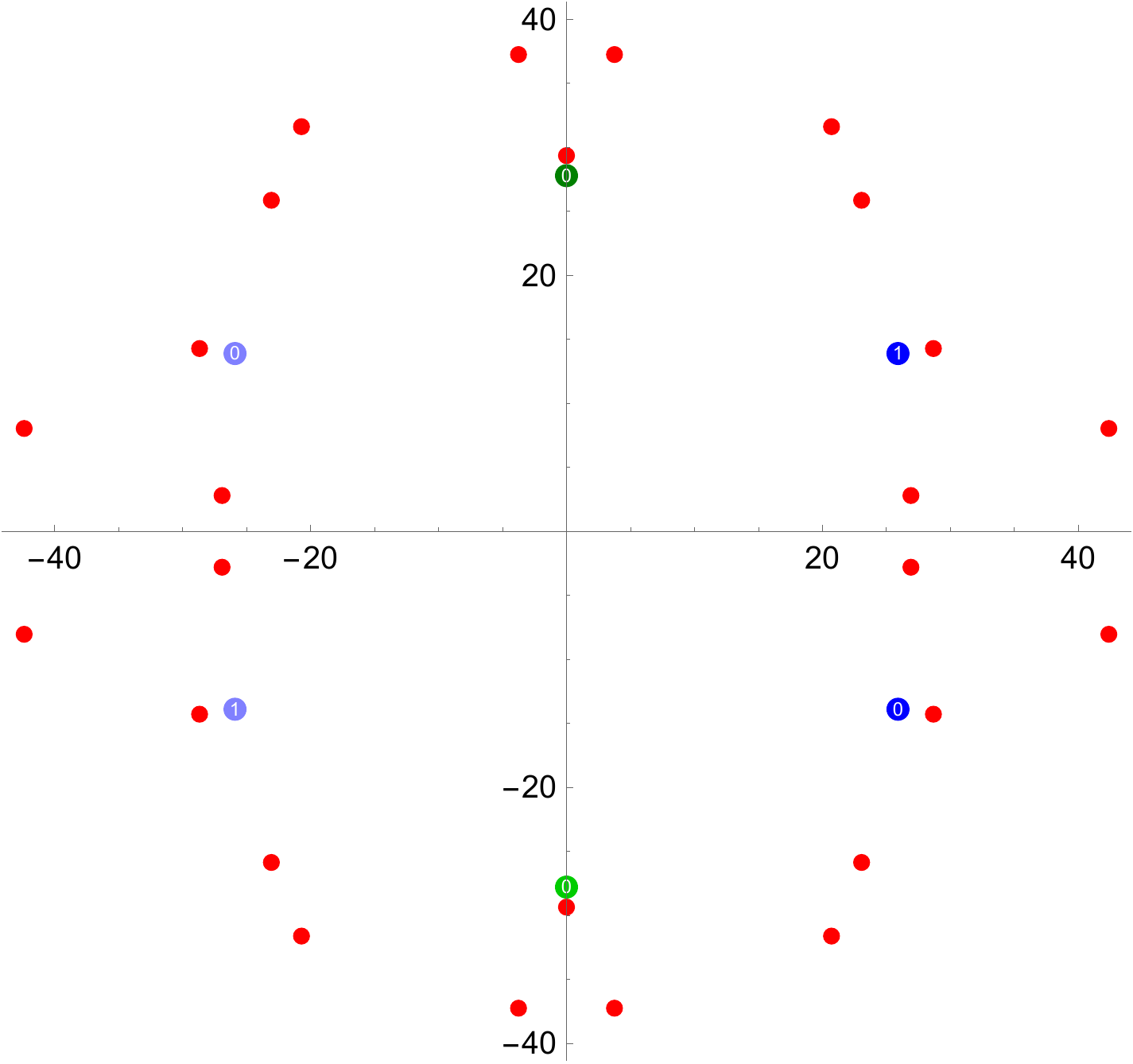}
     \caption{Borel plane of $X_{4,4}$ at $z=-100\mu$ in the frame $(-P_0+X^0,X^1)$. The (light) green dot indicates the point $(-)\aleph(-P_0+X^0)$, the (light) blue dots are the points ${(-)\aleph (-1,0,1,1).M_{\infty}^i\cdot\periodV},\, {i=0,1}$.}
     \label{fig:X44 -100mu}
 \end{figure}

$X_{4,3}$ possesses an attractor point at $z_*=-\frac{1}{432} =-4 \mu$, at which 
\begin{equation}
   \Lambda^\perp=\mathbb{Z} (0,1,3,0)+\mathbb{Z}(2,0,-1,-1)\,.
\end{equation}
The Borel plane at $z_*$ is depicted in Figure \ref{fig: X43 attractor}, and the leading Borel singularities and their associated Stokes constants are listed in Table \ref{table : X43 table in between point}.
The vanishing periods do not appear to play any special role in the Borel plane at this point, or slightly away from it. The generalized DT invariants associated to the lattice $\Lambda^\perp$ are of rank 2, hence not available, or of MSW type, which do not appear to predict Stokes constants away from the MUM point. 
  \begin{table}[H]
    \begin{center}
        \begin{tabular}{|c| c | c | c |  }
           \hline && $\pi$ &$-\pi$ \\ \hline
          $(-1,0,1,0)$ &charge& $(0,-1)_{\DT}$&$(0,1)_{\PT}$\\
          &gen. DT invariant&0&$-\chi=156$\\ 
          &Stokes constant &$-\chi$&$-\chi$\\ \hline
    $(-1,0,1,1)$ &charge& $(1,-1)_{\text{DT}}$&$(1,1)_{\text{PT}}$\\ 
          &gen. DT invariant&0&$n_{0,1}=1944$\\
          &Stokes constant &$n_{0,1}$&$n_{0,1}$\\\hline
           $(0,0,0,1)$ &charge& $(0,1)_{\text{D2D0}}$&$(0,-1)_{\text{D2D0}}$\\
          &gen. DT invariant&$n_{0,1}$&$n_{0,1}$\\
          &Stokes constant &$n_{0,1}$&$n_{0,1}$\\\hline
        \end{tabular}
        \caption[Borel singularities of $X_{4,3}$ at attractor point]{ Leading Borel singularities of $X_{4,3}$ at $z=-4\mu$ and their corresponding invariants.}
        \label{table : X43 table in between point}
    \end{center}
\end{table}
To verify that the attractor point indeed does not leave an imprint on the Borel plane, we studied several other models at similar points in moduli space, between $10\mu$ and $100\mu$. We find a similar pattern of Borel singularities as for $X_{4,3}$. For instance, for $X_{4,4}$ at $z=-100\mu$, we find the same  Borel singularities as $X_{4,3}$, see Figure \ref{fig:X44 -100mu}, and Table \ref{tab:X44beyond} for the associated Stokes constants.
  \begin{table}[H]
    \begin{center}
        \begin{tabular}{|c| c | c | c |   }
            \hline&& $\pi$&$-\pi$  \\ \hline
          $(-1,0,1,0)$ &charge& $(0,-1)_{\DT}$&$(0,1)_{\PT}$\\
          &gen. DT invariant&0&$144=-\chi$\\ 
          &Stokes constant &144&144\\ \hline
    $(-1,0,1,1)$ &charge& $(1,-1)_{\text{DT}}$&$(1,1)_{\text{PT}}$\\ 
          &gen. DT invariant&0&$3712=n_{0,1}$\\
          &Stokes constant &3712&3712\\\hline
           $(0,0,0,1)$ &charge& $(0,1)_{\text{D2D0}}$&$(0,-1)_{\text{D2D0}}$\\ 
          &gen. DT invariant&$3712=n_{0,1}$&$3712=n_{0,1}$\\
          &Stokes constant &3712&3712\\\hline
        \end{tabular}
        \caption[Borel singularities of $X_{4,4}$ between $\mu$ and $\infty$]{ Leading Borel singularities of $X_{4,4}$ at $z=-100\mu$ and their corresponding invariants.}
        \label{tab:X44beyond}
    \end{center}
\end{table}
For the Calabi-Yau threefolds $X_{3,3}$ and $X_{6,6}$, we record the positions of singularities and the associated Stokes constants in Table \ref{tab:X33beyond} and Table \ref{tab:X66beyond}.
  \begin{table}[H]
    \label{table : X33 table in between point}
    \begin{center}
        \begin{tabular}{|c| c | c |  c | }
          \hline  && $\pi$ &$-\pi$ \\ \hline
          $(0,0,-1,1)$ &charge& $(-1,1)_{\text{D0D2}}$&$(1,-1)_{\text{D0D2}}$\\
          &gen. DT invariant&$1053=n_{0,1}$&1053\\ 
          &Stokes constant &1053&1053\\ \hline
    $(-1,0,1,1)$ &charge& $(1,-1)_{\text{DT}}$&$(1,1)_{\text{PT}}$\\ 
          &gen. DT invariant&0&$1053=n_{0,1}$\\
          &Stokes constant &1053&1053\\\hline
           $(0,0,0,1)$ &charge& $(0,1)_{\text{D0D2}}$&$(0,-1)_{\text{D0D2}}$\\ 
          &gen. DT invariant&$1053=n_{0,1}$&1053\\
          &Stokes constant &1053&1053\\\hline
        \end{tabular}
        \caption[Borel singularities of $X_{3,3}$ between $\mu$ and $\infty$]{ Leading Borel singularities of $X_{3,3}$ at $z=100\mu$ and their corresponding invariants.}
        \label{tab:X33beyond}
    \end{center}
    \end{table}
      \begin{table}[h!]
    \begin{center}
        \begin{tabular}{|c| c | c |c |  }
          \hline  && $\pi$&$-\pi$  \\ \hline
          $(1,0,0,0)$ &charge& $(0,0)_{\PT}$&$(0,0)_{\DT}$\\
          &gen. DT invariant&$1$&1\\ 
          &Stokes constant &1&1\\ \hline
    $(1,1,-2,-1)$ &charge& $(1,1)_{\text{PT}}$&$(1,-1)_{\DT}$\\ 
          &gen. DT invariant&$67105$&1\\
          &Stokes constant &1&1\\\hline
        \end{tabular}
        \caption[Borel singularities of $X_{6,6}$ between $\mu$ and $\infty$]{ Leading Borel singularities of $X_{6,6}$ at $z=-10\mu$ and their corresponding invariants.}
         \label{tab:X66beyond}
    \end{center}
    \end{table}

The rank 2 attractor point of the Calabi-Yau threefold $X_{3,3}$ lies at $z_*=-\frac{1}{5832}= -\frac{1}{8}\mu$, at which
\begin{equation}
   \Lambda^\perp=\mathbb{Z} (1,0,3,0)+\mathbb{Z}(4,1,0,4)\,.
\end{equation}
Again, this lattice does not appear to play a special role in the Borel plane. Note that the charge associated to the first generator of the lattice is $(0,-3)_{\PT}$, with vanishing associated generalized DT invariant.

$z_*$ lies close to the MUM point, and indeed, the singularities we observe in the Borel plane are those predicted by equation \eqref{eq:MumAsymptotics} for singularities near the MUM point.

\newpage

\section{Conclusions} \label{s:conclusions}
Our numerical results clearly demonstrate that there is a close relationship between generalized DT invariants, at least at rank $\pm 1$ and at rank $0$ with $0$ D4 brane charge, and the resurgence data governing topological string amplitudes.

A necessary next step is to investigate whether discrepancies between Stokes constants in the vicinity of $z=\infty$ and generalized DT invariants computed at $z=0$ are indeed due to wall crossing phenomena, or whether a more intricate map between BPS data and the value of Stokes constants is at play. We are currently attempting to track singularities along paths from $z=0$ to $z=\infty$ to shine light on this question. If the relation between BPS invariants and Stokes constants proves direct or can be unraveled, then studying the asymptotics of topological string amplitudes could serve as a computational approach towards determining generalized DT invariants anywhere on moduli space. Whether or not monodromies are a symmetry of the Borel plane should serve as a useful indication regarding the existence of walls of marginal stability.

The ultimate goal, of course, is to arrive at a precise mathematical derivation of the relation between BPS invariants and resurgence data. A better understanding of the anholomorphic topological string amplitudes and the still somewhat mysterious choice of holomorphic limits will perhaps need to precede any progress in this direction. 

\section*{Acknowledgements}
We would like to thank Albrecht Klemm, Marcos Mari\~no and Boris Pioline for useful discussions.  
\vspace{0.3cm}

\noindent
The work of the authors is supported under ANR grant ANR-21-CE31-0021.

\bibliographystyle{JHEP}
\bibliography{biblio}
\end{document}